\documentclass[preprint,pteplogo]{ptephy_v2}

\preprintnumber{XXXX-XXXX} 
\usepackage{hyperref}

\usepackage{amssymb,amsmath}
\usepackage{color}
\usepackage{colortbl}
\usepackage{enumitem}
\usepackage{array}
\usepackage{braket}
\usepackage{wrapfig}
\usepackage{cancel}
\usepackage{float}
\usepackage{slashed}
\usepackage{subcaption}
\usepackage{soul}
\usepackage{cases}

\usepackage{xcolor}
\usepackage[normalem]{ulem}

\begin{document}

\title{Investigating New Physics through the Observables of Semileptonic $B_{s}\to K^{\ast}(\to K \pi)\mu^{+}\mu^{-}$ Decay}

\author[1]{Zohaib Aarfi}
\affil{Department of Physics, School of Natural Sciences, National University of Sciences and Technology, H-12, Islamabad, Pakistan}
\author[2,3]{Qazi Maaz Us Salam}
\affil{School of Science and Engineering, Lahore University of Management Sciences (LUMS),
 Opposite Sector U, D.H.A, Lahore 54792, Pakistan}
\author[3]{Ishtiaq Ahmed}
\affil{National Centre for Physics, Shahdra Valley Road, Islamabad 44000, Pakistan \email{zohaib.phdphy22sns@student.nust.edu.pk,
qazi.salam@lums.edu.pk, ishtiaq.ahmed@ncp.edu.pk,
faisal.munir@sns.nust.edu.pk, rizwan\_khalid@lums.edu.pk,
aliparacha@sns.nust.edu.pk}}
\author[1]{Faisal Munir Bhutta}
 \author[2]{Rizwan Khalid}
 \author[1]{M.Ali Paracha}

\begin{abstract}%
The rare four-fold decay $B_s \to K^*(\to K\pi)\mu^+\mu^- $, governed by the flavor-changing neutral current transition $b \to d\mu^+\mu^-$, provides a sensitive probe for testing the Standard Model (SM) and investigating signatures of new physics (NP). This work presents a comprehensive model-independent analysis of the decay using the framework of the weak effective field theory. We compute a set of key physical observables, including the differential branching ratio, forward-backward asymmetry, longitudinal polarization fraction, and several normalized angular coefficients $\langle I_i\rangle$, as a function of the dilepton invariant mass squared $q^2$. The impact of NP is explored via both one-dimensional (1D) and two-dimensional (2D) scenarios involving NP Wilson coefficients $C_7^{\text{NP}}$, $C_9^{(\prime)\text{NP}}$, and $C_{10}^{(\prime)\text{NP}}$. Our findings reveal notable deviations from the SM predictions across multiple observables. Furthermore, we analyze the correlations between different observables and their 2D contour plots which would be useful to further constrain the parametric space of possible NP contributions. This study reinforces the potential of $B_s \to K^* \mu^+ \mu^-$ decay as a complementary channel in the search for physics beyond the SM.
\end{abstract}

\subjectindex{xxxx, xxx}

\maketitle
\section{Introduction}
\raggedbottom
The study of rare decays mediated by flavor-changing neutral current~(FCNC) transitions plays a pivotal role not only to test the 
standard model of particle physics~(SM) but also to analyze the signatures of NP effects beyond the SM. Most of the research in this sector focuses on the detailed analysis of $B$ meson decays as these exhibit a rich phenomenology. Furthermore, from the NP perspective the rare decays of $B$ mesons, in particular the decays with quark level transitions $b\to s,d$ have attracted significant attention, as these decays are suppressed via the GIM mechanism in the SM \cite{Glashow:1970gm,Descotes-Genon:2013wba, Descotes-Genon:2015uva, Alok:2011gv, Ghosh:2014awa, Buras:2010zm, Fleischer:2017ltw}. In particular, rare semileptonic $B$ meson decays can serve as a sensitive probe to explore the structure of NP through lepton flavor dependent (LFD) and lepton flavor universal (LFU) observables~\cite{Dib:2023nln, Blake:2016olu}.

In this context, FCNC processes such as $b\to s\ell^{+}\ell^{-}$ have been extensively investigated both experimentally and theoretically \cite{Hiller:2003js, Bobeth:2011ka, LHCb:2022qnv, LHCb:2014set, LHCb:2021lvy, BaBar:2012mrf, BELLE:2019xld, Capdevila:2018jhy, Ali:2025xkw,Fukae:1998qy,Fukae:1999ww}. Several observables such as LFU ratios, isospin asymmetries, forward-backward asymmetries ($A_{\text{FB}}$)~\cite{Ali:1991is}, branching fractions, and angular coefficients have been studied in detail. In addition, the helicity fraction and lepton polarization asymmetries in rare B decays have been extensively explored in both inclusive and exclusive channels~\cite{Hewett:1995dk,Geng:1996az}. Notably, the transition $ b \to s \mu^+ \mu^- $ exhibits persistent tensions with SM predictions, often referred to as flavor anomalies~\cite{Descotes-Genon:2015uva,Descotes-Genon:2020buf,London:2021lfn,Hurth:2023jwr}.  For instance, angular observables in $ B \to K^* \mu^+ \mu^- $ decays show significant deviations from SM expectations~\cite{Descotes-Genon:2012isb, Descotes-Genon:2013vna}, and global fits to Wilson coefficients (WCs) suggest possible NP contributions affecting LFD observables. Additional decay modes such as $ B \to K_1(1270,1400)\mu^+\mu^- $~\cite{Ishaq:2013toa,Huang:2018rys,Munir:2015gsp,MunirBhutta:2020ber,Bhutta:2024zwj}, $ B \to K_2^* \mu^+\mu^- $~\cite{Das:2018orb,Mohapatra:2021izl}, $ B_s \to f^\prime_2 \mu^+\mu^- $~\cite{Rajeev:2020aut}, and $ B_c \to D_s^{(*)} \mu^+\mu^- $~\cite{Dutta:2019wxo,Mohapatra:2021ynn,Zaki:2023mcw,Li:2023mrj,Salam:2024nfv} have also been explored to further constrain NP effects.

Despite their potential, $ b \to d \ell^+ \ell^- $ decays remain comparatively less studied due to stronger CKM suppression by the factor $ |V_{td}/V_{ts}|^2 \approx 0.04 $ \cite{ParticleDataGroup:2024cfk} and significant experimental challenges arising from the lower expected branching ratios~\cite{BaBar:2013qaj,Belle:2024cis}. Theoretical predictions place these branching fractions at the level of $ \mathcal{O}(10^{-8}) $ or smaller~\cite{Wang:2007sp, Song:2008zzc, Wang:2012ab,Farooq:2024owx}, posing a formidable challenge for current detectors. However, the $b \to d$ transitions receive contributions from both $c\bar{c}$ and $u\bar{u}$ loop amplitudes with CKM factors $V_{cb}V^{\ast}_{cd}$ and $V_{ub}V^{\ast}_{ud}$ of order $\lambda^{3}$~\cite{Kruger:1996dt}. Because these amplitudes have comparable magnitudes but distinct weak phases, the interference between them can induce sizable CP-violating effects. Consequently, CP violation in $b \to d$ decays is expected to be  larger than in the corresponding $b \to s$ transitions. Hence the decays involving $b \to d\ell^{+}\ell^{-}$ transitions are valuable compared to that of $b\to s\ell^{+}\ell^{-}$ \cite{Paracha:2020gws}.

In addition, studying $b \to d\,\ell^{+}\ell^{-}$ processes provides a complementary probe of the flavour sector. 
Since both channels are governed by the same short-distance operators, a combined analysis of 
$b \to s$ and $b \to d$ transitions enables stringent tests of the CKM unitarity and potential sources 
of flavour non-universality. Furthermore, the stronger interference between up- and charm-quark loops in 
$b \to d$ decays enhances sensitivity to new CP-violating phases beyond the Standard Model. 
Therefore, our study not only extends the precision framework developed for 
$b \to s\,\ell^{+}\ell^{-}$ to the more suppressed 
$b \to d\,\ell^{+}\ell^{-}$ sector, but also highlights its unique role as a complementary and potentially more 
sensitive probe of CP violation and new physics in flavour-changing neutral currents \cite{Biswas:2022lhu, Bause:2022rrs, Handoko:1997gd}.

The advent of high-luminosity experiments such as Belle II and the ongoing upgrades at LHCb significantly enhance the prospects for observing these rare decays and performing precision measurements~\cite{Belle2WhitePaper2018}. LHCb has reported the evidence for the $B_s \to \bar{K}^* \mu^+ \mu^-$ with a branching ratio of about $(2.9\pm 1.1) \times 10^{-8}$ \cite{LHCb:2018rym}. Despite the experimental challenges, rare $ b \to d \ell^+ \ell^- $ decays can provide valuable complementary information to $ b \to s \ell^+ \ell^- $ modes in testing the physical observables to explore the potential NP contributions~\cite{Bause:2022rrs, Alok:2019xub, Becirevic:2012fy}. Furthermore, analyzing different decay channels provides complementary information that helps to resolve degeneracies in global fits of the WCs, hence improving the precision and robustness of constraints on NP models~\cite{Becirevic:2012fy, Hurth:2016fbr}. 

Experimentally, the Belle, BABAR, and LHCb collaborations have made contributions to the study of $ b \to d \ell^+ \ell^- $ transitions~\cite{BaBar:2013qaj, Belle:2008tjs}by reporting the branching fractions for the decay modes: $ B^+ \to \pi^+ \mu^+ \mu^- $, $ B^0 \to \pi^+ \pi^- \mu^+ \mu^- $, and $ B^0 \to \rho^0 \mu^+ \mu^- $, with measured values $ (1.78 \pm 0.23) \times 10^{-8} $, $ (2.11 \pm 0.52) \times 10^{-8} $, and $ (1.98 \pm 0.53) \times 10^{-8} $, respectively~\cite{LHCb:2015hsa}. Recently, Belle has extended the search to $ B \to (\eta, \omega, \rho, \pi) \ell^+ \ell^- $ modes, setting upper limits in the range of $ (3.8~\text{-}~47) \times 10^{-8}$ at the 90\% confidence level~\cite{Belle:2024cis}. Future measurements at Belle II and LHCb are expected to significantly reduce uncertainties in angular observables and branching ratios~\cite{Belle2WhitePaper2018}.

Model-independent analyses of effective field theories (EFT) facilitate the study of various $b \to d$ observables, providing deeper insight into the flavor structure of NP in both the quark and lepton sectors. The goal of this work is to use the model-independent Hamiltonian and perform the four-fold angular analysis of $B_s \to K^{\ast}(\to K\pi) \mu^+ \mu^-$ decays. For this decay, we use the fit results from simplified series expansion (SSE) coefficients in the fit to light cone sum rules (LCSR) and lattice form factors \cite{Bharucha:2015bzk}. The decay is analyzed in the low $q^{2}$ region of the spectrum. In this work, we choose the fitted values of the WCs from \cite{Bause:2022rrs} and give the predictions of different physical observables such as differential branching fractions, lepton forward-backward asymmetry, longitudinal helicity fractions, and normalized angular coefficients within the SM and model-independent NP scenarios. Moreover, to constrain the NP scenarios we also give the plots of physical observables versus the NP WCs, as well as correlation plots between observables in both 1D and 2D NP scenarios.

We would like to point out, similar to $b\to s$ transition, that the nearly 100\% branching ratio of $K^*(892) \to K\pi$ provides significant statistical advantages in the study of $b \to d$ transitions, such as $B_s \to K^* \ell^+ \ell^-$ decays, by maximizing the reconstruction efficiency and signal yield due to the dominance of this decay channel. This high branching fraction ensures that almost every $K^*$ produced in $B$-meson decays yields a detectable $K\pi$ pair, improving the precision of measurements for branching fractions, angular observables and lepton universality ratios ($R_{K^*}$) in rare $b \to d$ processes, which are inherently suppressed compared to $b \to s$ transitions by CKM factors \cite{LHCb:2015svh, Belle2016}. The narrow width of the $K^*(892)$ ($\sim$50 MeV) further enhances the signal-to-background ratio by allowing a clean invariant mass selection, reducing combinatorial backgrounds from unrelated $K\pi$ pairs \cite{PDG2023}. Experimentally, this has been crucial in LHCb and Belle II analyses, where high-precision measurements rely on large $K^* \to K\pi$ samples to constrain SM predictions and search for NP \cite{LHCb2022,BelleII2021}. Theoretically, the dominance of this decay channel simplifies amplitude analyses and reduces systematic uncertainties in form factor determinations, as discussed in QCD factorization and lattice QCD studies \cite{Khodjamirian2010,HPQCD2016}. Thus, the near-unity branching ratio of $K^* \to K\pi$ plays a pivotal role in enhancing the sensitivity of $b \to d$ transition studies, making it indispensable for tests of flavor physics and beyond standard model (BSM) and NP scenarios.

The organization of this work is as follows. In Section~\ref{framework}, we present the effective Hamiltonian in the model-independent framework which is used to compute the amplitude of the decay in terms of the hadronic matrix elements. Further, employing the helicity formalism, we have given the four-fold angular distribution of the $B_{s}\to K^{\ast}(\to K\pi)\mu^{+}\mu^{-}$ decay in terms of the angular coefficients which depend on the helicity amplitudes. Finally, the semi-analytical expressions of the physical observables are given in terms of the angular coefficients, which can be written in terms of NP WCs. In Section~\ref{NA}, we present the phenomenological analysis of the physical observables in the SM and in selected 1D and 2D model-independent NP scenarios. Finally, in Section~\ref{conl}, we summarize the main findings of our study.

{
\section{ Theoretical Framework}\label{framework}
In this section, we present the weak EFT framework to study the $B_{s}\to K^{\ast}\mu^{+}\mu^{-}$ decay, driven by the quark level $b\to d\mu^{+}\mu^{-}$ transition. The EFT is described by an effective Hamiltonian which can be used to compute the four-fold angular distribution of the $B_{s}\to K^{\ast}(\to K\pi)\mu^{+}\mu^{-}$ decay. From the angular distribution, we extract the $q^{2}$ dependent physical observables which are then used to analyze the signatures of NP in the model-independent scenarios.

\subsection[\texorpdfstring{Effective Hamiltonian and Decay Amplitude for \(B_s \to K^{\ast} \mu^{+} \mu^{-}\)}{Effective Hamiltonian and Decay Amplitude for Bs to K* mu+ mu-}]
{Effective Hamiltonian and Decay Amplitude for \(B_{s}\to K^{\ast}\mu^{+}\mu^{-}\) Decay}
The low energy effective Hamiltonian for $B^{0}_{s}\to K^{\ast}\mu^{+}\mu^{-}$ decay mode, whose quark level transition is $b\to d\mu^{+}\mu^{-}$ in the framework of SM and beyond can be expressed as~\cite{Bause:2022rrs},
\begin{eqnarray}
\mathcal{H}_{\text{eff}}=-\frac{4G_{F}}{\sqrt{2}}V_{tb}V^{\ast}_{td}\left[\mathcal{H}^{(t)}_{\text{eff}}+\lambda_{u}^{(d)}\mathcal{H}^{(u)}_{\text{eff}}\right]+\text{h.c.},\label{Heff}
\end{eqnarray}
where  $G_{F}$ is the Fermi coupling constant, $V_{tb}V^{\ast}_{td}$ are the corresponding CKM factors and $\lambda_{u}^{(d)}=\frac{V_{ub}V^{\ast}_{ud}}{V_{tb}V^{\ast}_{td}}$ is the ratio of the CKM factors. The explicit form of $\mathcal{H}^{(t)}_{\text{eff}}$ and $\mathcal{H}^{(u)}_{\text{eff}}$ can be written as,
\begin{eqnarray}
\mathcal{H}^{(t)}_{\text{eff}}=C_{1}O^{c}_{1}+C_{2}O^{c}_{2}+\sum_{i=3}^{6}C_{i}O_{i}+
\sum_{i=7,9,10}\left((C_{i}+C_{i}^{NP})O_{i}+C_{i}^{\,'NP}O_{i}^{'}\right),\label{Heff1}
\end{eqnarray}
and
\begin{eqnarray}
\mathcal{H}^{(u)}_{\text{eff}}=C_{1}(O^{c}_{1}-O^{u}_{1})+C_{2}(O^{c}_{2}-O^{u}_{2}).\label{Heff2}
\end{eqnarray}
In Eqs.~(\ref{Heff1}) and (\ref{Heff2}), the short distance contributions are encoded in the SM WCs $(C_{i})$ and the NP WCs $(C_{i}^{'NP})$, whereas the long distance effects are hidden in the local 4-fermion operators. The explicit form of the 4-fermion operators that contribute to the decay under consideration can be written as,
\begin{align}\label{op1}
O_{7} &=\frac{e}{16\pi ^{2}}m_{b}\left( \bar{d}\sigma _{\mu \nu }P_{R}b\right) F^{\mu \nu },
&  O_{7}^{\,\prime} &=\frac{e}{16\pi ^{2}}m_{b}\left( \bar{d}\sigma _{\mu \nu }P_{L}b\right) F^{\mu \nu },\notag\\
O_{9} &=\frac{e^{2}}{16\pi ^{2}}(\bar{d}\gamma _{\mu }P_{L}b)(\bar{\mu}\gamma^{\mu }\mu),
&  O_{9}^{\,\prime} &=\frac{e^{2}}{16\pi ^{2}}(\bar{d}\gamma _{\mu }P_{R}b)(\bar{\mu}\gamma^{\mu }\mu),\notag\\
O_{10} &=\frac{e^{2}}{16\pi ^{2}}(\bar{d}\gamma _{\mu }P_{L}b)(\bar{\mu} \gamma ^{\mu }\gamma _{5}\mu),
&  O_{10}^{\,\prime} &=\frac{e^{2}}{16\pi ^{2}}(\bar{d}\gamma _{\mu }P_{R}b)(\bar{\mu} \gamma ^{\mu }\gamma _{5} \mu).
\end{align}
In Eq. (\ref{op1}), $e$ is the electromagnetic coupling constant, $O_{i}^{\,\prime}$ with ($i=7,9,10$) are the chirality flipped operators, and $F^{\mu\nu}$ is the electromagnetic field strength tensor. It is important to mention here that $m_{b}$ appearing in the definition of the electromagnetic operator $O_{7}$ and its chirality flipped part is the running $b$ quark mass and is evaluated in the $\overline{\text{MS}}$ scheme. 

Using the effective Hamiltonian given in Eq.~(\ref{Heff1}), the amplitude for the $B_{s}\to K^{\ast}\mu^{+}\mu^{-}$ decay can be expressed as,
\begin{eqnarray}
\mathcal{M}\left(B_{s}\to K^{\ast}\mu^{+}\mu^{-}\right)=\frac{G_{F}\alpha}{2\sqrt{2}\pi}V_{tb}V^{\ast}_{td}\Big\{T^{1,K^{\ast}}_{\mu}(\bar{\mu}\gamma^{\mu}\mu)
+T^{2,K^{\ast}}_{\mu}(\bar{\mu}\gamma^{\mu}\gamma_{5}\mu)\Big\},\label{Amp1}
\end{eqnarray}
where
\begin{eqnarray}
T^{1,K^{\ast}}_{\mu}&=&(C_{9}^{\,eff}+C_{9}^{NP})\Big\langle K^{\ast}(k,\varepsilon)|\bar d\gamma_{\mu}(1-\gamma_{5})b|B_{s}(p)\Big\rangle\notag\\
&+&C_{9}^{\,'NP}\Big\langle K^{\ast}(k,\varepsilon)|\bar d\gamma_{\mu}(1+\gamma_{5})b|B_{s}(p)\Big\rangle\notag\\
&-&\frac{2m_{b}}{q^{2}}(C_{7}^{\,eff}+C_{7}^{NP})
\Big\langle K^{\ast}(k,\varepsilon)|\bar d i\sigma_{\mu\nu}q^{\nu}(1+\gamma_{5})b|B_{s}(p)\Big\rangle,\label{Amp1a}
\\
T^{2,K^{\ast}}_{\mu}&=&(C_{10}+C_{10}^{NP})\Big\langle K^{\ast}(k,\varepsilon)|\bar d\gamma_{\mu}(1-\gamma_{5})b|B_{s}(p)\Big\rangle\notag\\
&+&C_{10}^{\,'NP}\Big\langle K^{\ast}(k,\varepsilon)|\bar d\gamma_{\mu}(1+\gamma_{5})b|B_{s}(p)\Big\rangle.
\label{Amp1b}
\end{eqnarray}
The explicit form of WCs $C_{7}^{eff}(q^{2})$ and $C_{9}^{eff}(q^{2})$ \cite{Bobeth:1999mk,Beneke:2001at,Asatrian:2001de,Asatryan:2001zw,Greub:2008cy,Du:2015tda}, that contain the factorizable contributions from current-current, QCD penguins and chromomagnetic dipole operators $O_{1-6,8}$ is given in Appendix \ref{append123}.   

 It is important to mention here that in our current analysis, we did not incorporate the long-distance contributions from the $\rho,\omega,\phi,\psi$ and $\psi^{\prime}$resonances, as their effects are minimal below $q^{2}\approx (6-7)\text{GeV}^{2}$ \cite{PhysRevD.93.014028, Corley:2001hg,Deshpande:1988bd, Lim:1988yu, Kruger:1996cv}.The impact of these resonances on the branching ratio of the decay $B_{s}\to K^{\ast}(K\pi)\mu^{+}\mu^{-}$ in the SM is shown in Fig.~\ref{reseff} in Appendix \ref{append123}, which complements the work presented in \cite{Khodjamirian:2010vf}. Hence the effect of the above mentioned resonances can safely be neglected in the angular observables as well \cite{PhysRevD.93.014028, Jager:2012uw}.

\subsection{Matrix Elements for the \texorpdfstring{$B_{s}\to K^{\ast}\mu^{+}\mu^{-}$}{Bs to K* mu+ mu-} decay}\label{ffP}
For the decay $B_{s}\to K^{\ast}\mu^{+}\mu^{-}$, the transition matrix elements given in  Eqs. (\ref{Amp1a}) and (\ref{Amp1b}) can be expressed in terms of form factors as,
\begin{align}
\left\langle K^\ast(k,\overline\epsilon)\left\vert \bar{d}\gamma
_{\mu }b\right\vert B_s(p)\right\rangle &=\frac{2\epsilon_{\mu\nu\alpha\beta}}
{m_{B_s}+m_{K^\ast}}\overline\epsilon^{\,\ast\nu}p^{\alpha}k^{\beta}V(q^{2}),\label{2.13a}
\\
\left\langle K^\ast(k,\overline\epsilon)\left\vert \bar{d}\gamma_{\mu}\gamma_{5}b\right\vert
B_s(p)\right\rangle &=i\left(m_{B_s}+m_{K^\ast}\right)g_{\mu\nu}\overline\epsilon^{\,\ast\nu}A_{1}(q^{2})
\notag\\
&-iP_{\mu}(\overline\epsilon^{\ast}\cdot q)\frac{A_{2}(q^{2})}{\left(m_{B_s}+m_{K^\ast}\right)}\notag\\
&-i\frac{2m_K{\ast}}{q^{2}}q_{\mu}(\overline\epsilon^{\,\ast}\cdot q)
\left[A_{3}(q^{2})-A_{0}(q^{2})\right],\label{2.13b}
\end{align}
where $P_{\mu}=p_{\mu}+k_{\mu}$, $q_{\mu}=p_{\mu}-k_{\mu}$, and 
\begin{eqnarray}
A_{3}(q^{2})&=&\frac{m_{B_{s}}+m_{K^\ast}}{2m_{K^\ast}}A_{1}(q^{2})
-\frac{m_{B_{s}}-m_{K^\ast}}{2m_{K^\ast}}A_{2}(q^{2}),\label{A3}
\end{eqnarray}
with $A_3(0)=A_0(0)$. We have used the $\epsilon_{0123}=+1$ convention throughout the study. The tensor form factors also contribute in this process and the matrix elements associated with tensor operators can be written as,
\begin{align}
\left\langle K^\ast(k,\overline\epsilon)\left\vert \bar{d}i\sigma
_{\mu \nu }q^{\nu }b\right\vert B_{s}(p)\right\rangle
&=-2\epsilon _{\mu\nu\alpha\beta}\overline\epsilon^{\,\ast\nu}p^{\alpha}k^{\beta}T_{1}(q^{2}),\label{FF11}\\
\left\langle K^\ast(k,\overline\epsilon )\left\vert \bar{d}i\sigma
_{\mu \nu }q^{\nu}\gamma_{5}b\right\vert B_{s}(p)\right\rangle
&=i\Big[\left(m^2_{B_s}-m^2_{K^\ast}\right)g_{\mu\nu}\overline\epsilon^{\,\ast\nu}\notag\\
&-(\overline\epsilon^{\,\ast }\cdot q)P_{\mu}\Big]T_{2}(q^{2})+i(\overline\epsilon^{\,\ast}\cdot q)\notag\\
&\times\left[q_{\mu}-\frac{q^{2}}{m^2_{B_{s}}-m^2_{K^\ast}}P_{\mu}
\right]T_{3}(q^{2}).\label{F3}
\end{align}
To analyze the effects of  NP, we use the fits to light cone sum rules (LCSR) and lattice results for $B_{s}\to K^{\ast}$ transition form factors. 
The transition form factors  associated with the combined fit of the simplified series expansion (SSE) parametrization for our case can be expressed as~\cite{Bharucha:2015bzk},
\begin{eqnarray}
F_{i}(q^{2})=P_{i}(q^{2})\sum_{k}\alpha^{i}_{k}[z(q^{2})-z(0)]^{k},\label{FF}
\end{eqnarray}
where,
\begin{eqnarray}
z(t)\equiv\frac{\sqrt{t_{+}-t}-\sqrt{t_{+}-t_{0}}}{\sqrt{t_{+}-t}+\sqrt{t_{+}-t_{0}}},\label{SSE}
\end{eqnarray}
and, $t_{\pm}\equiv (m_{B_{s}}\pm m_{K^{\ast}})^{2}$, $t_{0}\equiv t_{+}\left(1-\sqrt{1-{t_{-}}/{t_{+}}}\right)$, 
and $P_{i}(q^{2})=\frac{1}{(1-{q^{2}}/{m^{2}_{R,i}})}$ is a simple pole corresponding to the first resonance in the spectrum. The resonance masses and the fit results for the SSE expansion coefficients in the fit to the LCSR computation for the decay $B_{s}\to K^{\ast}$  are given in Table \ref{tab:bestfitWC} and Table \ref{FF table11}.
\begin{table*}[!htbp]
\begin{center}
			\begin{tabular}{|c|c|c|}
				\hline
				$F_{i}$  & $J^{P}$ &  $m^{b\to d}_{R,i}$ \\
				\hline
                $A_{0}$ &  $0^{-}$                                         &$5.279$       \\
                $T_{1},V$ & $1^{-}$                  &$5.325$        \\
                $T_{2},T_{23},A_{1},A_{12}$ & $1^{+}$      & $5.274$        \\
                \hline
            \end{tabular}
\caption{Masses of resonances with quantum numbers $J^{P}$ used in the parametrization of the form factors $F_{i}$ for the $b\to d$ transition \cite{Bharucha:2015bzk}.}\label{tab:bestfitWC}            
	\end{center}
\end{table*}

\begin{table*}[!htbp]
\centering
 \renewcommand{\arraystretch}{1.5}
    \scalebox{0.8}{ 
\begin{tabular}{|c|c|c|c|c|c|c|c|}
\hline
&$A_{0}$&$A_{1}$&$A_{12}$&$V$&$T_{1}$&$T_{2}$&$T_{23}$
\\ \hline
$\alpha_{0}$&$0.36\pm 0.02$&$0.22\pm 0.01$&$0.27\pm 0.02$&$0.28\pm 0.02$&$0.24\pm 0.01$&$0.24\pm 0.01$&$0.60\pm 0.04$\\ \hline
$\alpha_{1}$&$-0.36\pm 0.20$&$0.24\pm 0.16$&$1.12\pm 0.11$&$-0.82\pm 0.19$&$-0.75\pm 0.15$&$0.31\pm 0.15$&$2.40\pm 0.27$\\ \hline
$\alpha_{2}$&$8.03\pm 2.027$&$1.77\pm 0.85$&$3.43\pm 0.78$&$5.08\pm 2.49$& $2.49\pm 1.37$&$1.58\pm 0.93$&$9.64\pm 2.03$ \\
 \hline
\end{tabular}}
\caption{Fit results for the SSE expansion coefficients in the fit to the LCSR $+$ lattice computation for the $B_{s}\to K^{\ast}$ decay \cite{Bharucha:2015bzk}.}\label{FF table11}
\end{table*}

\subsection{Four-fold Angular Distribution for the \texorpdfstring{$B_{s}\to K^{\ast}(\to K\pi)\mu^{+}\mu^{-}$}~~decay}
In an EFT framework, NP effects are incorporated through modified WCs and new operators given in Eq. (\ref{Heff}). We give our results for the four-dimensional differential decay~\cite{Kruger:1999xa,Kim:2000dq} in terms of the square of the momentum transfer $q^{2}$, and the kinematic angles (see Fig.~\ref{cascadeDecay}) $\theta_{\ell}$, $\theta_{V}$, and $\phi$  as,
\begin{eqnarray}
 \frac{d^4\Gamma\left(B_{s}\to K^{\ast}\,(\to (K\pi)\mu^+\mu^-\right)}{dq^2 \ d\cos{\theta_{l}} \ d\cos {\theta}_{V} \ d\phi} &=& \frac{9}{32 \pi} \mathcal{B}(K^{\ast}\to K\pi)\notag
 \\
&\times&\bigg[I_{1s}\sin^2\theta_{V}+I_{1c}\cos^2\theta_{V}\notag\\
&+&\Big(I_{2s}\sin^2\theta_{V}+I_{2c}\cos^2\theta_{V}\Big)\cos{2\theta_{l}}
\notag\\
&+&\Big(I_{6s}\sin^2\theta_{V}+I_{6c}\cos^2\theta_{V}\Big)\cos{\theta_{l}}\notag\\
&+&\Big(I_{3}\cos{2\phi}
+I_{9}\sin{2\phi}\Big)\sin^2\theta_{V}\sin^2\theta_{l}\notag
\\
&+&\Big(I_{4}\cos{\phi}+I_{8}\sin{\phi}\Big)\sin2\theta_{V}\sin2\theta_{l}\notag
\\
&+&\Big(I_{5}\cos{\phi}+I_{7}\sin{\phi}\Big)\sin2\theta_{V}\sin\theta_{l}\bigg],
\label{fullad}
\end{eqnarray}
\begin{figure}
    \centering
    \includegraphics[width=0.5\linewidth]{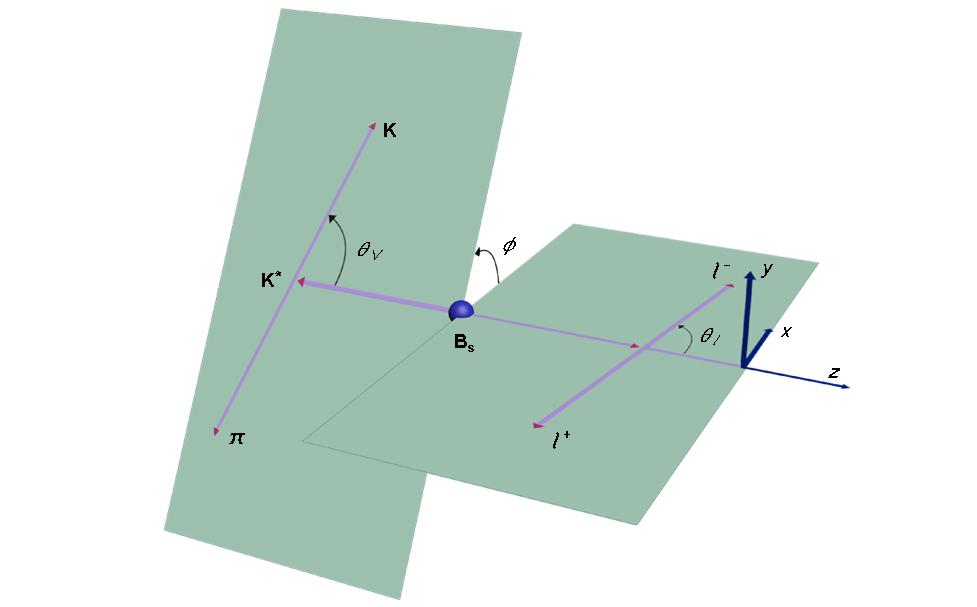}
    \caption{Kinematics of the  $B_{s}\to K^{\ast}(\to K\pi)l^{+}l^{-}$ decay.}
    \label{cascadeDecay}
\end{figure}
where the $I_{i}$ are angular coefficients which, in terms of helicity amplitudes are,
\begin{eqnarray}
&I_{1s}& = \frac{(2+\beta_l^2)}{2}N^2\left(|H_+^1|^2+|H_+^2|^2+|H_-^1|^2+|H_-^2|^2\right)\notag
\\&+&\frac{4m_l^2}{q^2}N^2\left(|H_+^1|^2-|H_+^2|^2+|H_-^1|^2-|H_-^2|^2\right),\label{I1s}\\
&I_{1c}& = 2N^2\left(|H_0^1|^2+|H_0^2|^2\right)+\frac{8m_l^2}{q^2}N^2\left(|H_0^1|^2-|H_0^2|^2+2|H_t^2|^2\right),\\
&I_{2s}& = \frac{\beta_l^2}{2}N^2\left(|H_+^1|^2+|H_+^2|^2+|H_-^1|^2+|H_-^2|^2\right),\\
&I_{2c}& = -2\beta_l^2N^2\left(|H_0^1|^2+|H_0^2|^2\right),\\
&I_{3}&=-2\beta_l^2N^2\bigg[\mathcal{R}e\left(H_+^{1}H_-^{1\ast}+H_+^{2}H_-^{2\ast}\right)\bigg],\\
&I_{4}&=\beta_l^2N^2\bigg[\mathcal{R}e\left(H_+^{1}H_0^{1\ast}+H_-^{1}H_0^{1\ast}\right)
+\mathcal{R}e\left(H_+^{2}H_0^{2\ast}+H_-^{2}H_0^{2\ast}\right)\bigg],\\
&I_{5}&=-2\beta_lN^2\bigg[\mathcal{R}e\left(H_+^{1}H_0^{2\ast}-H_-^{1}H_0^{2\ast}\right)
+\mathcal{R}e\left(H_+^{2}H_0^{1\ast}-H_-^{2}H_0^{1\ast}\right)\bigg],\\
&I_{6s}&=-4\beta_lN^2\bigg[\mathcal{R}e\left(H_+^{1}H_+^{2\ast}-H_-^{1}H_-^{2\ast}\right)\bigg],\label{I6s}\\
&I_{6c}&=0,\\
&I_{7}&=-2\beta_lN^2\bigg[\mathcal{I}m\left(H_0^{1}H_+^{2\ast}+H_0^{1}H_-^{2\ast}\right)
+\mathcal{I}m\left(H_0^{2}H_+^{1\ast}+H_0^{2}H_-^{1\ast}\right)\bigg],\\
&I_{8}&=\beta_l^2N^2\bigg[\mathcal{I}m\left(H_0^{1}H_+^{1\ast}-H_0^{1}H_-^{1\ast}\right)
+\mathcal{I}m\left(H_0^{2}H_+^{2\ast}-H_0^{2}H_-^{2\ast}\right)\bigg],\\
&I_{9}&=2\beta_l^2N^2\bigg[\mathcal{I}m\left(H_+^{1}H_-^{1\ast}+H_+^{2}H_-^{2\ast}\right)\bigg]. 
\end{eqnarray}
Here, 
\begin{eqnarray}\label{Norm}
N=V_{tb}V^{\ast}_{td}\Bigg[\frac{G_{F}^2\alpha^2}{3.2^{10} \pi^5 M_{B_s}^{3}} q^2\sqrt{\lambda}\beta_l\Bigg]^{1/2},
\end{eqnarray}
with $\lambda\equiv \lambda(M^2_{B_s}, M^2_{K^{\ast}}, q^2)$ and $\beta_l=\sqrt{1-4m_l^2/q^2}$. The expressions of helicity amplitudes in terms of SM WCs, NP WCs as well as transition form factors for the process $B_{s}\to K^{\ast}$ are,
\begin{align}
H^{1,K^{\ast}}_t&=-i\sqrt{\frac{\lambda}{q^2}}(C_{9}^{eff}+C_{9}^{NP}-C_{9}^{'NP})A_0,\notag
\\
H^{2,K^{\ast}}_t&=-i\sqrt{\frac{\lambda}{q^2}}(C_{10}+C_{10}^{NP}-C_{10}^{'NP})A_0,\notag
\\
H^{1,K^{\ast}}_{\pm}&=-i\left(M^2_{B_s}-M^2_{K^\ast}\right)\Big[(C_{9}^{eff}+C_{9}^{NP}-C_{9}^{'NP})
\frac{A_{1}}{\left(M_{B_s}-M_{K^\ast}\right)}\notag
\\
&+\frac{2m_{b}}{q^{2}}(C_{7}^{eff}+C_{7}^{NP})T_{2}\Big]
\pm i\sqrt{\lambda}\Big[(C_{9}^{eff}+C_{9}^{NP}+C_{9}^{'NP})
\frac{V}{\left(M_{B_s}+M_{K^\ast}\right)} \notag \\
&+\frac{2m_{b}}{q^{2}}(C_{7}^{eff}+C_{7}^{NP})T_{1}\Big],\notag
\\
H^{2,K^\ast}_{\pm}&=-i(C_{10}+C_{10}^{NP}-C_{10}^{'NP})\left(M_{B_s}+M_{K^\ast}\right)
A_{1}\notag
\\
&\pm i\sqrt{\lambda}(C_{10}+C_{10}^{NP}+C_{10}^{'NP})
\frac{V}{\left(M_{B_s}+M_{K^\ast}\right)},\notag
\\
H^{1,K^\ast}_0&=-\frac{i}{2M_{K^\ast}\sqrt{q^2}}\Bigg[(C_{9}^{eff}+C_{9}^{NP}-C_{9}^{'NP})
\Big\{(M^2_{B_s}-M^2_{K^\ast}-q^2)\left(M_{B_s}+M_{K^\ast}\right)A_{1}\notag
\\
&-\frac{\lambda}{M_{B_s}+M_{K^\ast}}A_{2}\Big\}+2m_b (C_{7}^{eff}+C_{7}^{NP})\Big\{(M^2_{B_s}+3M^2_{K^\ast}-q^2)T_{2} \notag \\
&-\frac{\lambda}{M^2_{B_s}-M^2_{K^\ast}}T_{3}\Big\}\Bigg],\notag \\
H^{2,K^\ast}_0&=-\frac{i}{2M_{K^\ast}\sqrt{q^2}}(C_{10}+C_{10}^{NP}-C_{10}^{'NP})
\Bigg[(M^2_{B_s}-M^2_{K^\ast}-q^2)\left(M_{B_s}+M_{K^\ast}\right)A_{1}\notag
\\
&-\frac{\lambda}{M_{B_s}+M_{K^\ast}}A_{2}\Bigg].\label{HA6}
\end{align}
The angular coefficients given in Eqs.~(\ref{I1s})-(\ref{I6s}) can be used to compute the physical observables such as branching ratio ${d\mathcal{B}(B_{s}\to K^{\ast}\mu^{+}\mu^{-})}/{dq^{2}}$, the forward-backward asymmetry $A_{\text{FB}}$, the longitudinal polarization of the $K^{\ast}$ meson $f_{L}$ and the normalized angular coefficients $\langle I_{i}\rangle$. 

We now present the formulas used for computing the physical observables. 
\begin{itemize}
\item Differential decay rate: In terms of angular coefficients, the differential decay rate for the decay $B_{s}\to K^{\ast}_{s}(\to K\pi)\mu^{+}\mu^{-}$
     can be expressed as,
    \begin{equation}
        \frac { \mathrm { d } \Gamma ( B_s \rightarrow K^* ( \rightarrow K \pi ) \mu ^ { + } \mu ^ { - } ) } { \mathrm { d } q ^ { 2 } } = B ( K^* \rightarrow K \pi ) \frac { 1 } { 4 } ( 3 I _ { 1c } + 6 I _ { 1s } - I _ {2c} - 2 I _ { 2s }). \label{Br1}
    \end{equation}

    \item {Lepton forward-backward asymmetry:} The lepton forward-backward asymmetry for the $B_{s}\to K^{\ast}_{s}(\to K\pi)\mu^{+}\mu^{-}$ decay in terms of angular coefficients $I_{i}$ can be written as,
    \begin{equation}
        A_{\mathrm{FB}}(q^{2}) = \frac{6I_{6s}}{2(3I_{1c} + 6I_{1s} - I_{2c}- 2I_{2s})}.\label{AFB1}
    \end{equation}

    \item {Longitudinal helicity fraction:} The longitudinal helicity fraction for the $B_{s}\to K^{\ast}_{s}(\to K\pi)\mu^{+}\mu^{-}$ decay  in terms of angular coefficients  $I_i$ can be expressed as,
    \begin{equation}
        f_L (q^2) = \frac{3 I_{1c} - I_{2c}}{3 I_{1c} + 6 I_{1s} - I_{2c} - 2 I_{2s}}.\label{fL1}
    \end{equation}

    \item {Normalized angular coefficients:} The normalized angular coefficients are defined as,
    \begin{equation}
    \langle I_{i}\rangle = 
    \frac{
    \mathcal{B}(K^* \to K \pi) \, I_{i}
    }{
    \mathrm{d}\Gamma\left( B_s \to K^* (\to K \pi)\mu^+ \mu^- \right)/\mathrm{d}q^{2}
    }.\label{angobsorig}
    \end{equation}
\end{itemize}

We have also computed semi-analytical expressions for the physical observables after integrating over $q^2\epsilon [q^2_{\text{min}}, 6]$ region, where $q^2_{\text{min}}\equiv4m_\mu^2\simeq0.045$ GeV$^2$ in 
terms of NP WCs. These expressions are given in Appendix~\ref{AppSemiAnalytical}. These angular observables can be obtained from the data through the maximum likelihood fit or via method of total and partial moments (MoM) \cite{LHCb:2018jna, Gratrex:2015hna, James:2006zz, Beaujean:2015xea}. In experiment the angular information about $B \to K^* \ell \bar{\ell}$ is extracted at the level of $I_i$'s \cite{LHCb:2015svh}, and also suggested for the analysis at the amplitude level \cite{Egede:2015kha}.

\section{Numerical Analysis}\label{NA}
In this section, we explore potential NP effects via the observables associated with the decay $B_{s}\to K^{\ast}(\to K\pi)\mu^{+}\mu^{-}$. The computation of the various observables depends on  the properties of the  mesons involved and other SM parameters Table~\ref{input} in addition to the SM WCs computed at the scale $\mu\sim m_b$.
Crucial to the numerical analysis are the form factors whose parametrization has been discussed in Section~\ref{ffP}, and their values at $q^{2}=0$  are listed in Table \ref{FF table11}.
\begin{table}[ht]
\centering
\begin{tabular}{|c|c|}
\hline
$M_{B_{s}}=5.36692$ GeV & $m_{b}=4.18$ GeV \\
$m_{d}=0.0048$ GeV & $m_{\mu}=0.105$ GeV \\
$\mathcal{B}(K^{\ast}\to K\pi)\sim 100$ $\%$ & $|V_{tb}V_{td}^{\ast}|=8.1\times 10^{-3}$ \\
$\alpha^{-1}=137$ & $G_{F}=1.17\times 10^{-5}$ GeV$^{-2}$ \\
$\tau_{B_{s}}=1.51\times 10^{-12}$ sec & $M_{K^{\ast}}=0.89555$ GeV \\
\hline
\end{tabular}
\caption{Values of different input parameters used in the numerical analysis \cite{ParticleDataGroup:2024cfk}.}
\label{input}
\end{table}

\begin{table*}[H]
\centering
 \renewcommand{\arraystretch}{1.5}
    \scalebox{0.9}{ 
\begin{tabular}{|c|c|c|c|c|c|c|c|c|c|}
\hline
$C_{1}$&$C_{2}$&$C_{3}$&$C_{4}$&$C_{5}$&$C_{6}$&$C_{7}$&$C_{8}$&$C_{9}$&$C_{10}$
\\ \hline
  $-0.294$ &   $1.017$  & $0.0059$  &   $-0.087$  &
  $0.0004$  &  $0.0011$   &   $-0.324$  &  $-0.176$  &
    $4.114$  &  $-4.193$ \\
\hline
\end{tabular}}
\caption{The numerical values of the SM WCs up to NNLL accuracy, evaluated at the renormalization scale $\mu\sim m_{b}$ \cite{Blake:2016olu}.}
\label{wc table}
\end{table*}

In order to quantify the effects of NP, we consider various model-independent scenarios and analyze their impact on the angular coefficients associated with the  $B_{s}\to K^{\ast}(\to K\pi)\mu^{+}\mu^{-}$ decay. We present in Table~\ref{tableNPWCs} a description of the 1D (labeled as SI-SIV)  and  2D (labeled SV-SVIII) NP scenarios that we consider. We have taken the best fit values, and the $1\sigma$ and $2\sigma$ ranges of these NP WCs from the global fit analysis that is based on data from the  
$B^{+}\to\pi^{+}\mu^{+}\mu^{-},B^{0}_{s}\to K^{\ast 0}\mu^{+}\mu^{-},B^{0}\to\mu^{+}\mu^{-}$ and radiative $B\to X_{d}\gamma$ decays data~\cite{Bause:2022rrs}.

It is worthwhile to mention that the NP scenarios SI, SII, and SIII can be realized in simple NP models involving tree-level exchanges of a leptoquark or a $Z^{\prime}$ boson. In particular, scenarios SI and SII can arise in $Z^{\prime}$ models, whereas scenario SIII can be generated by scalar or vector leptoquarks, or by a $Z^{\prime}$ boson with purely left-handed couplings \cite{Crivellin:2018yvo,Bobeth:2016llm,DAlise:2024qmp}.  Moreover, our analysis focuses on CP-even observables and aims to study the magnitude of possible NP effects rather than CP-violating asymmetries. Hence we restrict our consideration to real NP WCs \cite{Altmannshofer:2014rta, Capdevila:2017bsm, Bobeth:2010wg, Descotes-Genon:2013wba}.

In Section~\ref{results-1D} we discuss the predictions of observables for 1D scenarios and do the same for 2D scenarios in Section~\ref{results-2D}. We discuss the differential branching ratio $dB/dq^{2}$, the lepton forward backward asymmetry $A_{\text{FB}}$ the longitudinal helicity fraction $f_{L}$  of the $K^{\ast}$ meson, and the angular coefficients $\langle I_{1s}\rangle, \langle I_{2s}\rangle,\langle I_{1c}\rangle,\langle I_{2c}\rangle,\langle I_{3}\rangle,\langle I_{4}\rangle,\langle I_{5}\rangle$ and $\langle I_{6s}\rangle$.

\begin{table*}[ht!]
 \centering
 \renewcommand*{\arraystretch}{1.6}
 \resizebox{0.95\textwidth}{!}{
 \begin{tabular}{|c|c|c|c|c|c|c|c|}
  \hline
  \hline
  Scenario  & Fit Parameters & Best Fit & $1\sigma$ & $2\sigma$ & $\chi^2_{H_i, min}$ & $\text{Pull}_{H_i}$ & p-value(\%)\\
  \hline
  \hline
   SI & $C^{NP}_9$ & -1.37 & [-2.97, -0.47] & [-7.65, 0.26] & 1.23 & 1.63 & 94 \\
   SII & $C^{NP}_{10}$ & 0.96 & [0.31, 1.74] & [-0.27, 2.88] & 1.55 & 1.53 & 90 \\
   SIII & $C^{NP}_9=-C^{NP}_{10}$ & -0.54 & [-0.90, -0.20] & [-1.29, 0.13] & 1.32 & 1.60 & 93  \\
   SIV & $C^{NP}_9=-C^{NP}_{10}=-C^{' NP}_9=-C^{' NP}_{10}$ & -0.58 & [-1.06, -0.20] & [-4.04, 0.12] & 1.28 & 1.61 & 93 \\
  \hline  
  SV & ($C^{NP}_7$,\,$C^{NP}_9$) & (0.11, -1.55) & ([-0.05, 0.34], [-3.05, -0.61]) & ([-0.18, 1.46], [-10.07, 0.18]) & 0.77 & 1.25 & 94 \\
  SVI & ($C^{NP}_9$,\,$C^{' NP}_9$) & (-2.22, 1.18) & ([-6.55, -0.63], [-2.99, 2.89]) & ([-7.58, 0.23], [-3.92, 3.81]) & 0.87 & 1.22 & 92 \\
  SVII & ($C^{NP}_9$,\,$C^{' NP}_{10}$) & (-1.83, -0.38) & ([-6.58, -0.6], [-1.2, 0.32]) & ([-7.6, 0.25], [-1.8, 0.99]) & 0.95 & 1.20 & 91 \\
  SVIII & $( C^{NP}_9=-C^{' NP}_9,\,C^{NP}_{10}=+C^{' NP}_{10} )$ & (-1.73, 0.44) & ([-3.34, -0.19], [0.04, 0.95]) & ([-4.1, 0.51], [-0.34, 4.52]) & 0.88 & 1.22 & 92 \\
  \hline
  \hline
  \end{tabular}
}
  \caption{Best fit, $1\sigma$, and $2\sigma$ values of NP WCs in 1D and 2D NP scenarios \cite{Bause:2022rrs}.}
  \label{tableNPWCs}
\end{table*}

\subsection{Probing NP signatures using 1D NP scenarios\label{results-1D}}
The predictions of the $dB/dq^{2}$, $A_{\text{FB}}$ and $f_{L}$ in the presence of 1D NP scenarios with their SM values are shown in Fig.~\ref{Fig1}. The first column of Fig.~\ref{Fig1}, represents the behavior of $dB/dq^{2}$, $A_{\text{FB}}$ and $f_{L}$ as a function of $q^2$ by using the best fit values of 1D NP scenarios (see legend), where the uncertainty comes from the form factors. We include the SM result with  the form factor uncertainties as a gray band and plot SI-SIV on top of it. The second column of Fig.~\ref{Fig1}, shows the behavior of these observables after integrating over low $q^2\epsilon [q^2_{\text{min}}, 6]$ region. In this column (as well as the third) the color is used for the 1$\sigma$ (2$\sigma$) ranges of 1D NP WCs, where the width of the bands is due to form factor uncertainties as before. In the third column we have shown the variation in the magnitude of these observables due to the 1$\sigma$ and 2$\sigma$ ranges of 1D NP WCs as the height of bars. The results are shown in sets of three bars each with the central bar corresponding to the best fit values of the form factors and the left (right) bar corresponding to the upper (lower) limit of the form factor uncertainties. We now discuss in detail the features seen in the following subsections.

\subsubsection{Branching Ratio} 
We can see from Fig.~\ref{Fig1} that the branching ratio is not a very good discriminant. In particular, we can see from the top left plot that even the differential branching ratio (as a function of $q^2$) shows regions of marked overlap between  the SM and the 1D scenarios S1-SIV, even though this plot does not take into account the $1\sigma$ uncertainty in the WCs. 
\begin{figure}[H]
\centering
\includegraphics[width=1.8in,height=1.25in]{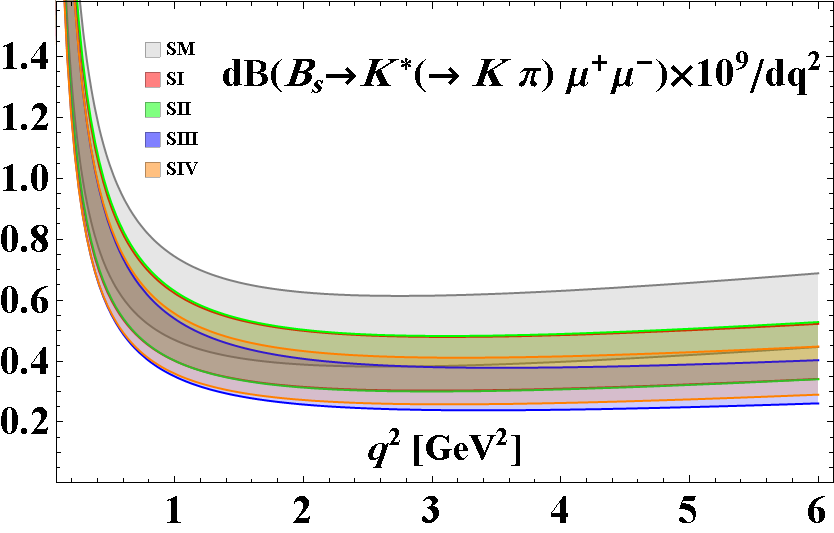}
\includegraphics[width=1.8in,height=1.25in]{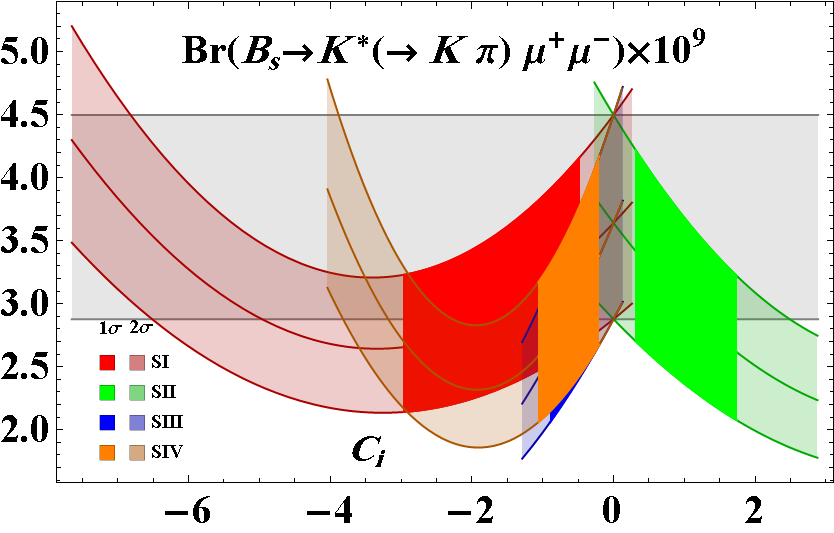}
\raisebox{0.08in}{\includegraphics[width=1.8in,height=1.17in]{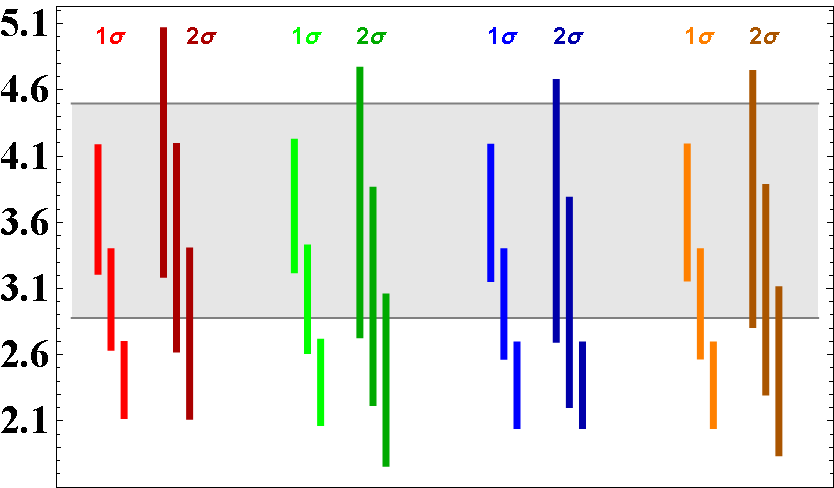}}
\includegraphics[width=1.8in,height=1.25in]{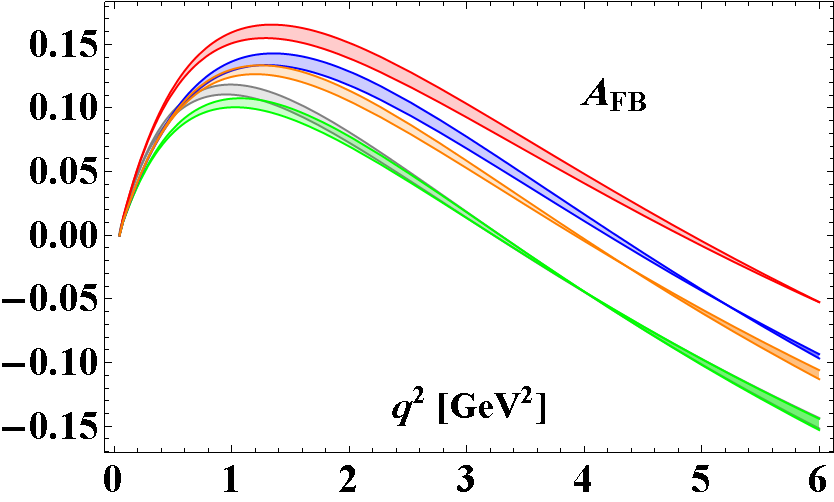}
\includegraphics[width=1.8in,height=1.25in]{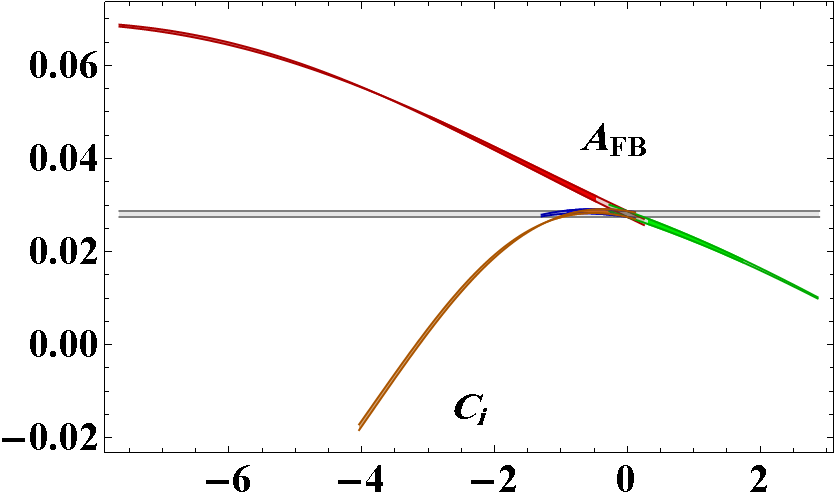}
\raisebox{0.08in}{\includegraphics[width=1.8in,height=1.17in]{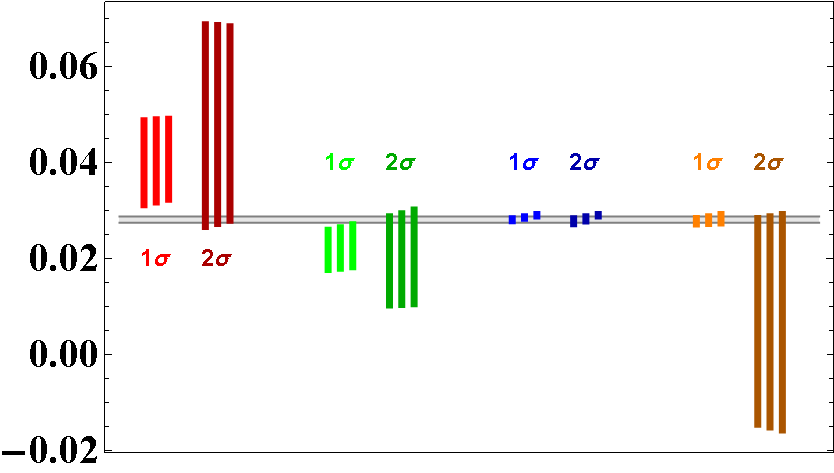}}
\includegraphics[width=1.8in,height=1.25in]{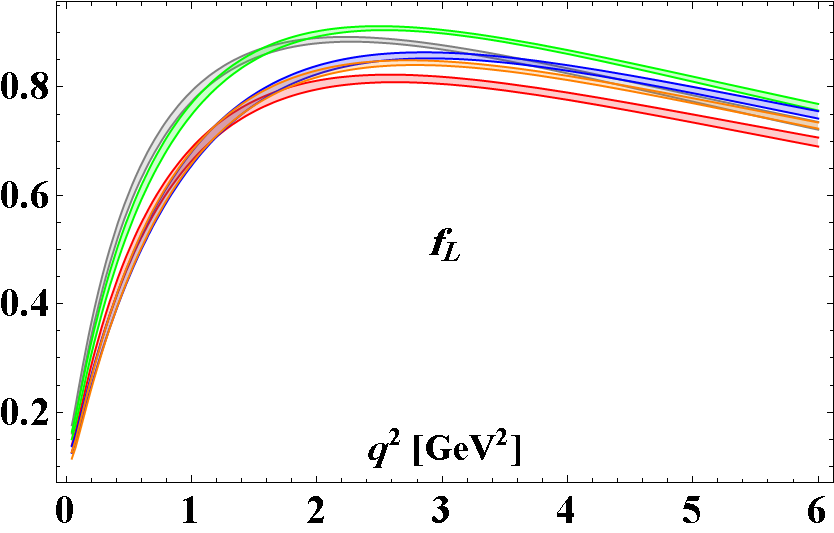}
\includegraphics[width=1.8in,height=1.25in]{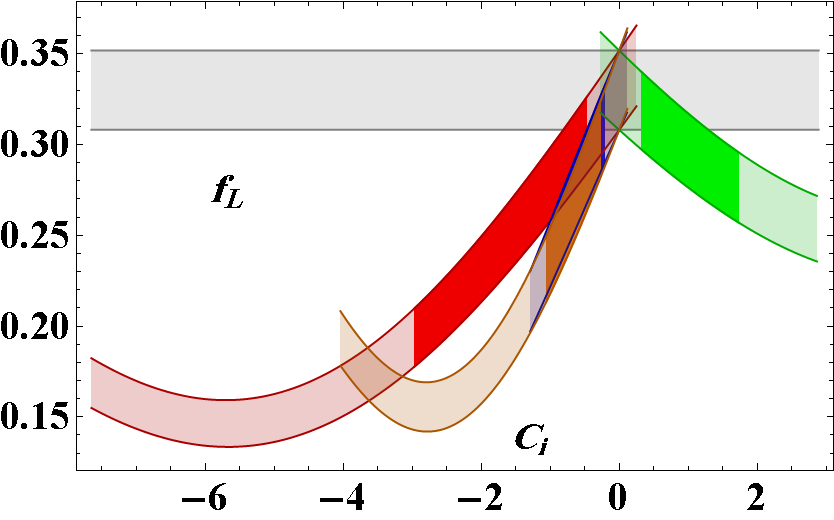}
\raisebox{0.08in}{\includegraphics[width=1.8in,height=1.17in]{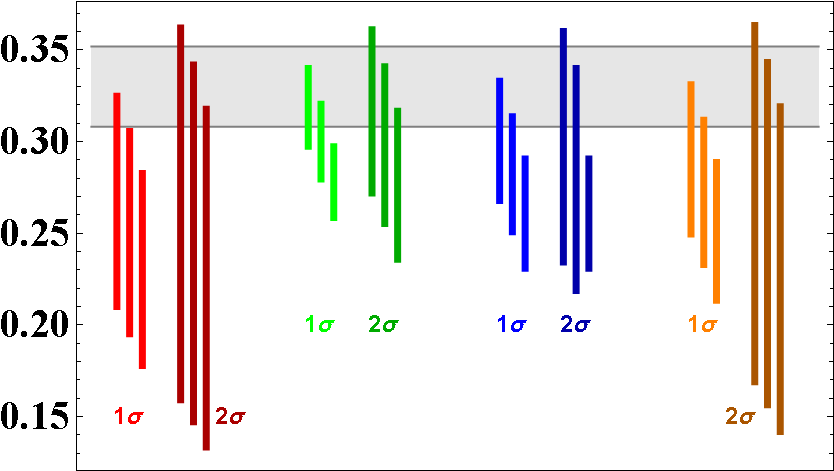}}
\caption{Different observables of $B_s \to K^*(\to K\pi)\mu^+\mu^- $ decay in the SM and in 1D NP scenarios. The first and the second columns show (from top to bottom) the differential branching ratio $\mathrm{d}B/\mathrm{d}q^2$, the $A_{\mathrm{FB}}$, $f_L$ as functions of the squared dilepton mass $q^2$ $(\text{GeV}^2)$ and of WCs ($C_i=C_{9,10}^{(')NP}$), respectively. While the third column displays the variation in their values  by different colors bar due to $1\sigma$ and $2\sigma$ parametric ranges of 1D NP WCs.}
\label{Fig1}
\end{figure}

It is clearer from the second and third plots in the first row of Fig.~\ref{Fig1} that the SM and the 1D NP scenarios cannot  be distinguished from the point of view of the branching ratio. One reason, of course, is the large form factor uncertainties.

\subsubsection{Forward-backward asymmetry} 
The best-fit plot of $A_{\text{FB}}$ (first plot in the second row of Fig.~\ref{Fig1}) shows the benefit of focusing  attention on QCD safe observables. 
We can see that three of the four 1D scenarios considered (SI, SIII, and SIV) show significant deviation from the SM considering the best-fit value of the WCs, whereas SII does show considerable overlap with the SM result. Among these scenarios SI exhibits the most prominent deviation, with the maximum deviation from the SM shown for $q^2\sim 1.3$~$\text{GeV}^2$. The second plot in the second row shows the explicit dependence of the values of $A_{\text{FB}}$ (integrated over $q^2$) on the NP WCs by using the $1\sigma$ and $2\sigma$ confidence regions. 
In this plot, we can see that scenario SIII remains well within the SM band across both the $1\sigma$ and $2\sigma$ ranges. This plot also depicts that scenario SI exhibits higher values than the SM, whereas scenarios SII and SIV yield lower values over most of the $2\sigma$ confidence interval for the WCs. 
Interestingly, SIII lies mostly within the SM band in this plot. These trends are further illustrated in the bar plot (third plot in the second row of Fig.~\ref{Fig1})  which highlights that scenarios SI and SII show a clear departure from the SM prediction within the $1\sigma$ range of WCs after taking into account form factor uncertainties. 

\subsubsection{Helicity fraction} 
 The best-fit plot of the longitudinal helicity fraction $f_{L}$ (first plot in the third row of Fig.~\ref{Fig1}) shows that scenarios SII and SIII are approximately consistent with the SM predictions for $q^{2}\epsilon[0.045,2]~(\text{GeV}^2)$ and $q^{2}\epsilon[3,6]~(\text{GeV}^2)$, respectively. Beyond these regions, both scenarios start to deviate from the SM values. In contrast, scenarios SI and SIV show significant deviations from the SM predictions across the given $q^{2}$ region. Most of the scenarios show a lower value of $f_{L}$ compared to that of SM. The second plot in the third row (Fig.~\ref{Fig1}) shows the integrated values of $f_{L}$ over $q^{2}\approx[0.045,6]~ (\text{GeV}^{2})$ using the $1\sigma$ and $2\sigma$ confidence intervals. This plot reveals notable deviations from the SM across all scenarios, with scenarios SIII and SIV exhibiting complete overlap within both confidence regions. A similar trend is also observed in the corresponding bar plot (third plot in the third row), which also illustrates the lower $f_{L}$ values among  all scenarios compared to that of  SM. 

\subsubsection{Angular coefficients} 
In Fig.~\ref{angularpnl1} and Fig.~\ref{angularpnl2} we present the plots for angular coefficients $\langle I_i \rangle$'s following the scheme used in Fig.~\ref{Fig1}. In the best-fit plots (first column) for $\langle I_i \rangle$'s, all scenarios form distinguishable bands, with partial overlap with the SM band and mutual overlap in various regions. The deviations of these bands are  particularly pronounced in the case of $\langle I_{2s}\rangle$, $\langle I_3 \rangle$, $\langle I_5 \rangle$, and $\langle I_{6s} \rangle$. Scenario SI produces the most distinct band in $\langle I_{1s} \rangle$, $\langle I_{1c} \rangle$, $\langle I_{2s} \rangle$, $\langle I_{2c} \rangle$, $\langle I_5 \rangle$, and $\langle I_{6s} \rangle$, while maintaining overlap with the SM band in the $\langle I_3 \rangle$ and $\langle I_4 \rangle$ plots. The behavior of SII aligns closely with the SM in several regions across all plots. However,  scenario SIV demonstrates a different trend than other scenarios in the $\langle I_3 \rangle$ plot. We note that  scenario SI has the most distinct behavior in $\langle I_{1s} \rangle$, $\langle I_{2s} \rangle$, $\langle I_{2c} \rangle$, $\langle I_{5} \rangle$ and $\langle I_{6s} \rangle$ plots.

\begin{figure}[H]
\centering
\includegraphics[width=1.8in,height=1.25in]{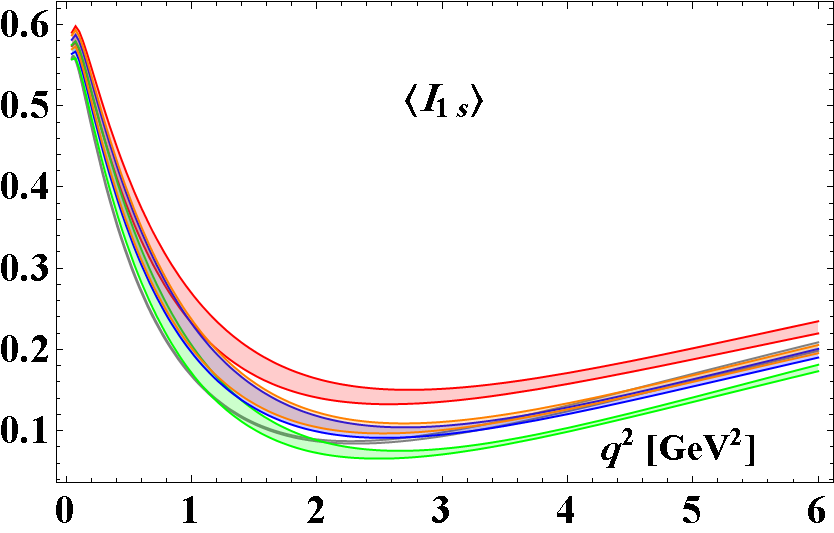}
\includegraphics[width=1.8in,height=1.25in]{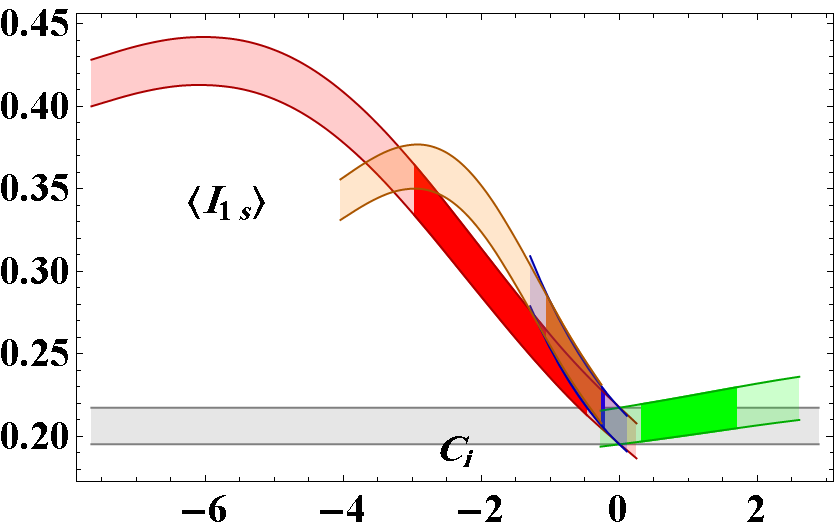}
\raisebox{0.08in}{\includegraphics[width=1.8in,height=1.17in]{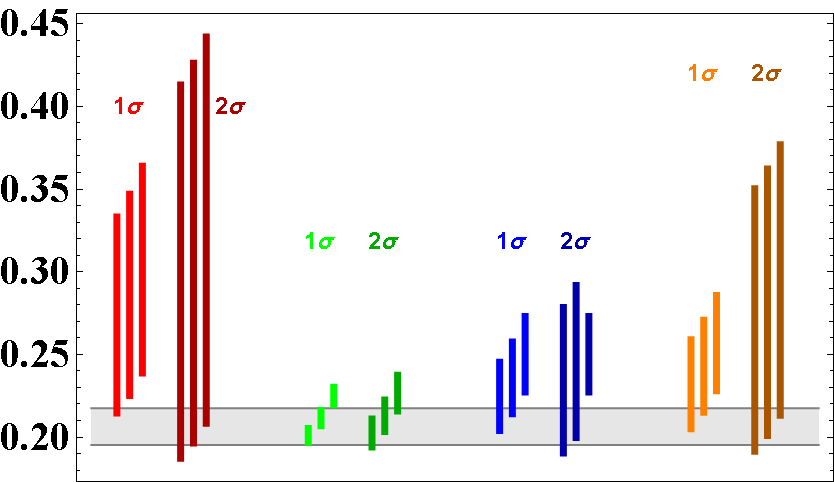}}
\\
\includegraphics[width=1.8in,height=1.25in]{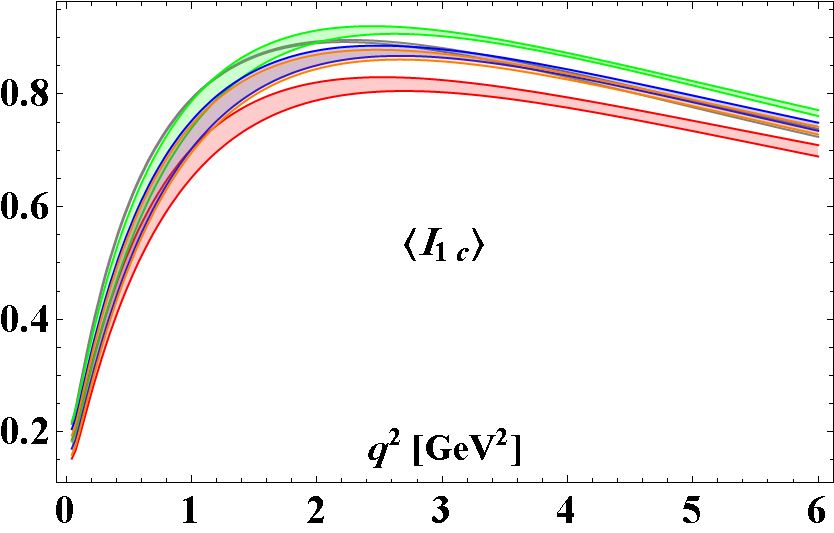}
\includegraphics[width=1.8in,height=1.25in]{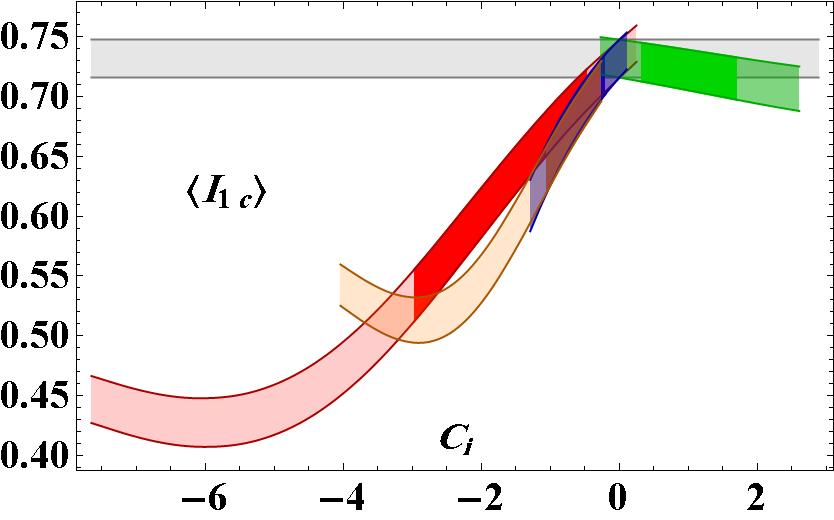}
\raisebox{0.08in}{\includegraphics[width=1.8in,height=1.17in]{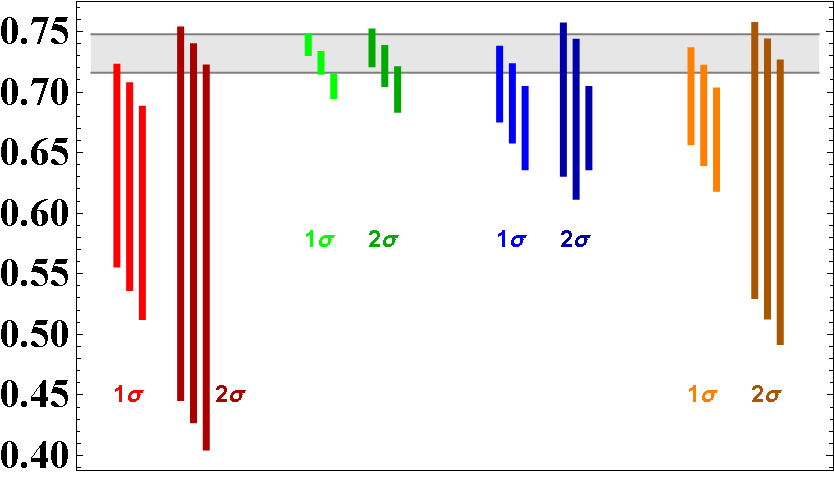}}
\\
\includegraphics[width=1.8in,height=1.25in]{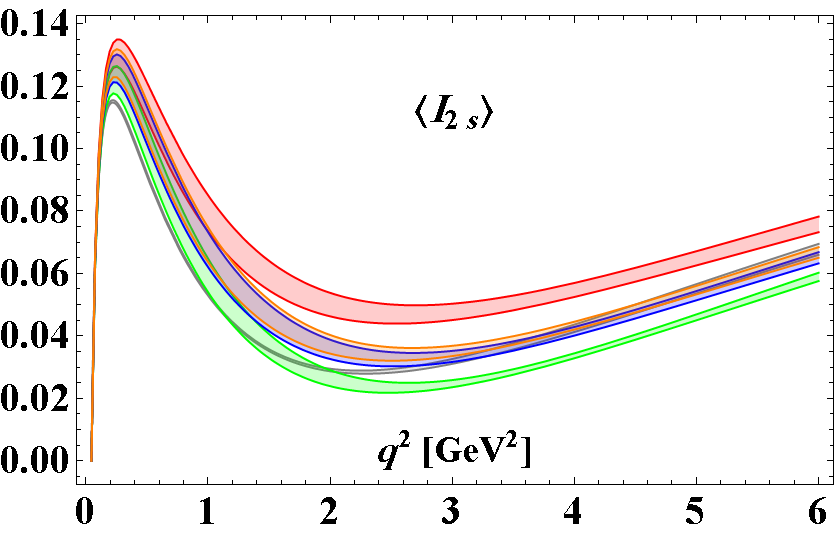}
\includegraphics[width=1.8in,height=1.25in]{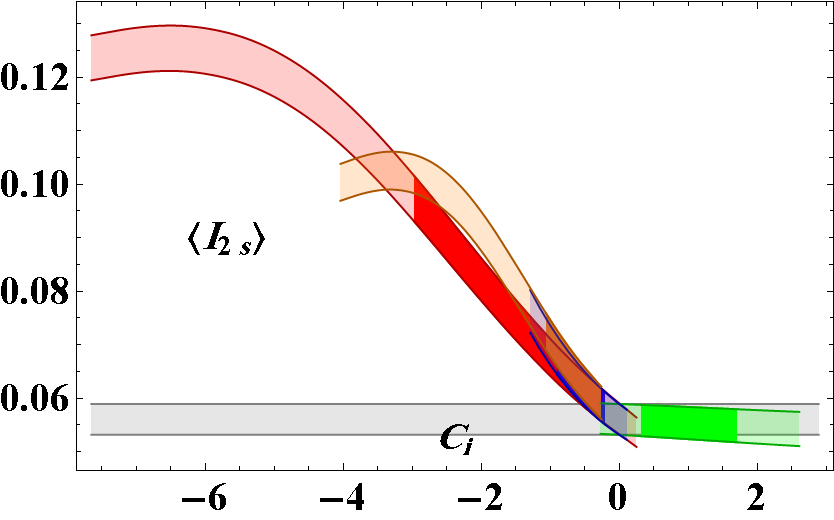}
\raisebox{0.08in}{\includegraphics[width=1.8in,height=1.17in]{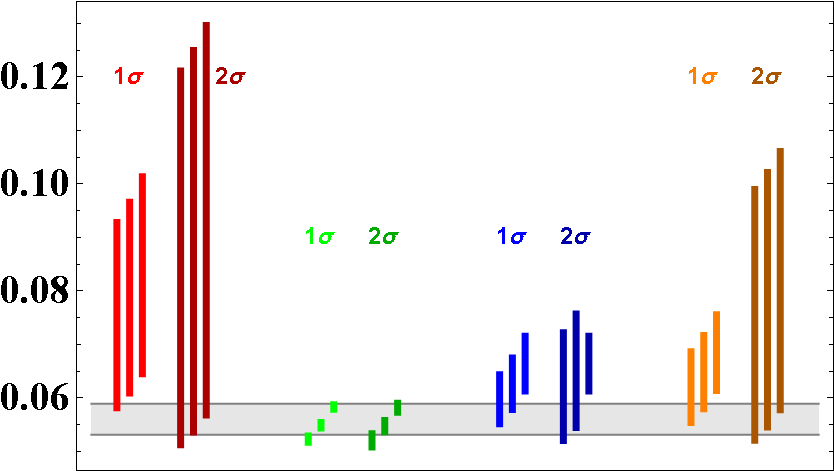}}
\\
\includegraphics[width=1.8in,height=1.25in]{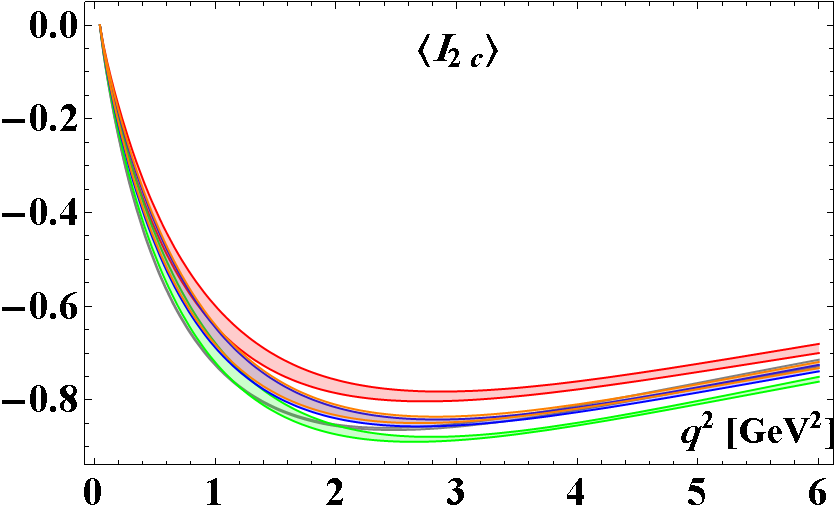}
\includegraphics[width=1.8in,height=1.25in]{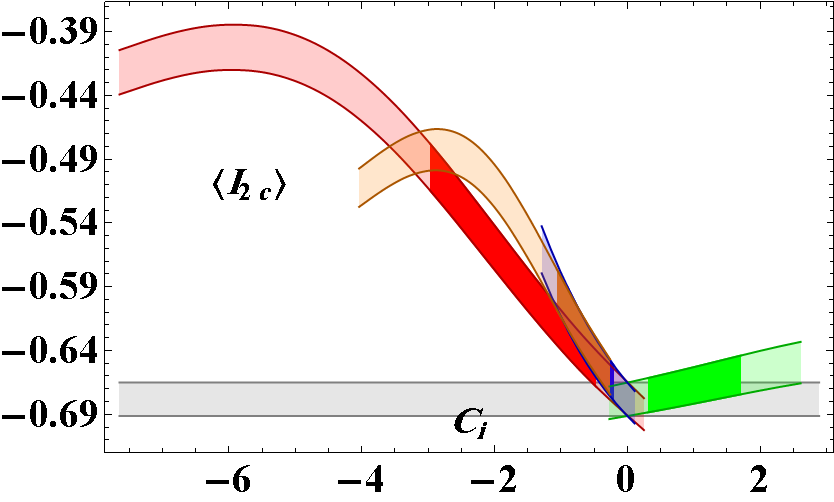}
\raisebox{0.08in}{\includegraphics[width=1.8in,height=1.17in]{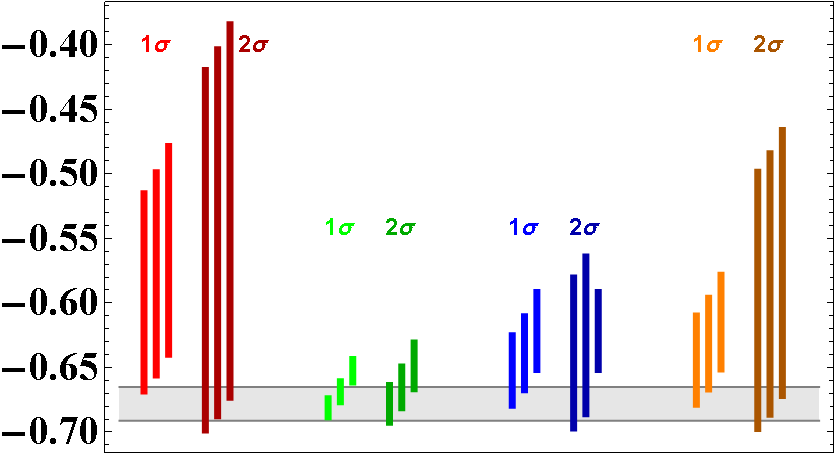}}
\caption{The angular coefficients  $\langle I_{1s} \rangle$, $\langle I_{1c} \rangle$, $\langle I_{2s} \rangle$, and $\langle I_{2c} \rangle$ for $B_s \to K^*(\to K\pi)\mu^+\mu^- $ where the remaining description is the same as describe in caption of Fig. \ref{Fig1}.}
\label{angularpnl1}
\end{figure}

 The plots in the second columns of Fig.~\ref{angularpnl1} and Fig.~\ref{angularpnl2} depict the angular coefficients (integrated over $q^2$) as functions of the WCs $C_i$. Scenario SII exhibits deviations in the $\langle I_4 \rangle$ and $\langle I_5 \rangle$ angular coefficients while remaining consistent with the SM predictions in other angular coefficients. This behavior of SII is consistent with observations from the best-fit plots. In contrast, scenarios SI, SIII and SIV display substantial deviations from the SM in all plots. While these three scenarios share mutually overlapping regions in the $\langle I_{1s} \rangle$, $\langle I_{1c} \rangle$, $\langle I_{2s} \rangle$ and $\langle I_{2c} \rangle$ plots, they diverge into distinct directions in the remaining plots, particularly in $\langle I_4 \rangle$. All scenarios give higher predictions for $\langle I_{1s} \rangle$ and $\langle I_{2c} \rangle$.

\begin{figure}[h!]
\centering
\includegraphics[width=1.8in,height=1.25in]{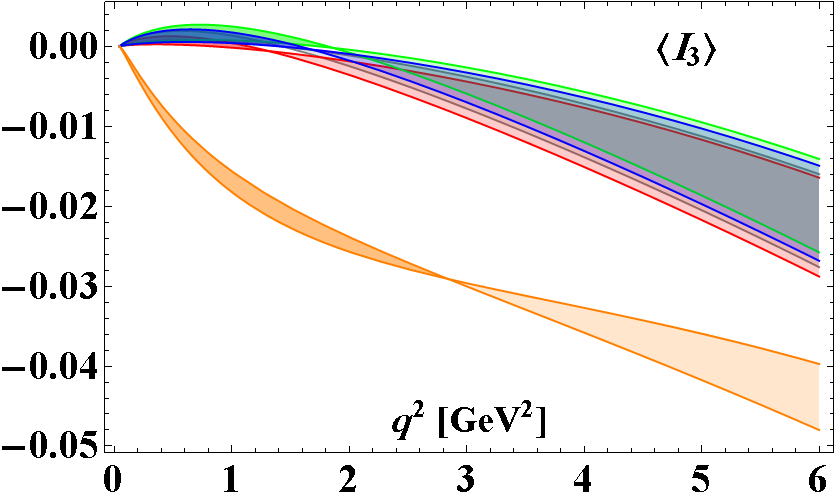}
\includegraphics[width=1.8in,height=1.25in]{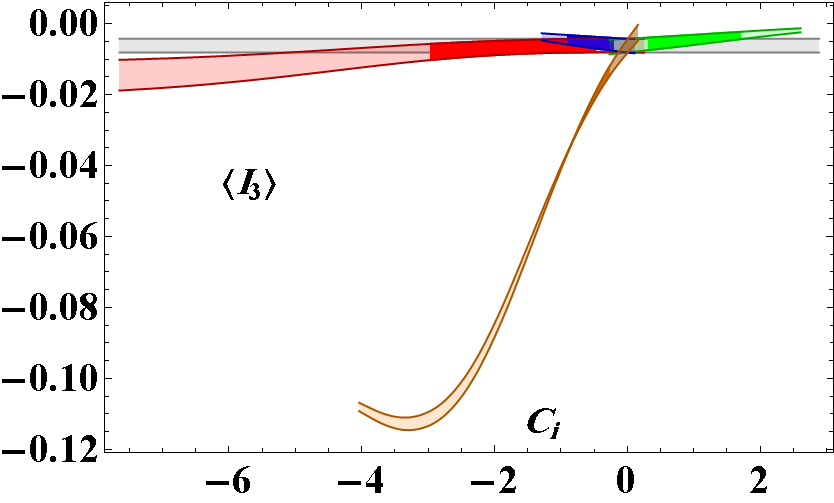}
\raisebox{0.08in}{\includegraphics[width=1.8in,height=1.17in]{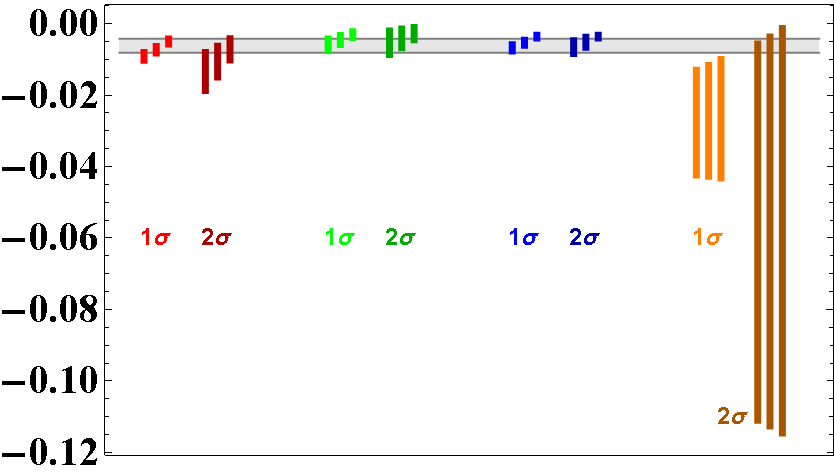}}
\\
\includegraphics[width=1.8in,height=1.25in]{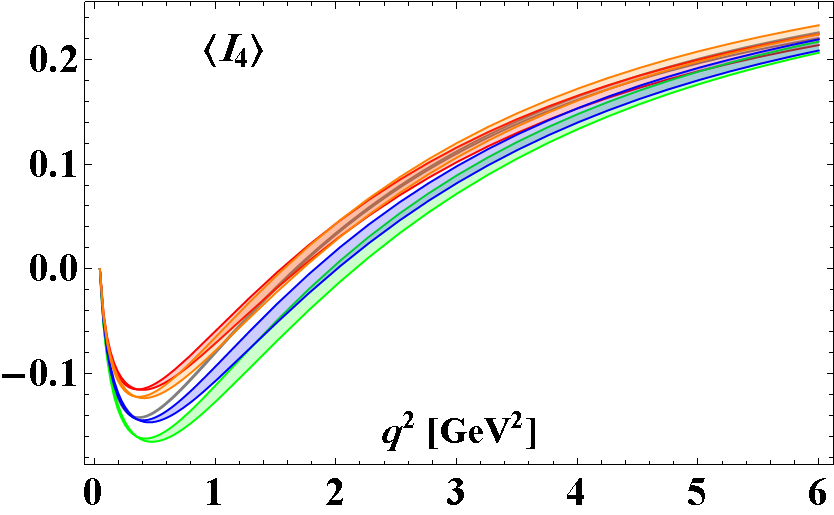}
\includegraphics[width=1.8in,height=1.25in]{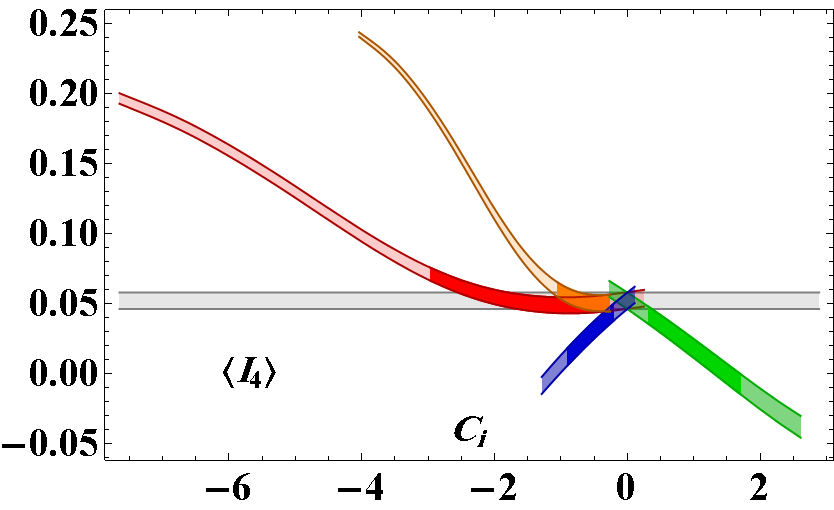}
\raisebox{0.08in}{\includegraphics[width=1.8in,height=1.17in]{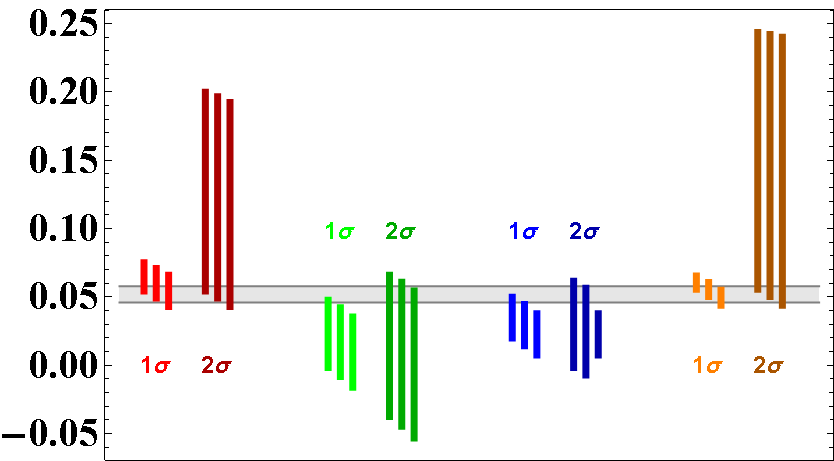}}
\\
\includegraphics[width=1.8in,height=1.25in]{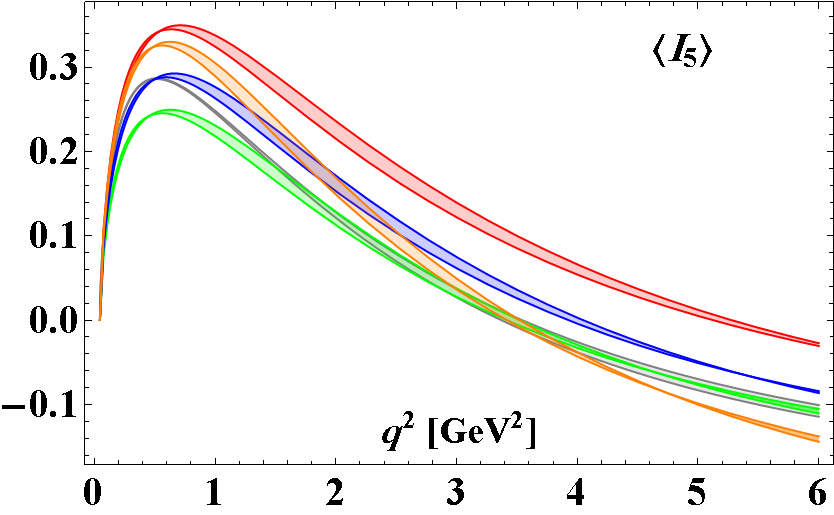}
\includegraphics[width=1.8in,height=1.25in]{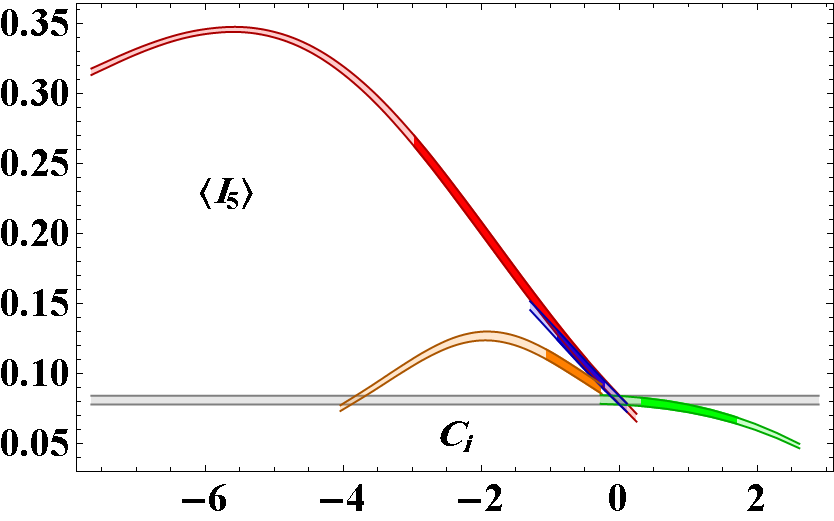}
\raisebox{0.08in}{\includegraphics[width=1.8in,height=1.17in]{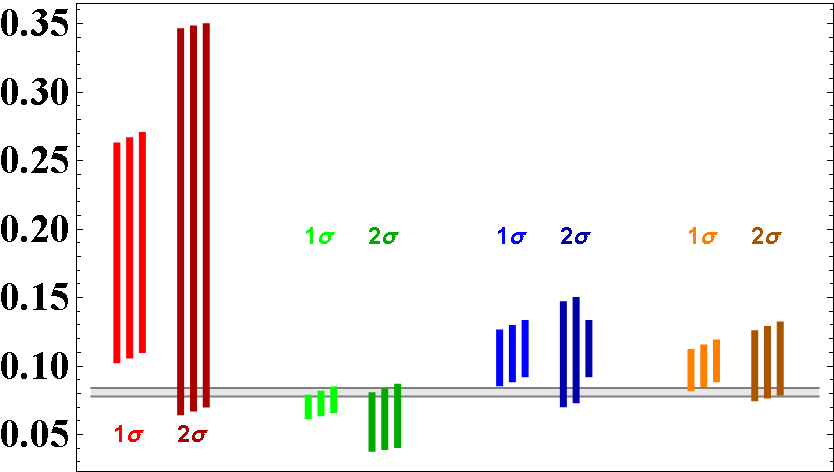}}
\\
\includegraphics[width=1.8in,height=1.25in]{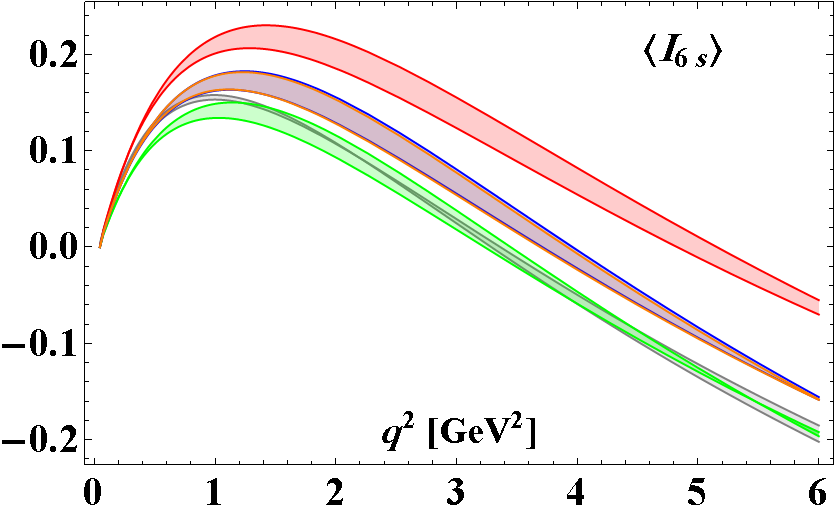}
\includegraphics[width=1.8in,height=1.25in]{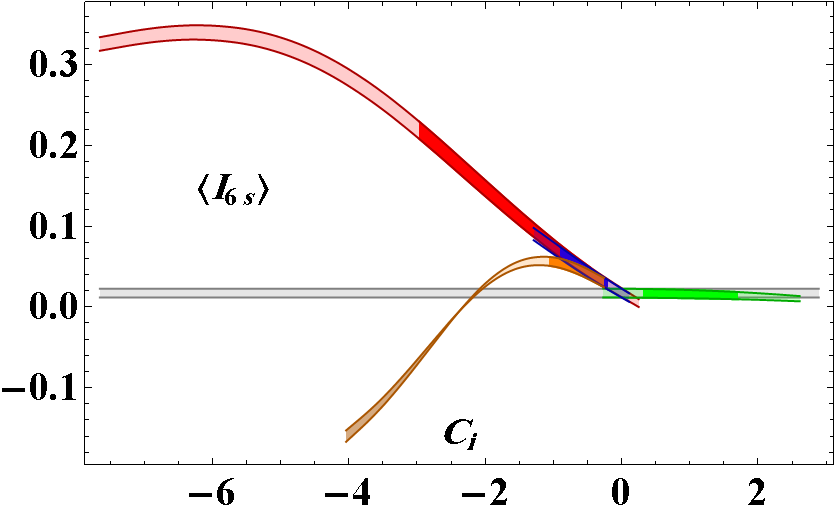}
\raisebox{0.08in}{\includegraphics[width=1.8in,height=1.17in]{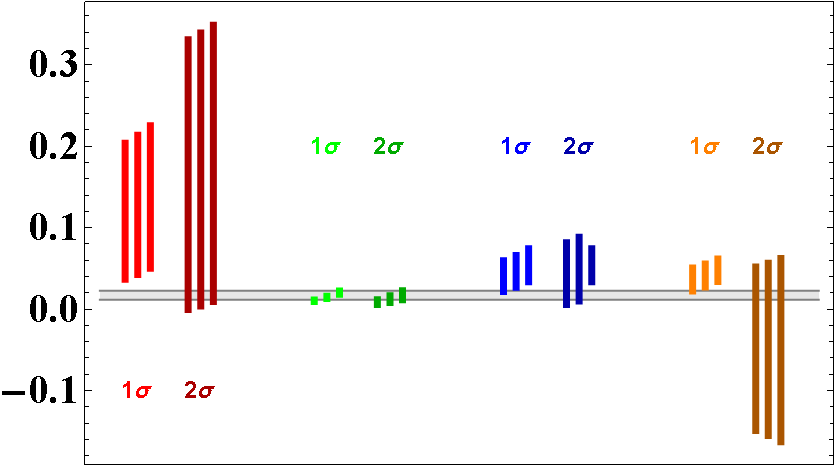}}
\caption{The  angular coefficients  $\langle I_{3} \rangle$, $\langle I_{4} \rangle$, $\langle I_{5} \rangle$, and $\langle I_{6s} \rangle$ for $B_s \to K^*(\to K\pi)\mu^+\mu^- $ where the remaining description is the same as describe in the caption of Fig. \ref{Fig1}.}
\label{angularpnl2}
\end{figure}
 The bar plots for the angular coefficients clearly illustrate the behavior of the discussed scenarios. We can see that while deviations from the SM can be seen for all the scenarios at the $1\sigma$ level is one or other angular coefficients $\langle I_i\rangle$, all scenarios are consistent with SM within $2\sigma$. Furthermore, 
 as the experimental uncertainties decrease in the form factors and the allowed WC ranges, we expect to be able to distinguish between 
 the various scenarios given their distinct character in the various bar plots.
 
\begin{table}[htbp!]
    \centering
    \renewcommand{\arraystretch}{1.5}
     \resizebox{\textwidth}{!}{
    {\fontsize{7}{10}
    \begin{tabular}{|c|c|c|c|c|c|}
\hline
\text{Dataset} & \text{SM} & SI & SII & SIII & SIV \\
\hline
$Br \times 10^{9} $ & (2.88, 4.50) & (2.14, 5.05) & (1.78, 4.75) & (2.07, 4.66) & (1.86, 4.73) \\
\hline
$A_{\text{FB}}$ & (0.03, 0.03) & (0.03, 0.07) & (0.01, 0.03) & (0.03, 0.03) & (-0.02, 0.03) \\
\hline
$f_{L}$ & (0.31, 0.35) & (0.13, 0.36) & (0.24, 0.36) & (0.22, 0.36) & (0.14, 0.36) \\
\hline
$\langle I_{1s} \rangle$ & (0.20, 0.22) & (0.19, 0.44) & (0.19, 0.24) & (0.19, 0.29) & (0.19, 0.38) \\
\hline
$\langle I_{1c} \rangle$ & (0.72, 0.75) & (0.41, 0.75) & (0.69, 0.75) & (0.61, 0.75) & (0.49, 0.75) \\
\hline
$\langle I_{2s} \rangle$ & (0.05, 0.06) & (0.05, 0.13) & (0.05, 0.06) & (0.05, 0.08) & (0.05, 0.11) \\
\hline
$\langle I_{2c} \rangle$ & (-0.69, -0.67) & (-0.70, -0.38) & (-0.69, -0.63) & (-0.70, -0.56) & (-0.70, -0.47) \\
\hline
$\langle I_{3} \rangle$ & (-0.01, 0.00) & (-0.02, 0.00) & (-0.01, 0.00) & (-0.01, 0.00) & (-0.12, 0.00) \\
\hline
$\langle I_{4} \rangle$ & (0.05, 0.06) & (0.04, 0.20) & (-0.05, 0.07) & (-0.01, 0.06) & (0.04, 0.24) \\
\hline
$\langle I_{5} \rangle$ & (0.08, 0.08) & (0.07, 0.35) & (0.04, 0.08) & (0.07, 0.15) & (0.08, 0.13) \\
\hline
$\langle I_{6s} \rangle$ & (0.01, 0.02) & (0.00, 0.35) & (0.01, 0.02) & (0.01, 0.09) & (-0.16, 0.06) \\
\hline
    \end{tabular}
    }}
\caption{Results for the $q^2$ bin [$q^2_{\text{min}}$, 6] GeV$^2$ in the SM as well as for 1D NP scenarios for different observables.}
    \label{1Dbarplots}
\end{table}

\begin{figure}[H]
\centering
\includegraphics[width=1.45in,height=1.2in,trim=0 0 0 0 ,clip]{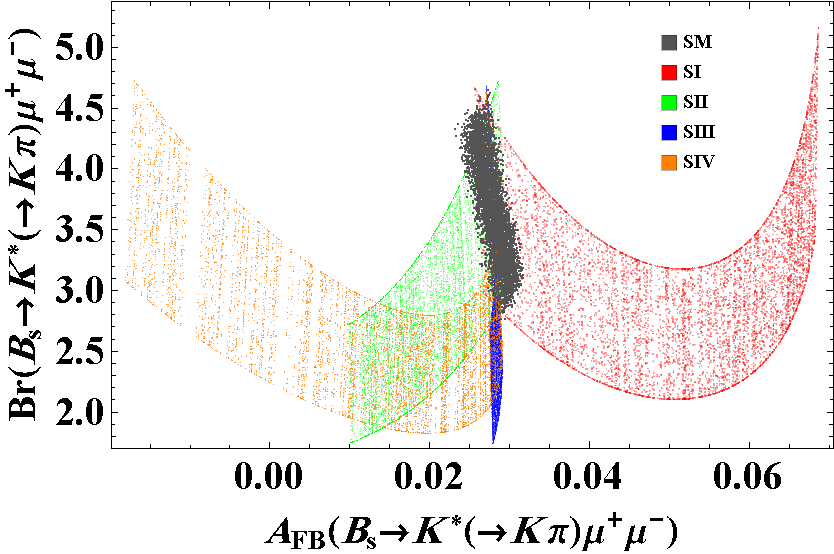}
\includegraphics[width=1.45in,height=1.2in,trim=0 0 0 0 ,clip]{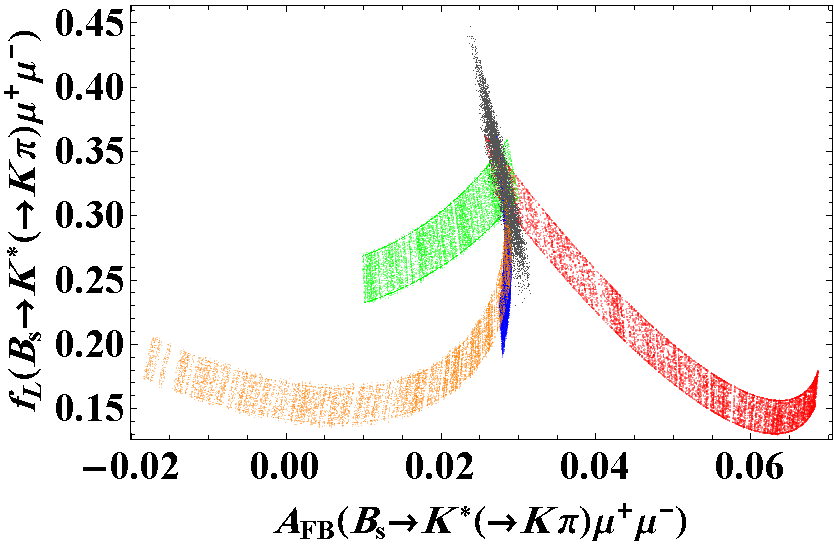}
\includegraphics[width=1.45in,height=1.2in,trim=0 0 0 0 ,clip]{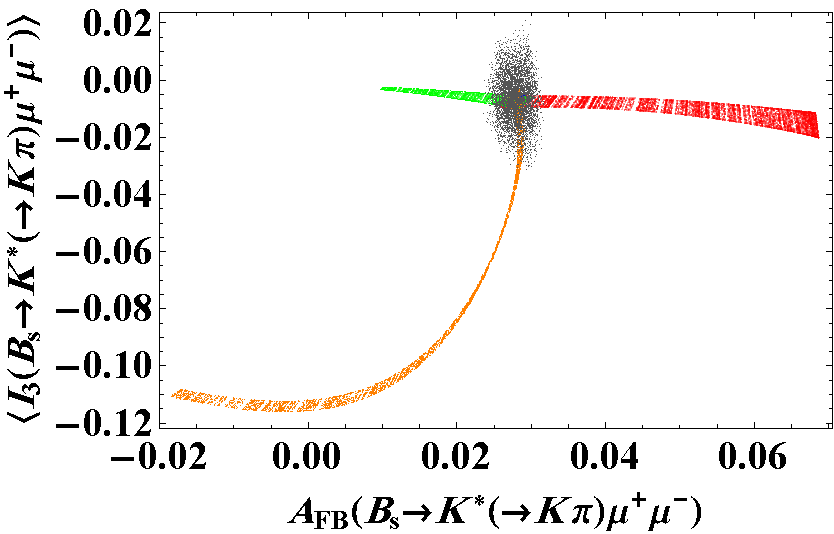}
\includegraphics[width=1.45in,height=1.2in,trim=0 0 0 0 ,clip]{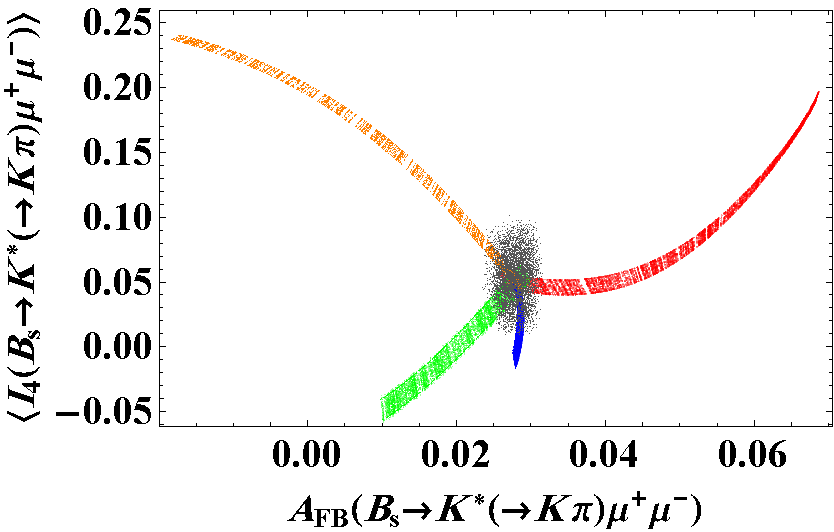}
\caption{Correlations between $A_{\text{FB}}$ and different observables of $B_s \to K^*(\to K\pi)\mu^+\mu^- $ decay in SM and in 1D NP scenarios.}
\label{1Dcorrelation}
\end{figure}

\subsubsection{Correlations between the physical observables}
In Fig.~\ref{1Dcorrelation} we show the correlation of a few of the observables that we have computed. Such correlations are valuable and a powerful tool to probe NP. Furthermore, they could also help to distinguish among different NP scenarios, as deviations in the correlations could indicate NP effects that differ across the two observables. In particular, we show in the first plot of Fig.~\ref{1Dcorrelation}, the correlation of the branching ratio with the forward-backward asymmetry $A_{\text{FB}}$. The SM appears as a narrow strip in this plot with NP effects easily distinguished from the SM 
for most of the parameter space. In the second plot the $f_L$ against the 
$A_{\text{FB}}$ and we can see the SM shrinking in this plane to an even narrower band. What is more interesting 
is that in this plane SI and SII separate out from each other and also from SIII and SIV. However, there is a substantial overlap 
between SIII and SIV. A similar trend can be seen in the correlation of the angular variable $\langle I_3\rangle$ and the $A_{\text{FB}}$ 
plotted in the third plot of Fig.~\ref{1Dcorrelation}. 
Finally, 
we can see that in the correlation of $\langle I_4\rangle$ and $A_{\text{FB}}$ all the four 1D NP scenarios separate out into 
distinct bands, thereby allowing one to clearly distinguish between them.

\subsection{Probing NP signatures using 2D NP scenarios\label{results-2D}}

We now focus our attention on the  2D scenarios, as presented in Table \ref{tableNPWCs}, which illustrate the dependence of 
the observables on pairs of WCs. Naturally, a broader uncertainty region is expected due to the combined uncertainty  of the two 
 WCs involved. This effect is reflected in the form of wider bands in the 2D scenario plots.

In Fig.~\ref{2Dbrafbfl} we show the results for the branching ratio, the forward-backward asymmetry and the longitudinal helicity 
fraction. Then, in Figs.~\ref{2Dangularpnl1}~and~\ref{2Dangularpnl2} we move on to discuss the corresponding results for the 
angular coefficients.

\subsubsection{Branching Ratio} The first plot in the top row of Fig.~\ref{2Dbrafbfl}  compares the branching ratio in the SM prediction with scenarios SV (red), SVI (green), SVII (blue), and SVIII (orange). All scenarios tend to lower the value to that of the SM  over the  given $q^2$ range. Among these, SV exhibits the largest deviation below $q^2=1$ $\text{GeV}^2$ and partially overlaps with the SM band over $q^2=1$ $\text{GeV}^2$.  
We can see from the  corresponding bar plot in the second row that while the four NP scenarios do agree with the SM within $1\sigma$, they can lead to large deviations in the predictions for this channel. 

\subsubsection{Forward-backward asymmetry} The $A_{\text{FB}}$ best-fit plot (first row, middle column of Fig.~\ref{2Dbrafbfl}) reveals distinct and well-separated bands. All scenarios give a higher prediction than the SM. Although SV approaches the SM over a small region, it subsequently diverges. Scenarios SVI, SVII, and SVIII form distinctly separate bands, highlighting clear deviations among them. 

The corresponding bar plot (middle column, second row) reinforces these findings, and shows in particular that the NP scenarios can lead to results quite distinct from the SM.

\subsubsection{Helicity fraction} Finally, in the $f_L$ best-fit plot, a behavior similar to that of the $A_{\text{FB}}$ best-fit plot is observed. However, in contrast to $A_{\text{FB}}$, where the scenarios predicted higher values than the SM, here  the scenarios predict systematically lower values. These deviating trends more or less persist as we allow the WCs to vary within $2\sigma$ and are consistently reflected in the associated bar plots (second row, last column).

\begin{figure}[H]
\centering
\includegraphics[width=1.8in,height=1.25in]{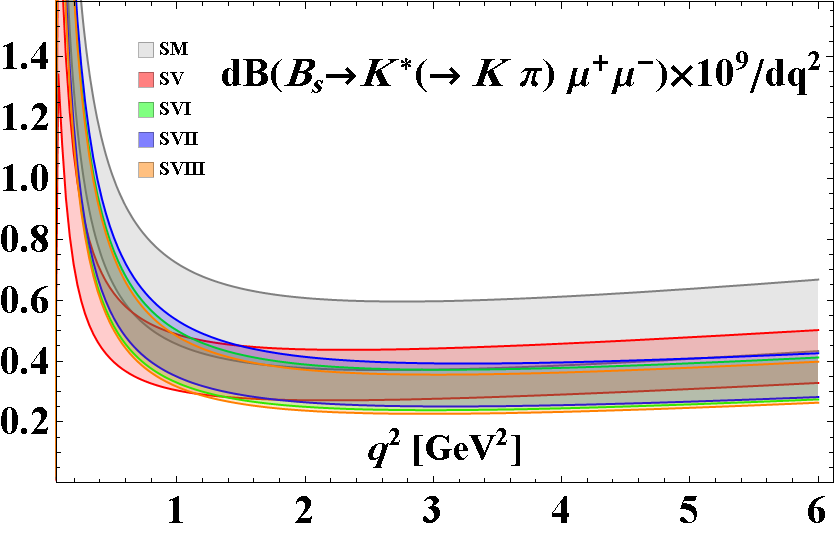}
\includegraphics[width=1.8in,height=1.25in]{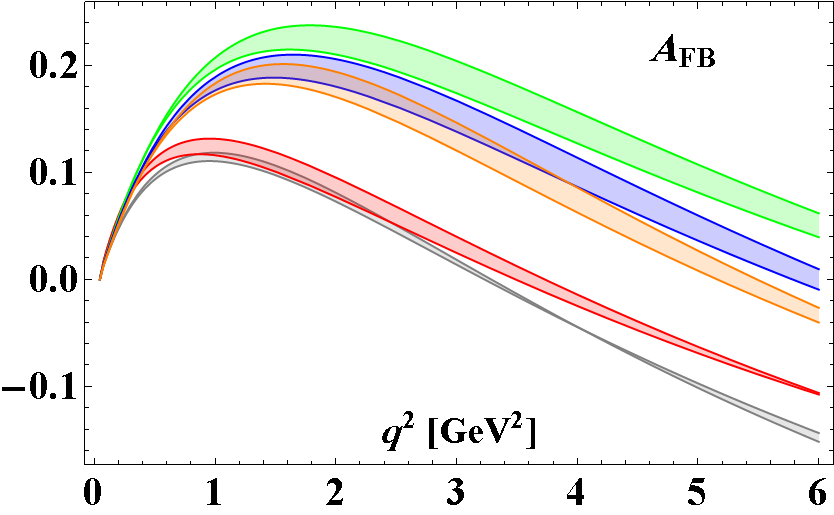}
\includegraphics[width=1.8in,height=1.25in]{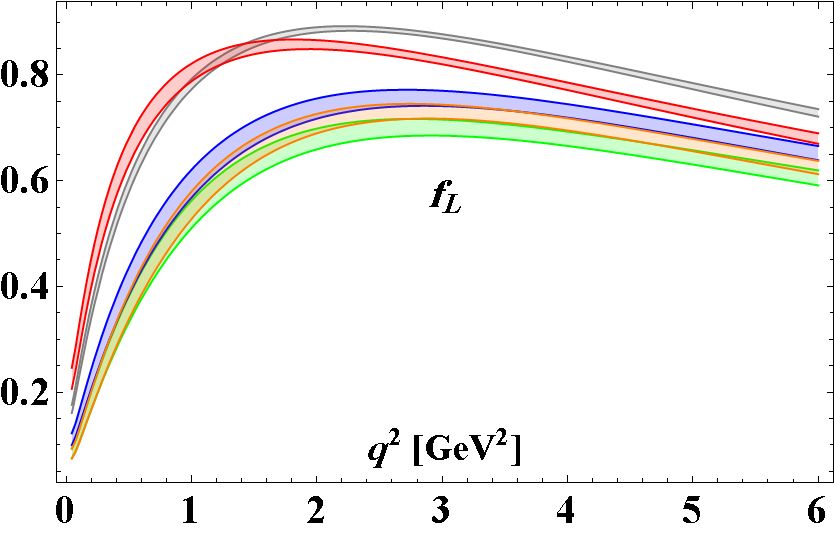}
\raisebox{0.08in}{\includegraphics[width=1.8in,height=1.17in]{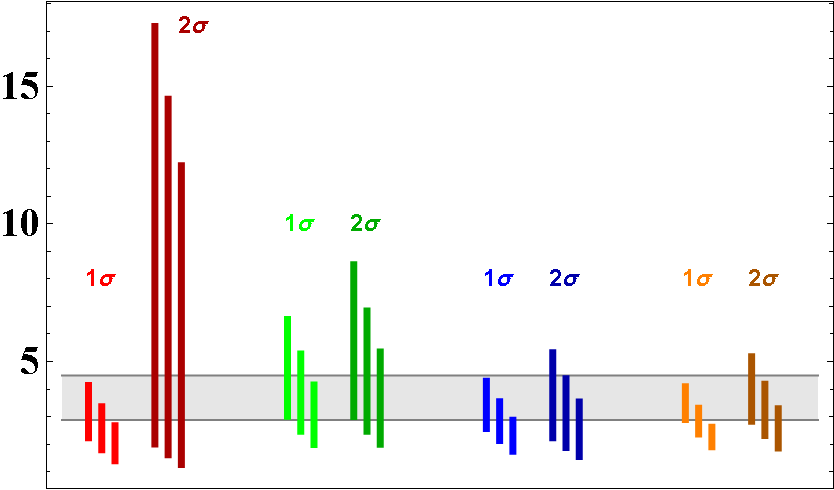}}
\raisebox{0.08in}{\includegraphics[width=1.8in,height=1.17in]{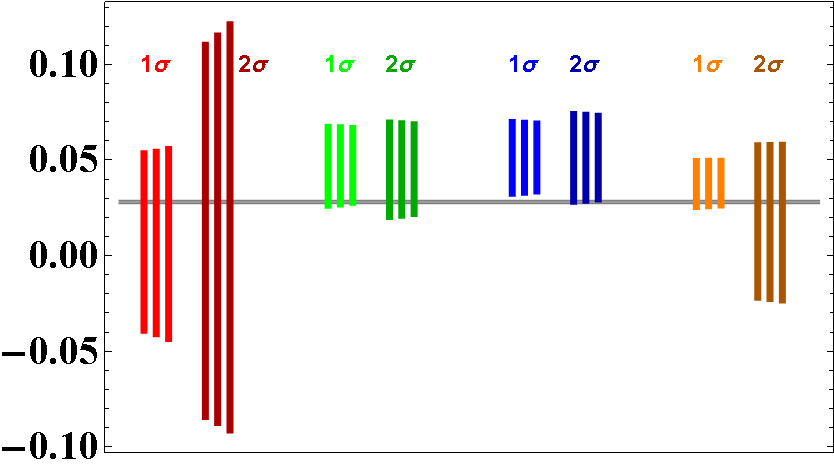}}
\raisebox{0.08in}{\includegraphics[width=1.8in,height=1.17in]{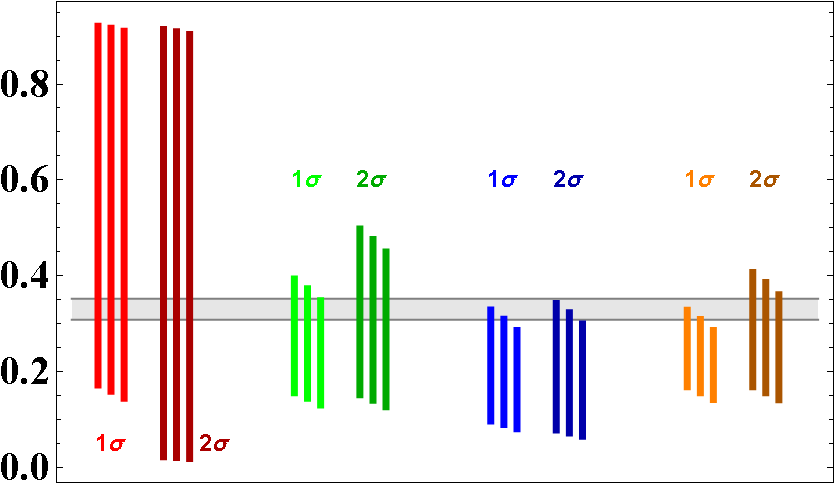}}
\caption{ The first row shows, the $\mathrm{d}B/\mathrm{d}q^2$, the $A_{\mathrm{FB}}$, and $f_L$ as a function of $q^2$. The second row displays the variation in their values  by different colors bar due to $1\sigma$ and $2\sigma$ parametric ranges of 2D NP WCs.}
\label{2Dbrafbfl}
\end{figure}

\subsubsection{Angular Coefficients} The 2D best-fit plots for the angular coefficients $\langle I_i\rangle$ against $q^2$ can be seen in the first row of Figs.~\ref{2Dangularpnl1}~and~\ref{2Dangularpnl2}. We can see distinct and well-separated bands for all scenarios in these plots. Scenario SV initially overlaps with the SM in certain regions before deviating from it. The remaining scenarios largely remain distinct from the SM band. Notably, in the $\langle I_3 \rangle$ plot (top row, first column of Fig.~\ref{2Dangularpnl2}), scenario SVIII exhibits a significant deviation. Interestingly, scenario SVII displays a trend comparable to SVIII in the $\langle I_3 \rangle$ plot, but predicts are value larger than the SM whereas SVIII predicts one much smaller. These distinct patterns can be seen in 
all the angular coefficients and provide strong motivation for further investigation of the discussed scenarios.

\begin{figure}[H]
\centering
\includegraphics[width=1.4in,height=1.25in]{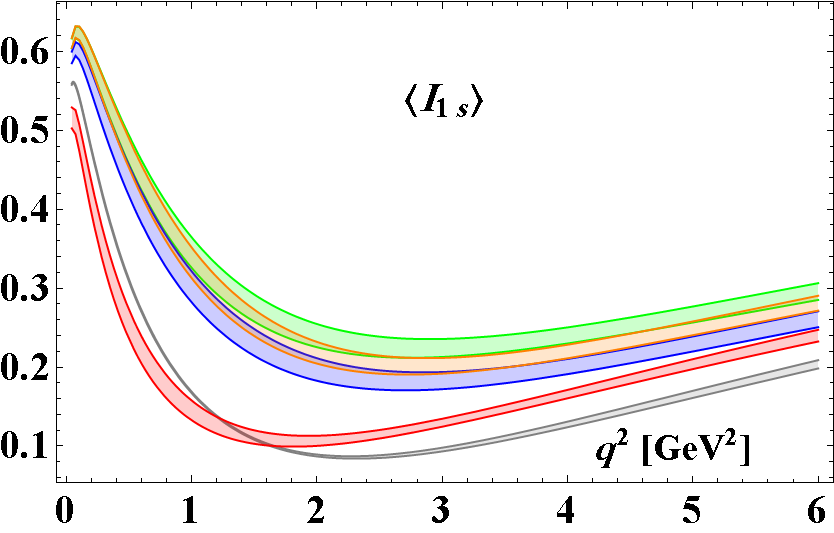}
\includegraphics[width=1.4in,height=1.25in]{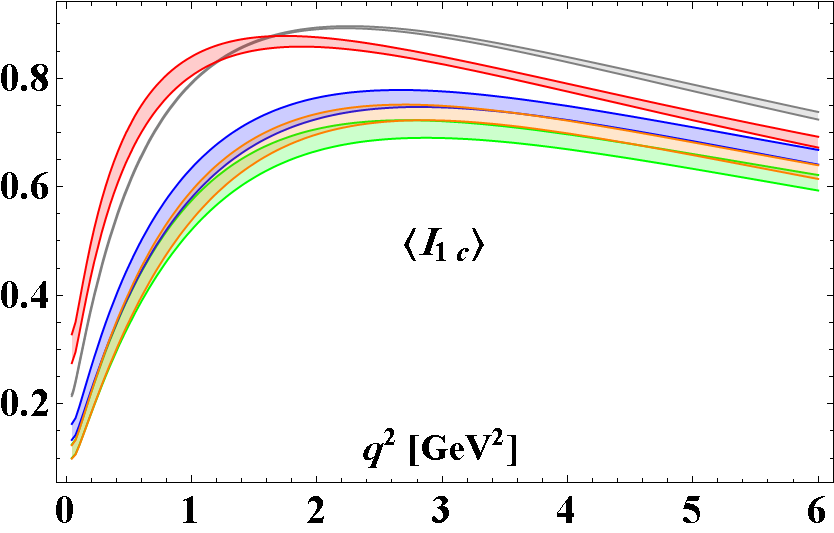}
\includegraphics[width=1.4in,height=1.25in]{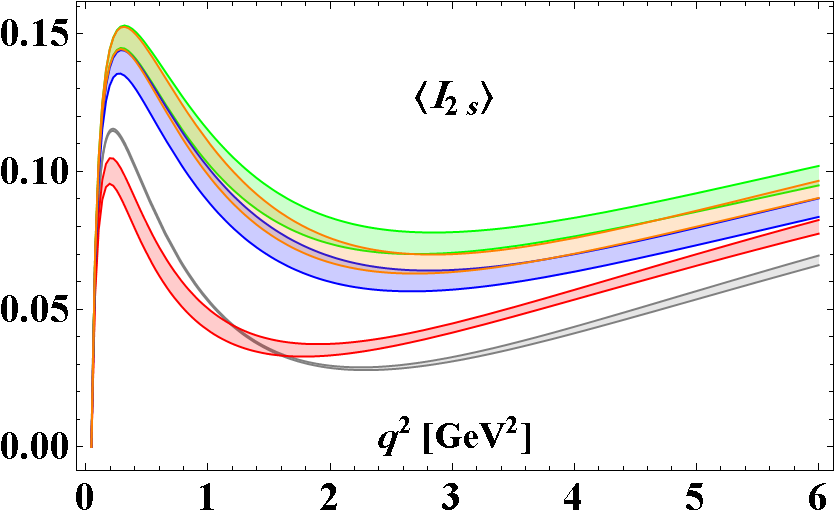}
\includegraphics[width=1.4in,height=1.25in]{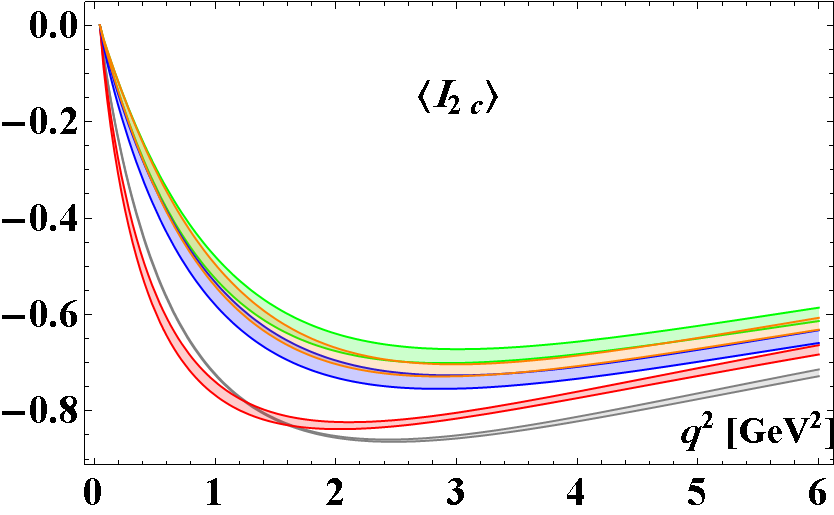}
\raisebox{0.08in}{\includegraphics[width=1.4in,height=1.17in]{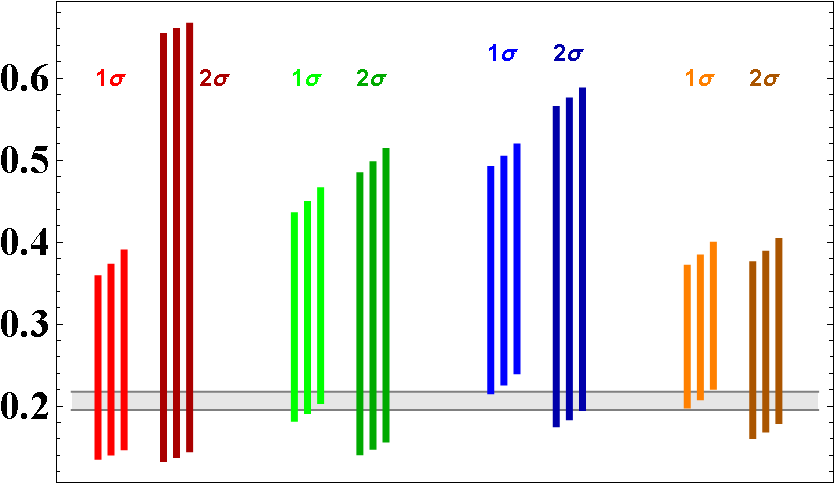}}
\raisebox{0.08in}{\includegraphics[width=1.4in,height=1.17in]{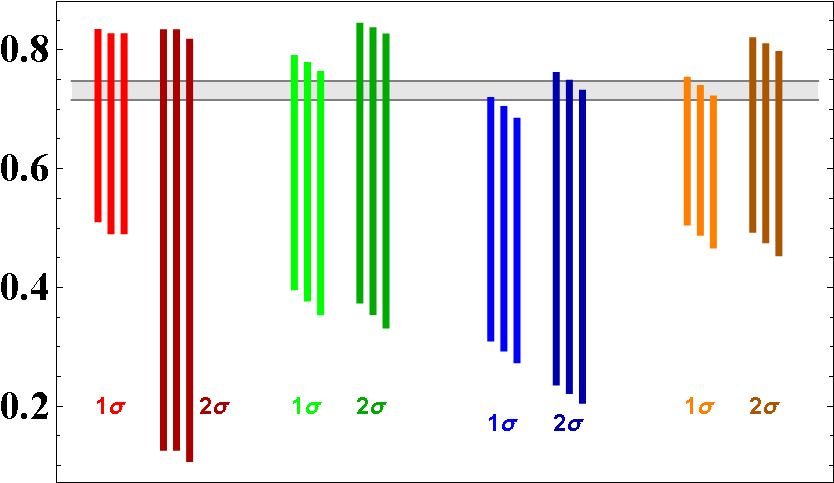}}
\raisebox{0.08in}{\includegraphics[width=1.4in,height=1.17in]{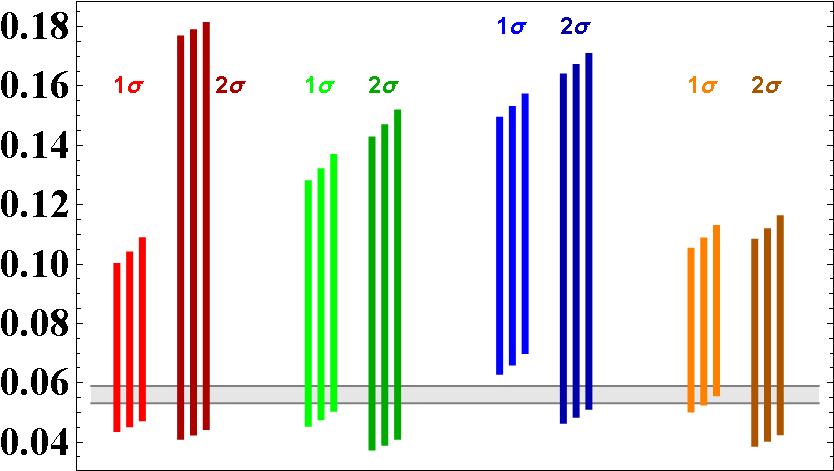}}
\raisebox{0.06in}{\includegraphics[width=1.4in,height=1.17in]{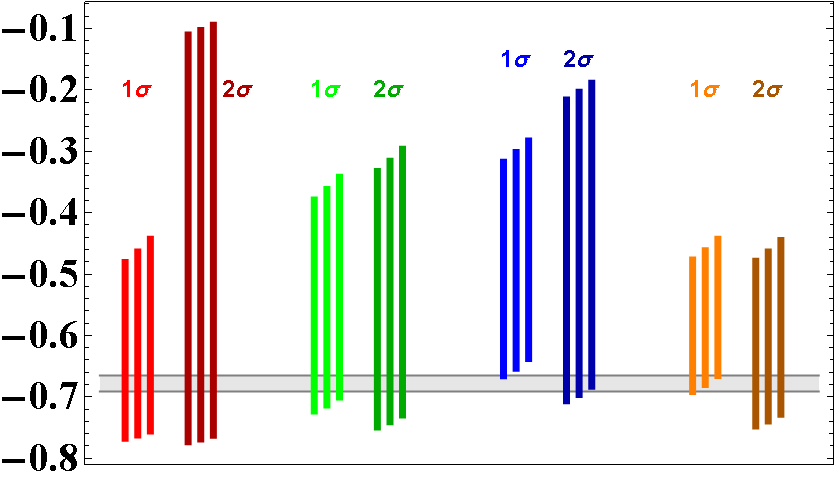}}
\caption{The angular coefficients  $\langle I_{1s} \rangle$, $\langle I_{1c} \rangle$, $\langle I_{2s} \rangle$, and $\langle I_{2c} \rangle$ for $B_s \to K^*(\to K\pi)\mu^+\mu^- $ where the remaining description is the same as describe in caption of Fig. \ref{2Dbrafbfl}.}
\label{2Dangularpnl1}
\end{figure}

\begin{figure}[H]
\centering
\includegraphics[width=1.4in,height=1.25in]{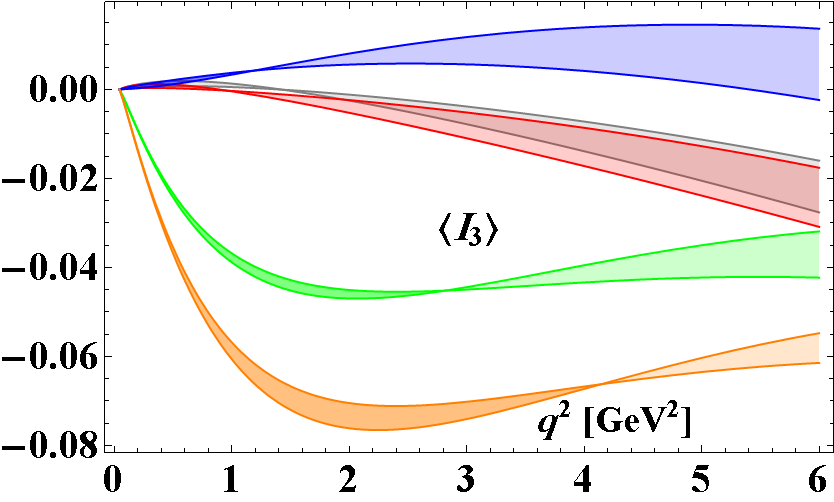}
\includegraphics[width=1.4in,height=1.25in]{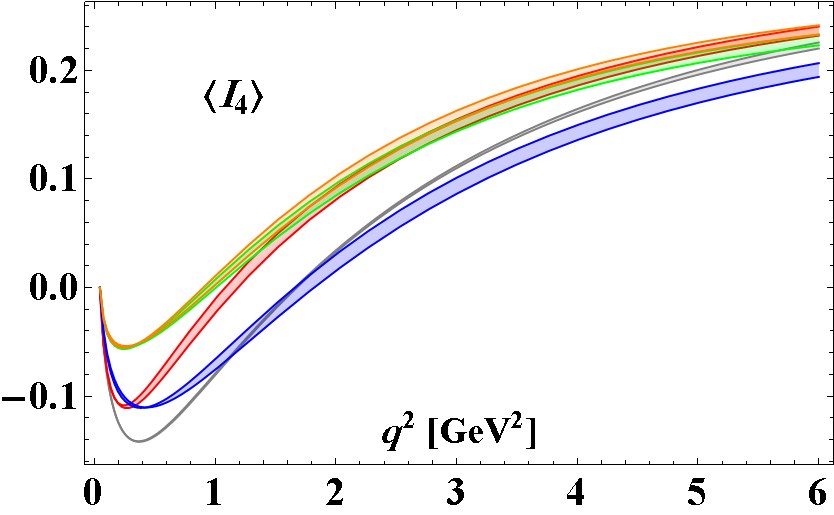}
\includegraphics[width=1.4in,height=1.25in]{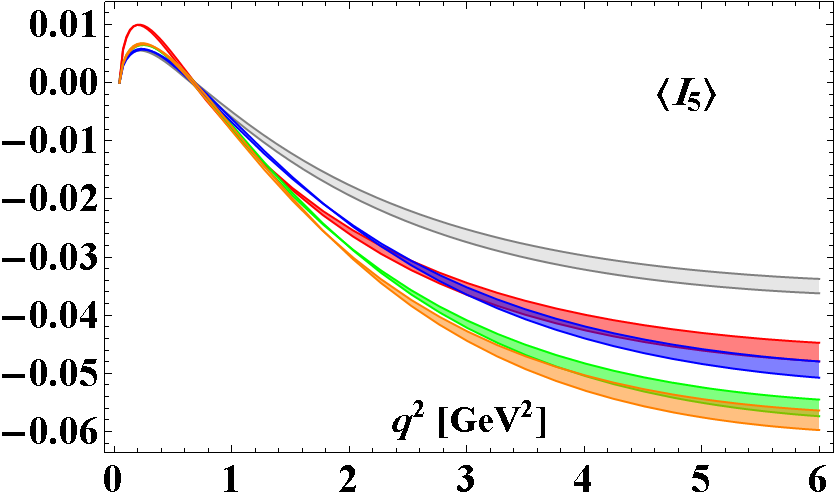}
\includegraphics[width=1.4in,height=1.25in]{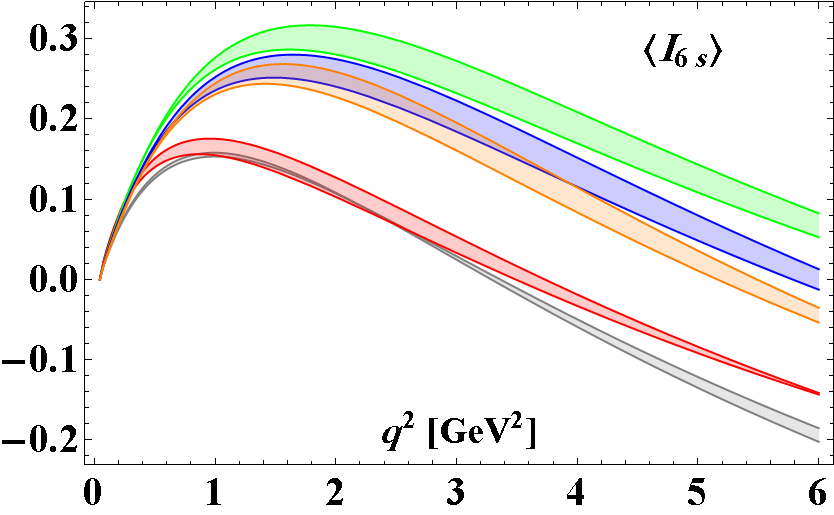}
\raisebox{0.08in}{\includegraphics[width=1.5in,height=1.17in]{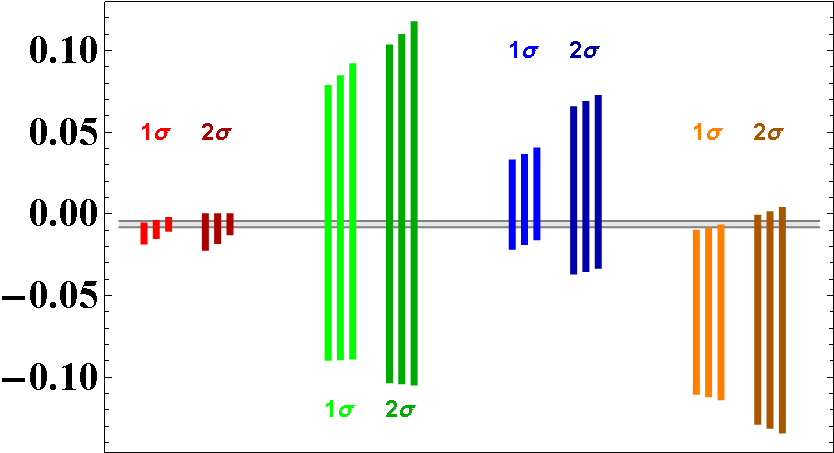}}
\raisebox{0.08in}{\includegraphics[width=1.4in,height=1.17in]{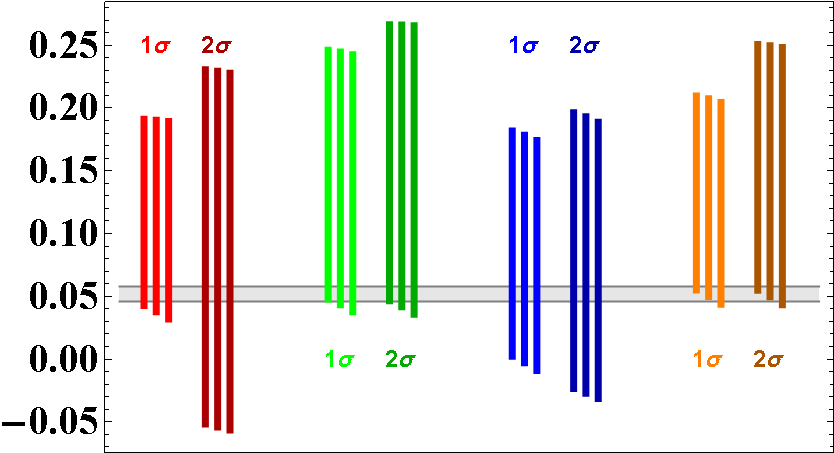}}
\raisebox{0.08in}{\includegraphics[width=1.4in,height=1.17in]{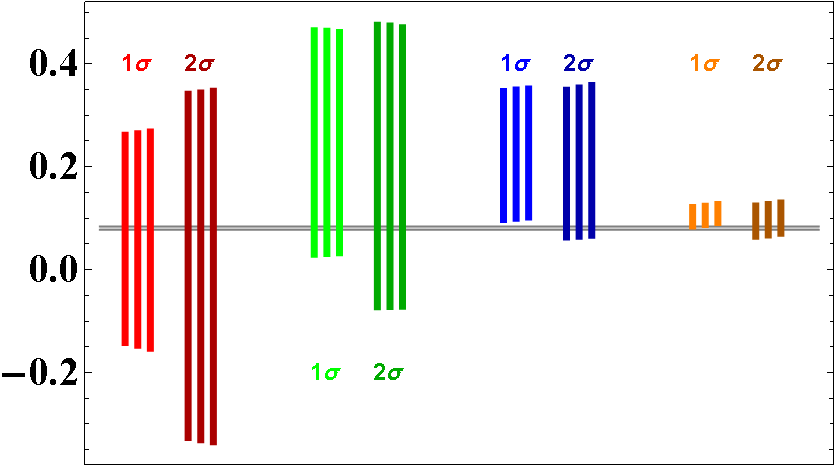}}
\raisebox{0.06in}{\includegraphics[width=1.4in,height=1.17in]{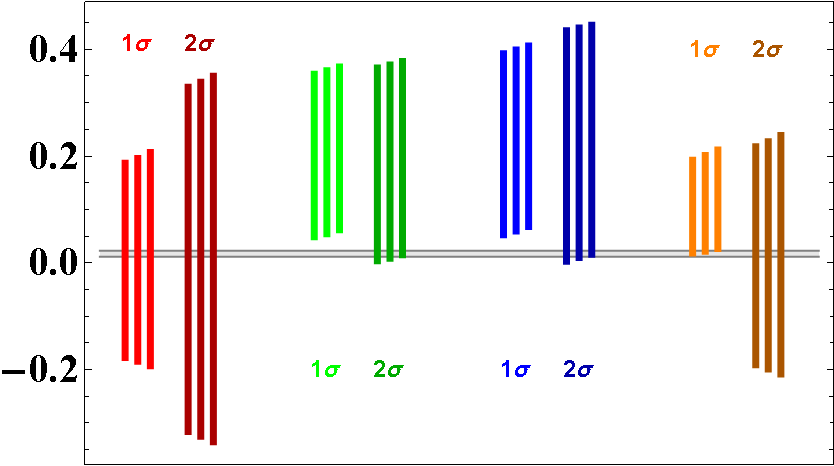}}
\caption{The  angular coefficients  $\langle I_{3} \rangle$, $\langle I_{4} \rangle$, $\langle I_{5} \rangle$, and $\langle I_{6s} \rangle$ for $B_s \to K^*(\to K\pi)\mu^+\mu^- $ where the remaining description is the same as describe in the caption of Fig. \ref{2Dbrafbfl}.}
\label{2Dangularpnl2}
\end{figure}

We present bar plots of the angular coefficients showing variation due to the full range of the WCs in the second row of Figs.~\ref{2Dangularpnl1}~and~\ref{2Dangularpnl2}. All scenarios show very large uncertainty bars as compared to the SM band. The scenarios SV and SVIII show some exceptionally small deviations only in $\langle I_3 \rangle$ and $\langle I_5 \rangle$ plots as compared to others. The over-all results give us a very pronounced picture of possible NP allowed scenarios which should be tested. 

\subsubsection{Correlations between observables}

As discussed earlier, correlations between the various physical observables can help better distinguish between different 
NP scenarios.

\begin{figure}[H]
\centering
\includegraphics[width=1.45in,height=1.2in,trim=0 0 0 0 ,clip]{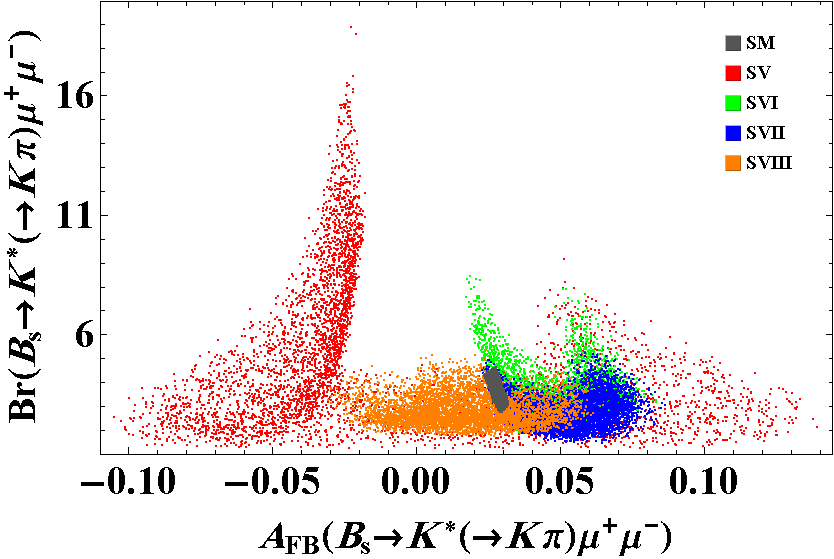}
\includegraphics[width=1.45in,height=1.2in,trim=0 0 0 0 ,clip]{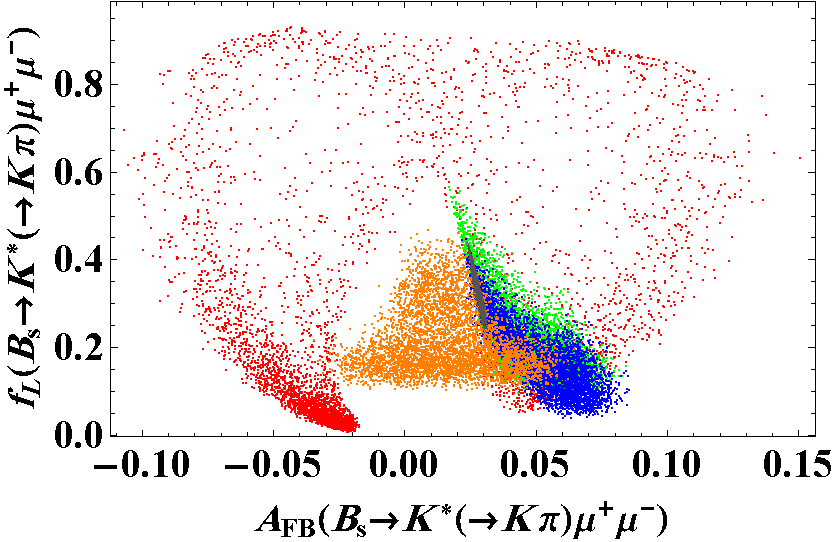}
\includegraphics[width=1.45in,height=1.2in,trim=0 0 0 0 ,clip]{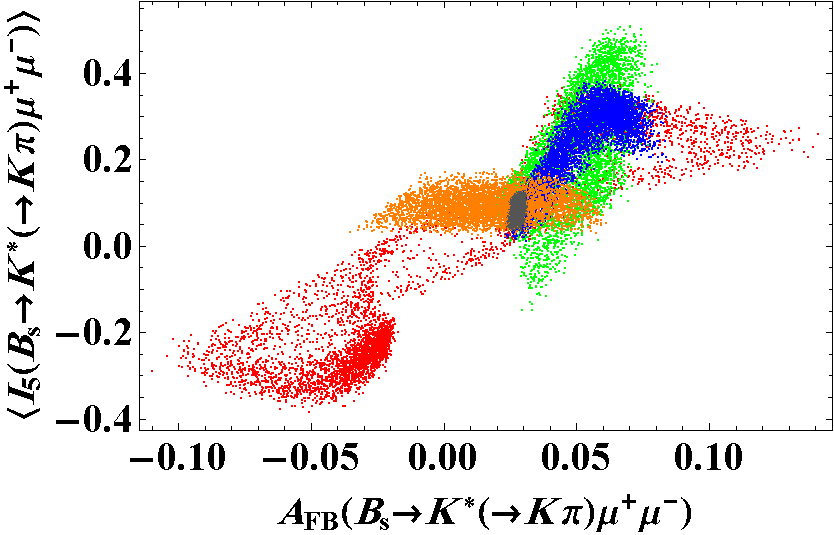}
\includegraphics[width=1.45in,height=1.2in,trim=0 0 0 0 ,clip]{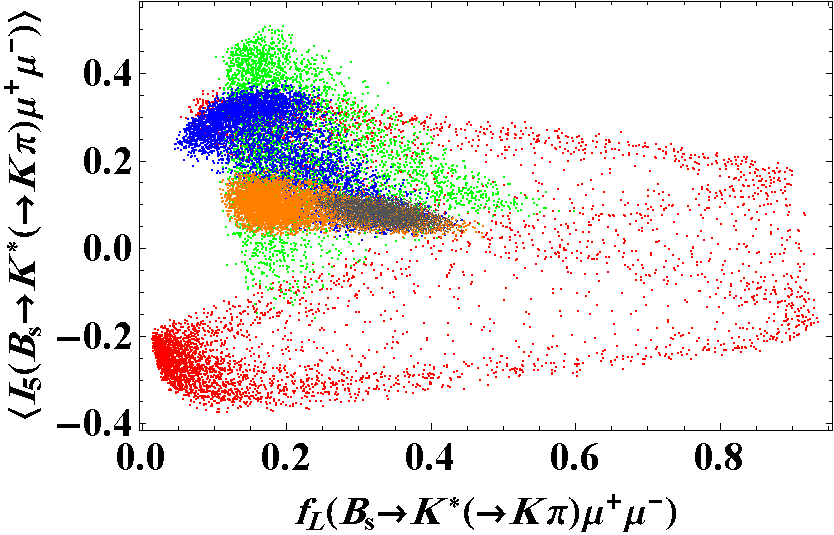}

\caption{Correlation between the different observables of $B_s \to K^*(\to K\pi)\mu^+\mu^- $ in SM and in 2D NP scenarios.}
\label{2Dcorrelation}
\end{figure}

With this in view, we present in Fig.~\ref{2Dcorrelation} representative correlations between physical observables. 

In the first plot, we show the correlation between the branching ratio and the forward-backward asymmetry where the SM result shows up as a narrow region but the different NP scenarios overlap in this plane, except for SV that clearly shows the deviation among others. In the second plot, we show how the longitudinal helicity fraction is correlated with the forward-backward asymmetry. In this plane, in addition to a region where SV is disentangled from the other scenarios, we can see SVIII occupy a distinctive region in the center of the `heart' that is formed in this plot. In the last two plots, we can see that in the correlations
between the angular observable $I_5$ against the forward-backward asymmetry and the helicity fraction, there are regions where SVI shows a distinctive character. These results further underscore the importance of a precision analysis in separating the various models from each other.

\begin{table}[H]
    \centering
    \renewcommand{\arraystretch}{1.5}
     \resizebox{\textwidth}{!}{
    {\fontsize{7}{10}
    \begin{tabular}{|c|c|c|c|c|c|}
\hline
\text{Dataset} & \text{SM} & SV & SVI & SVII & SVIII \\
\hline
$Br\times 10^{9}$ & (2.88, 4.50) & (1.27, 17.20) & (2.00, 8.52) & (1.56, 5.32) & (1.86, 5.18) \\
\hline
$A_{\text{FB}}$ & (0.03, 0.03) & (-0.09, 0.12) & (0.02, 0.07) & (0.03, 0.07) & (-0.02, 0.06) \\
\hline
$f_L$ & (0.31, 0.35) & (0.02, 0.91) & (0.13, 0.50) & (0.06, 0.34) & (0.14, 0.41) \\
\hline
$\langle I_{1s} \rangle$ & (0.20, 0.22) & (0.14, 0.66) & (0.15, 0.51) & (0.18, 0.58) & (0.16, 0.40) \\
\hline
$\langle I_{1c} \rangle$ & (0.75, 0.72) & (0.11, 0.83) & (0.34, 0.84) & (0.21, 0.76) & (0.46, 0.82) \\
\hline
$\langle I_{2s} \rangle$ & (0.05, 0.06) & (0.04, 0.18) & (0.04, 0.15) & (0.05, 0.17) & (0.04, 0.12) \\
\hline
$\langle I_{2c} \rangle$ & (-0.69, -0.67) & (-0.77, -0.09) & (-0.75, -0.30) & (-0.71, -0.19) & (-0.75, -0.45) \\
\hline
$\langle I_{3} \rangle$ & (-0.01, 0.00) & (-0.02, 0.00) & (-0.10, 0.12) & (-0.03, 0.07) & (-0.13, 0.00) \\
\hline
$\langle I_4 \rangle$ & (0.05, 0.06) & (-0.06, 0.23) & (0.04, 0.27) & (-0.03, 0.20) & (0.04, 0.25) \\
\hline
$\langle I_{5} \rangle$ & (0.08, 0.08) & (-0.33, 0.35) & (-0.07, 0.48) & (0.06, 0.36) & (0.07, 0.13) \\
\hline
$\langle I_{6s} \rangle$  & (0.01, 0.02) & (-0.34, 0.35) & (0.00, 0.38) & (0.00, 0.45) & (-0.21, 0.24) \\
\hline
    \end{tabular}
    }}
    \caption{Results for the $q^2$ bin [$q^2_{\text{min}}$, 6] GeV$^2$ in the SM as well as for 2D NP scenarios for different observables.}
    \label{tab:extended}
\end{table}

\subsubsection{Contour Plots}

In Fig.~\ref{2Ddensitybrafbfl} we show contour plots in three rows corresponding to observables; branching ratio, forward-backward asymmetry and longitudinal helicity fraction.
We depict the four 2D scenarios in the columns of Fig.~\ref{2Ddensitybrafbfl} with SV-SVIII plotted in sequence, with the axes showing the NP WCs in the corresponding scenario. Each contour plot shows ten contours with the variable magnitude bar shown to the right of each plot representing the percentage deviation of the NP result from the SM central value. To be precise, we are plotting the contours corresponding to:
\[\frac{O^{NP}_{i} - O^{SM}_{i}}{O_{SM}} \times100\%,\]
where $O^{NP}_i$ is the value of the $i^{\rm th}$ observable in some NP scenario and $O_i^{SM}$ is its value in the SM. The 
point correspondong to the SM central value always occurs for zero values of the NP coefficients and is shown as a dot in these 
plots. We also show for comparison the SM value of the observable with a dot and the best-fit value with an asterisk. Finally, we 
have drawn a rectangle that corresponds to the $1\sigma$ range of the NP WCs. Similarly, we show in Fig.~\ref{2Dcontangular1} 
and Fig.~\ref{2Dcontangular2} contour plots along these lines for the angular coefficients $\langle I_i\rangle$.

\begin{figure}[H]
\centering

\makebox[5.7in][c]{%
  \textbf{SV} \hspace{2.5cm} \textbf{SVI} \hspace{2.5cm} \textbf{SVII} \hspace{2.5cm} \textbf{SVIII}
}

\vspace{0.1cm} 

\includegraphics[width=5.7in,height=1.3in]{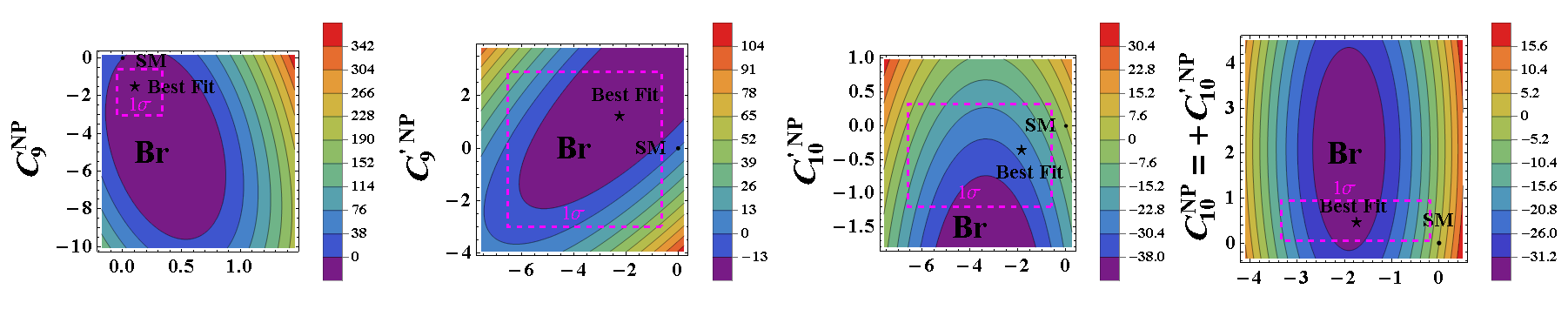}
\vspace{-0.35cm}

\includegraphics[width=5.7in,height=1.3in]{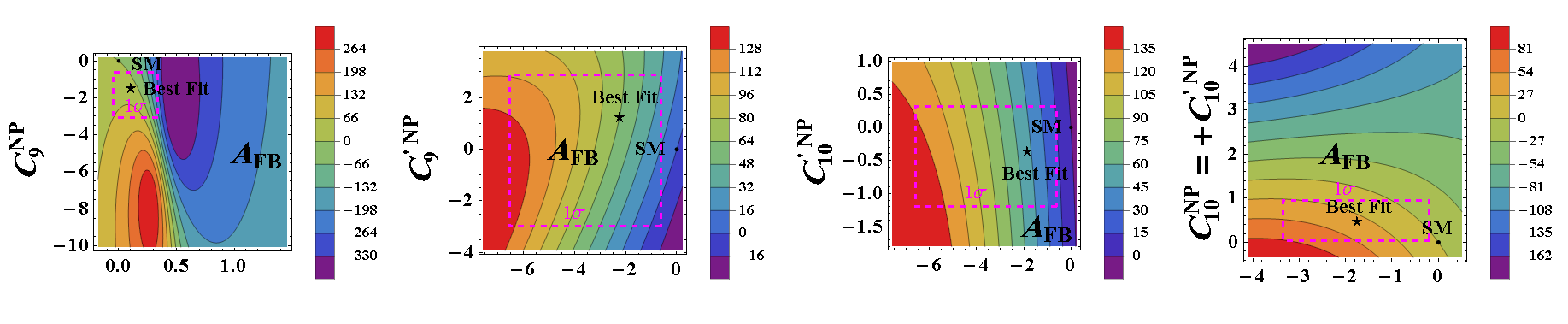}
\vspace{-0.3cm}

\includegraphics[width=5.7in,height=1.3in]{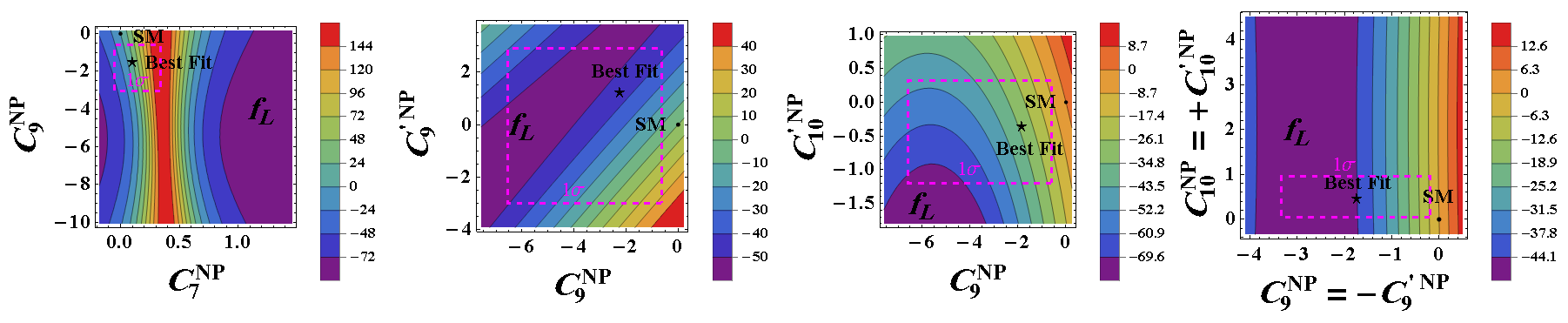}

\caption{The contour plots of Br, $A_{\text{FB}}$, and $f_L$ by using the $1\sigma$ and $2\sigma$ allowed regions of 2D WCs. The black dot and star are defined by the SM value and the value by using the best fit point of WCs. The dashed rectangular box represents the values of observables by using $1\sigma$ parametric space of WCs.}
\label{2Ddensitybrafbfl}
\end{figure}

The contour plots in Figs.~\ref{2Ddensitybrafbfl}-\ref{2Dcontangular2} show a useful visual presentation of the effects of the 
2D NP scenarios. In particular, we can see that SM result is always outside the best-fit $1\sigma$ region. Further, we note that 
large variations are predicted by various NP scenarios. In particular, we can see that the $Br$ can vary as much as 342\% in scenario 
SV. Likewise, the maximum variation in the $A_{\text{FB}}$ can be seen to be 264\% and -330\% in SV, while for $f_L$ it is 144\% corresponding to SV again. Focusing at the angular coefficients, we see that among these the $\langle I_{6s}\rangle$ shows the maximum variation from the SM result of 2400\% in scenario SVII and -1850\% in SV. The variation of $\langle I_{6s}\rangle$ is very high in the remaining scenarios as well. Similarly $\langle I_3\rangle$ shows the second largest deviation of 1050\% and -2100\% in SVI scenario. Other angular coefficients also depict the variations of a few hundred percents across different scenarios that can be seen in the plots.

\begin{figure}[H]
    \centering

    \makebox[5.7in][c]{%
  \textbf{SV} \hspace{2.5cm} \textbf{SVI} \hspace{2.5cm} \textbf{SVII} \hspace{2.5cm} \textbf{SVIII}
}

    \includegraphics[width=5.7in,height=1.3in]{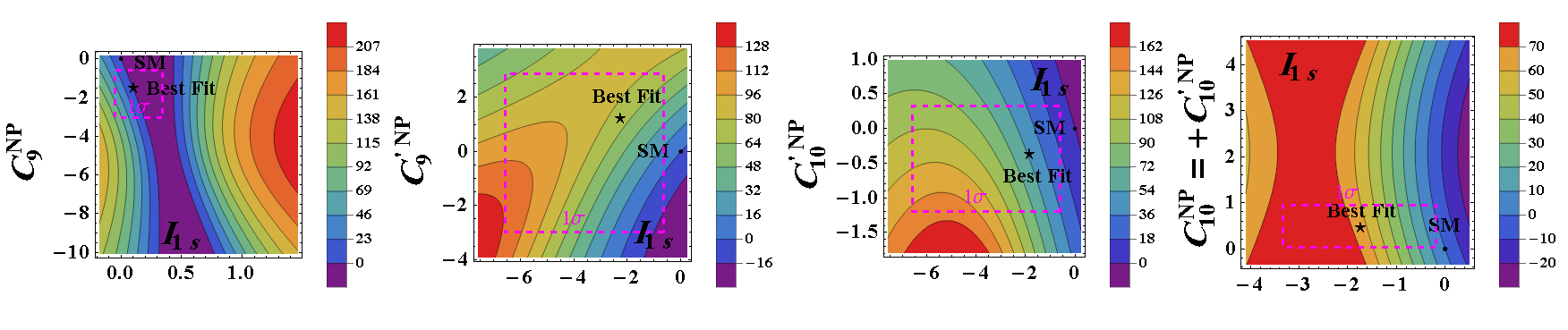}
    
    \vspace{-0.3cm} 
    \includegraphics[width=5.7in,height=1.3in]{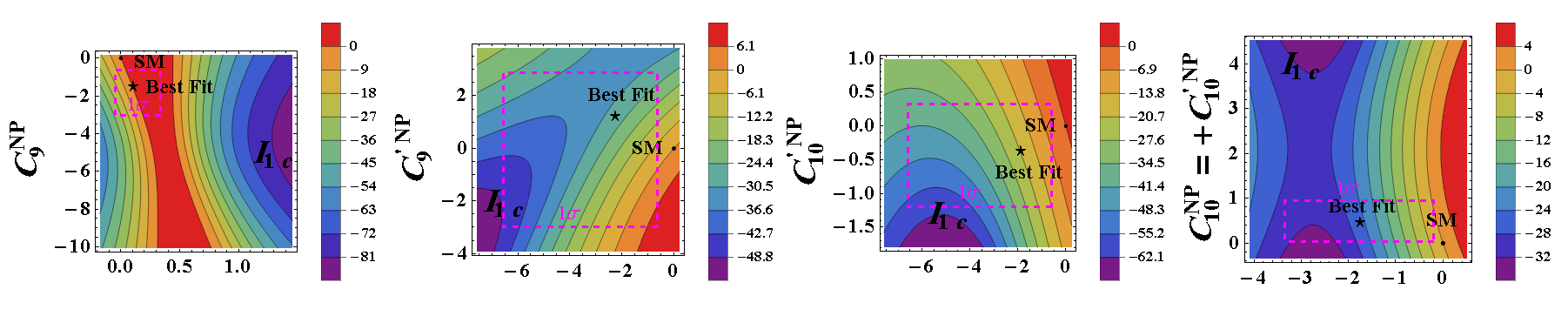}
    
    \vspace{-0.3cm} 
    \includegraphics[width=5.7in,height=1.3in]{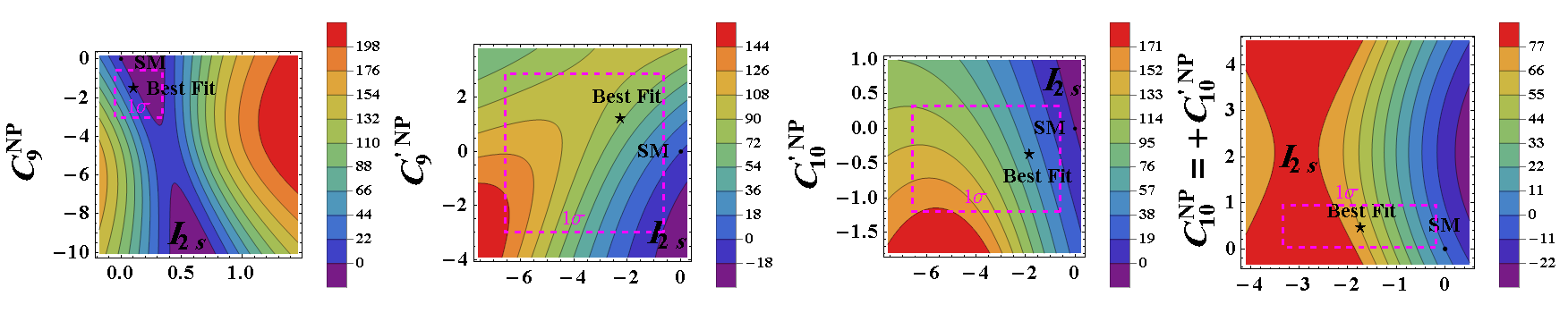}
    
    \vspace{-0.2cm} 
    \includegraphics[width=5.7in,height=1.3in]{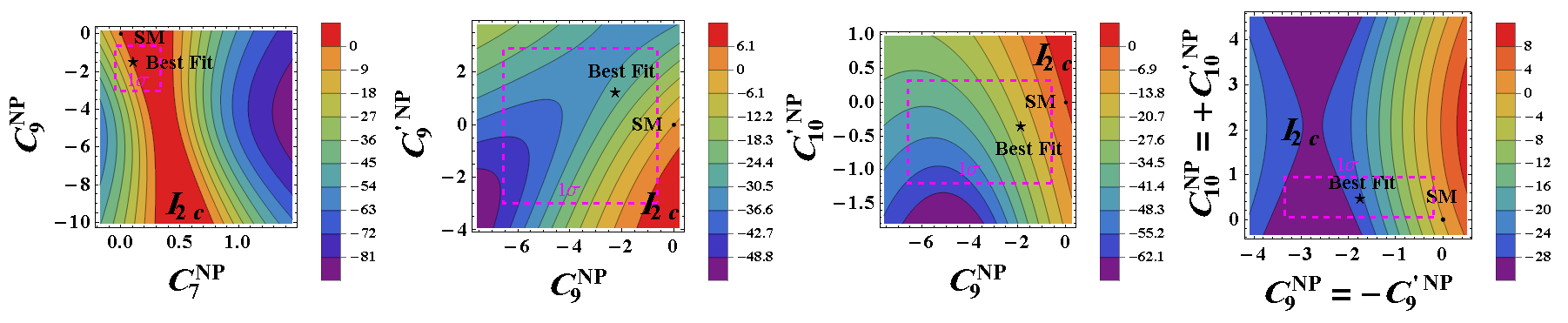}
    \caption{The contour plots of $\langle I_{1s} \rangle$, $\langle I_{1c} \rangle$, $\langle I_{2s} \rangle$, and $\langle I_{2c} \rangle$ by using the $1\sigma$ and $2\sigma$ allowed regions of 2D WCs. The black dot and star are defined by the SM value and the value by using the best fit point of WCs. The dashed rectangular box represents the values of observables by using $1\sigma$ parametric space of Wcs.}
    \label{2Dcontangular1}
\end{figure}
 
These contour plots further illustrate the main point of our analysis: a detailed experimental study of this channel 
can lead to interesting results from the point of view of testing the SM fit and constraining various NP scenarios. In particular, 
even within the current $2\sigma$ fit of the SM, we can see results that can differ to a large extent (upto 2400\%, as in the 
case of $\langle I_{6s}\rangle$) from the SM value.

\begin{figure}[H]
\centering

\makebox[5.7in][c]{%
  \textbf{SV} \hspace{2.5cm} \textbf{SVI} \hspace{2.5cm} \textbf{SVII} \hspace{2.5cm} \textbf{SVIII}
}

\includegraphics[width=5.7in,height=1.3in]{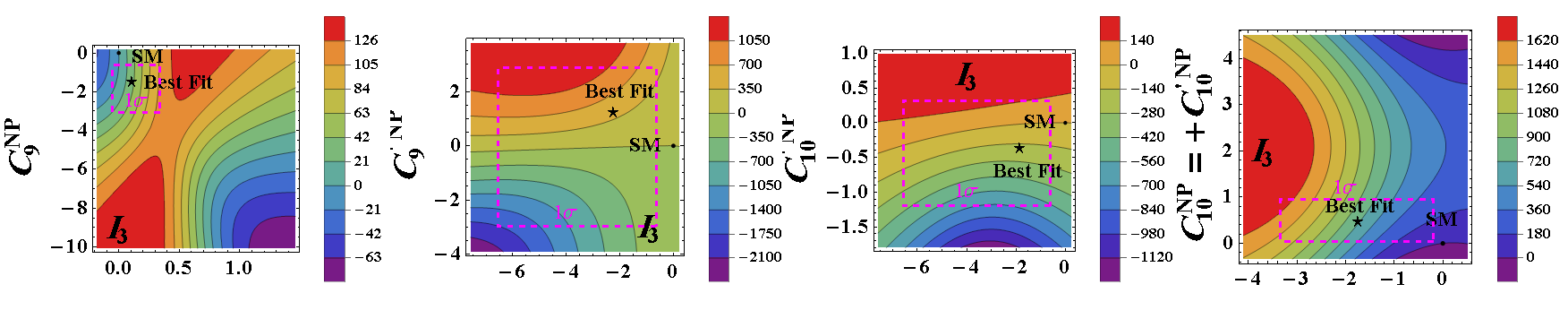}

\vspace{-0.3cm} 
\includegraphics[width=5.7in,height=1.3in]{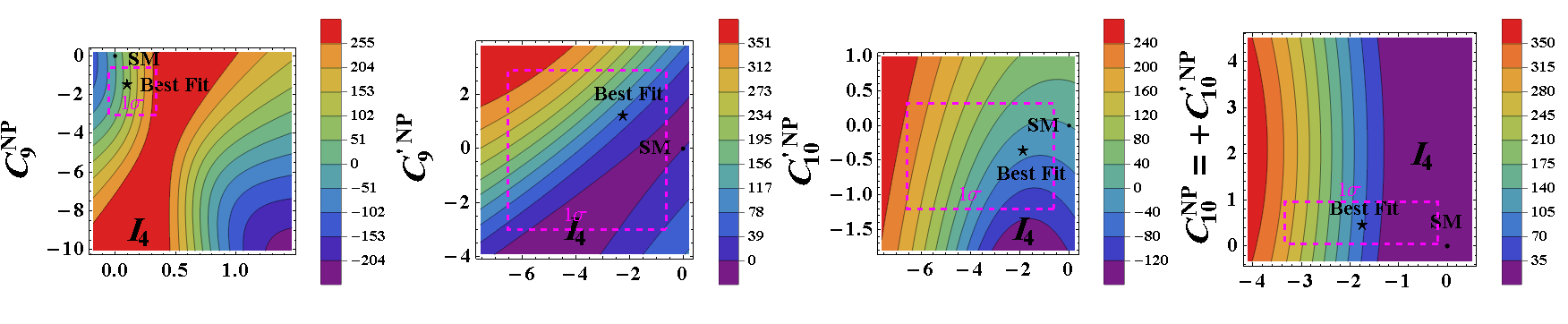}

\vspace{-0.3cm} 
\includegraphics[width=5.7in,height=1.3in]{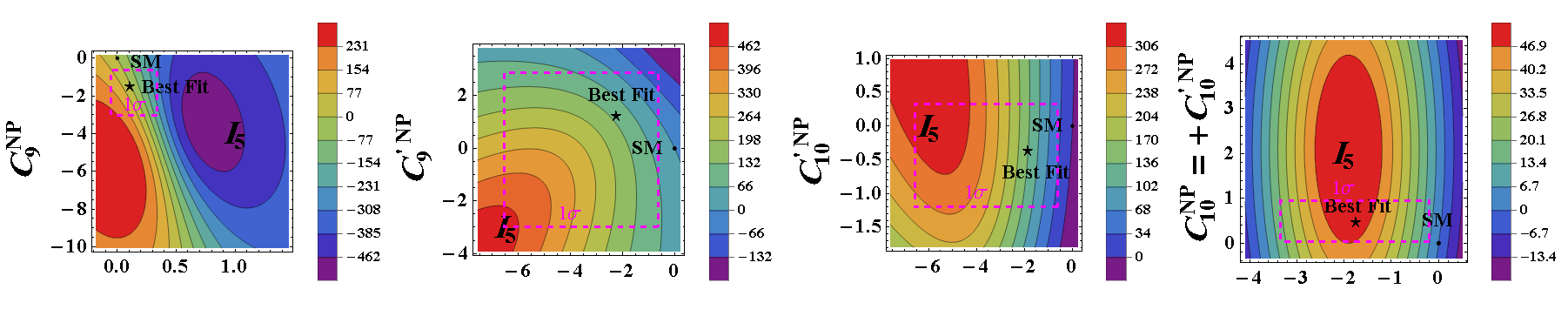}

\vspace{-0.2cm} 
\includegraphics[width=5.7in,height=1.3in]{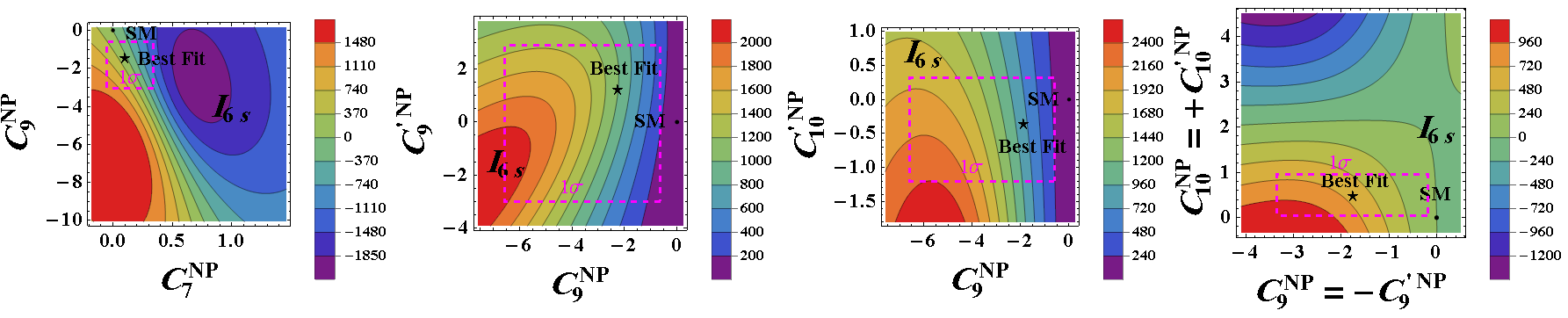}

\caption{The contour plots of $\langle I_{3} \rangle$, $\langle I_{4} \rangle$, $\langle I_{5} \rangle$, and $\langle I_{6s} \rangle$ by using the $1\sigma$ and $2\sigma$ allowed regions of 2D WCs. The black dot and star are defined by the SM value and the value by using the best fit point of WCs. The dashed rectangular box represents the values of observables by using $1\sigma$ parametric space of Wcs.}
\label{2Dcontangular2}
\end{figure}

\section{Conclusion}\label{conl}

In this work, we have carried out a model-independent study of the rare decay process $B_s \to K^*(\to K\pi)\mu^+\mu^-$ within the framework of weak 
effective field theory. 
This decay, governed by the suppressed flavor changing neutral current transition $b \to d\mu^+\mu^-$, offers a valuable opportunity to explore possible effects of  physics beyond the standard model.

We computed several physical observables including the differential branching ratio, the forward-backward asymmetry $A_{\text{FB}}$, the longitudinal polarization fraction $f_L$, and a set of normalized angular coefficients $\langle I_i \rangle$, within the context of the standard model and various 
NP scenarios. The impact of NP is explored in both one and two dimensional scenarios involving NP WCs $C_7^{\text{NP}}$, $C_9^{(\prime)\text{NP}}$, and $C_{10}^{(\prime)\text{NP}}$.

We have also explored correlations between different observables and presented contour plots to highlight the interplay 
between observables and the allowed ranges of WCs. These visual tools may serve as a useful complement for future analyses, 
especially as more precise experimental data become available. Additionally, focusing on correlation plots allows us to better 
distinguish between different NP scenarios.

Our findings reveal notable deviations from the SM predictions across multiple observables. Through correlations among observables and 2D contour plots of the decay $B_s \to K^*(\to K\pi)\mu^+\mu^- $, we identify parameter regions consistent with the current experimental measurements of the other $b\to d$ decay modes, thereby providing constraints on possible NP contributions. Although these deviations are intriguing, they should be interpreted with caution, as theoretical uncertainties, particularly those related to form factors, and current experimental limitations may affect their significance.
Overall, the current analysis suggests that the $B_s \to K^*\mu^+\mu^-$ decay could serve as a supplementary probe in the broader effort to test the SM and search for signs of NP.

\section*{Acknowledgements}

M.A.P and Z.A would like to acknowledge the financial support provided by the Higher Education Commission (HEC) of Pakistan
through Grant no. NRPU/20-15142 .


\appendix
\section*{Appendix}
\section{Expressions for Wilson coefficients}\label{append123}

The expressions used for the WCs are given as under \cite{Bobeth:1999mk,Beneke:2001at,Asatrian:2001de,Asatryan:2001zw,Greub:2008cy,Du:2015tda},
\begin{eqnarray}
C_{7}^{eff}(q^2)&=&C_{7}-\frac{1}{3}\left(C_{3}+\frac{4}{3}C_{4}+20C_{5}+\frac{80}{3}C_{6}\right)
-\frac{\alpha_{s}}{4\pi}\left[(C_{1}-6C_{2})F^{(7)}_{1,c}(q^2)+C_{8}F^{(7)}_{8}(q^2)\right]\notag\\&-&\frac{\alpha_{s}}{4\pi}\lambda^{(q)}_{u}\left(C_{1}-6C_{2}\right)\left(F^{7}_{1,c}-F^{7}_{1,u}\right),\notag\\
C_{9}^{eff}(q^2)&=&C_{9}+\frac{4}{3}\left(C_{3}+\frac{16}{3}C_{5}+\frac{16}{9}C_{6}\right)
-h(0, q^2)\left(\frac{1}{2}C_{3}+\frac{2}{3}C_{4}+8C_{5}+\frac{32}{3}C_{6}\right)\notag\\
&-&h(m_{b}^{\text{pole}}, q^2)\big(\frac{7}{2}C_{3}+\frac{2}{3}C_{4}+38C_{5}+\frac{32}{3}C_{6}\big)+h(m_{c}^{\text{pole}}, q^2)
\big(\frac{4}{3}C_{1}+C_{2}+6C_{3}\notag \\
&+&60C_{5}\big) +\lambda^{(q)}_{u}\left[h(m_{c},q^{2})-h(0,q^{2})\right]\left(\frac{4}{3}C_{1}+C_{2}\right)
-\frac{\alpha_{s}}{4\pi}\big[C_{1}F^{(9)}_{1,c}(q^2) \notag \\
&+& C_{2}F^{(9)}_{2,c}(q^2) + C_{8}F^{(9)}_{8}(q^2)\big]-\frac{\alpha_{s}}{4\pi}\lambda^{(q)}_{u}\left[C_{1}(F^{(9)}_{1,c}-F^{(9)}_{1,u})+C_{2}(F^{(9)}_{2,c}-F^{(9)}_{2,u})\right],\label{WC3}
\end{eqnarray}
where the functions $h(m_{q}^{\text{pole}}, q^2)$ with $q=c, b$ and the functions $F^{(7,9)}_{8}(q^2)$ are
defined in \cite{Beneke:2001at}, while the functions $F^{(7,9)}_{1,c}(q^2)$, $F^{(7,9)}_{2,c}(q^2)$ are
given in \cite{Asatryan:2001zw} for low $q^{2}$ and in \cite{Greub:2008cy} for high $q^{2}$.

\begin{figure}[H]
    \centering
    \includegraphics[width=0.4\linewidth]{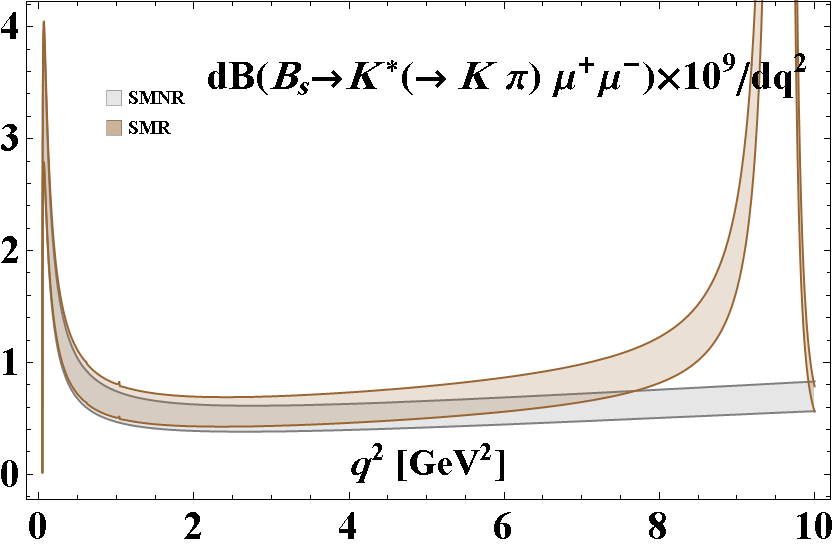}
    \caption{Resonance effects: The gray band (SMNR) presents the SM behavior without the resonance effects, while the brown band (SMR) presents the effects of resonance on the branchin ratio.}
    \label{reseff}
\end{figure}

\section{Semi-analytical expressions for the physical observables\label{AppSemiAnalytical}}

The general binned semi-analytical expression for the different observables given in Eqs. (\ref{Br1})-(\ref{angobsorig}) after integrating over the kinematical 
region $q^{2}=\left[q^2_{\text{min}}, 6\right]$ GeV$^2$ are given as,
\begin{equation}
    \mathcal{O}^{\text{tot}}_{i} = \frac{\mathcal{O}^{\text{N}}_{i}}{\mathcal{O}^{\text{D}}_{i}},
\end{equation}
where $\mathcal{O}^{\text{tot}}_{i}$ is the observable under consideration containing the SM as well as the NP contributions. $\mathcal{O}^{\text{N}}_{i}$ and $\mathcal{O}^{\text{D}}_{i}$ represent the numerator and denominator of the corresponding observable (the branching ratio ${d\mathcal{B}(B_{s}\to K^{\ast}\mu^{+}\mu^{-})}/{dq^{2}}$, the forward-backward asymmetry $A_{\text{FB}}$, the longitudinal polarization of the $K^{\ast}$ meson $f_{L}$, and the normalized angular coefficients $\langle I_{i}\rangle$). The expressions for the numerator $\mathcal{O}^{N}_{i}$ and denominator $\mathcal{O}^{D}_{i}$ can be written as,
\begin{align}
\mathcal{O}_{\text{N}}^i = &\mathcal{A}_1 + \mathcal{A}_2 C_{10}^{NP} + \mathcal{A}_3 (C_{10}^{NP})^2 + \mathcal{A}_4 C_{10}^{'NP} + \mathcal{A}_5 C_{10}^{NP} C_{10}^{'NP} + \mathcal{A}_6 (C_{10}^{'NP})^2 + \mathcal{A}_7 C_{7}^{NP} \nonumber \\
&+ \mathcal{A}_8 (C_7^{NP})^2 + \mathcal{A}_9 C_7^{'NP} + \mathcal{A}_{10} C_{7}^{NP} C_7^{'NP} + \mathcal{A}_{11} (C_7^{'NP})^2 + \mathcal{A}_{12} C_9^{NP} + \mathcal{A}_{13} C_{7}^{NP} C_9^{NP} \notag\\
&+ \mathcal{A}_{14} C_7^{'NP} C_9^{NP} + \mathcal{A}_{15} (C_{9}^{NP})^2 + \mathcal{A}_{16} C_9^{'NP} + \mathcal{A}_{17} C_{7}^{NP} C_9^{'NP} + \mathcal{A}_{18} C_7^{'NP} C_9^{'NP} \notag \\
&+ \mathcal{A}_{19} C_9^{NP} C_9^{'NP} + \mathcal{A}_{20} (C_9^{'NP})^2 
+ \mathcal{A}_{21} C_{10}^{NP} C_{7}^{NP} + \mathcal{A}_{22} C_{10}^{'NP} C_{7}^{NP} + \mathcal{A}_{23} C_{10}^{NP} C_7^{'NP} \notag \\
&+ \mathcal{A}_{24} C_{10}^{'NP} C_7^{'NP} + \mathcal{A}_{25} C_{10}^{NP} C_9^{NP} + \mathcal{A}_{26} C_{10}^{'NP} C_9^{NP} + \mathcal{A}_{27} C_{10}^{NP} C_9^{'NP}\notag\\
&+ \mathcal{A}_{28} C_{10}^{'NP} C_9^{'NP}\notag
\end{align}
and
\begin{align}
\mathcal{O}_{\text{D}}^i=&\mathcal{B}_1 + \mathcal{B}_2 C_{10}^{NP} + \mathcal{B}_3 (C_{10}^{NP})^2 
+ \mathcal{B}_4 C_{10}^{'NP} + \mathcal{B}_5 C_{10}^{NP} C_{10}^{'NP} + \mathcal{B}_6 (C_{10}^{'NP})^2 + \mathcal{B}_7 C_{7}^{NP} \notag \\
&+ \mathcal{B}_8 (C_7^{NP})^2
+ \mathcal{B}_9 C_7^{'NP} + \mathcal{B}_{10} C_{7}^{NP} C_7^{'NP} + \mathcal{B}_{11} (C_7^{'NP})^2 + \mathcal{B}_{12} C_9^{NP} + \mathcal{B}_{13} C_{7}^{NP} C_9^{NP}\notag \\
&+\mathcal{B}_{14} C_7^{'NP} C_9^{NP}
+ \mathcal{B}_{15} (C_{9}^{NP})^2+ \mathcal{B}_{16} C_9^{'NP} + \mathcal{B}_{17} C_{7}^{NP} C_9^{'NP} + \mathcal{B}_{18} C_7^{'NP} C_9^{'NP}\notag\\
&+ \mathcal{B}_{19} C_9^{NP} C_9^{'NP} + \mathcal{B}_{20} (C_9^{'NP})^2+ \mathcal{B}_{21} C_{10}^{NP} C_{7}^{NP} + \mathcal{B}_{22} C_{10}^{'NP} C_{7}^{NP} + \mathcal{B}_{23} C_{10}^{NP} C_7^{'NP}\notag\\
&+ \mathcal{B}_{24} C_{10}^{'NP} C_7^{'NP} + \mathcal{B}_{25} C_{10}^{NP} C_9^{NP}+ \mathcal{B}_{26} C_{10}^{'NP} C_9^{NP} + \mathcal{B}_{27} C_{10}^{NP} C_9^{'NP}\notag\\
&+ \mathcal{B}_{28} C_{10}^{'NP} C_9^{'NP},
\label{Expressions}
\end{align}
where the numerical values of the coefficients $\mathcal{A}_i$ and $\mathcal{B}_i$ are given in Tables \ref{tabcof1} and \ref{tabcof2}. The input parameters used in the evaluation of the above-mentioned physical observables include the masses and lifetimes of the particles, as well as the elements of the CKM matrix which are listed in Table~\ref{input}.

\begin{table}[H]
    \centering
    \renewcommand{\arraystretch}{1.5}
     \resizebox{\textwidth}{!}{
    \begin{tabular}{||>{\columncolor{green!5}}c|c|c|c|c|c|c|c|c|c|}
        \hline
        \textbf{Coeff.} & $Br \times 10^{9}$ & $A_{FB} \times 10^{-3}$ & $f_{l} \times 10^{-3}$ & $\langle I_{1s} \rangle $ & $\langle I_{1c} \rangle $ & $\langle I_{2s} \rangle $ & $\langle I_{2c} \rangle $ & $\langle I_{3} \rangle $ \\
        \hline
        $\mathcal{A}_1$ & $3.64^{+4.49}_{+2.87}$ & $28.7^{+5.14}_{-4.73}$ & $170^{+47.0}_{-41.3}$ & $91.6^{+16.2}_{-14.7}$ & $328^{+84.8}_{-75.0}$ & $24.9^{+4.47}_{-4.06}$ & $-77.8^{-77.8}_{-68.9}$ & $-2.83^{-1.63}_{+1.35}$ \\
        \hline
        $\mathcal{A}_2$ & $-0.74^{-0.92}_{-0.58}$ & $-6.85^{-1.23}_{-1.12}$ & $-29.1^{-8.36}_{+7.29}$ & $-16.1^{-3.23}_{+2.92}$ & $-70.9^{-18.4}_{+16.2}$ & $-5.37^{-1.08}_{-0.98}$ & $17.0^{+17.0}_{-15.1}$ & $1.29^{+0.72}_{-0.62}$ \\
        \hline
        $\mathcal{A}_3$ & $0.08^{+0.11}_{+0.07}$ & - & $3.47^{+0.99}_{-0.87}$ & $1.92^{+0.38}_{-0.34}$ & $8.45^{+2.19}_{-1.94}$ & $0.64^{+0.13}_{-0.12}$ & $-2.03^{+2.03}_{-1.80}$ & $-0.15^{-0.09}_{+0.07}$ \\
        \hline
        $\mathcal{A}_4$ & $0.58^{+0.74}_{+0.44}$ & $0.45^{+0.29}_{-0.26}$ & $29.1^{+8.36}_{-7.29}$ & $1.93^{+1.08}_{-0.93}$ & $70.9^{+18.4}_{-16.2}$ & $0.64^{+0.36}_{-0.31}$ & $-17.0^{-17.0}_{-15.1}$ & $-10.7^{-2.16}_{+1.95}$ \\
        \hline
        $\mathcal{A}_5$ & $-0.14^{-0.18}_{-0.10}$ & - & $-6.95^{-1.99}_{+1.74}$ & $-0.46^{-0.26}_{+0.22}$ & $-16.9^{-4.39}_{+3.88}$ & $-0.154^{-0.09}_{-0.07}$ & $4.05^{+4.05}_{-3.59}$ & $2.56^{+0.52}_{-0.47}$ \\
        \hline
        $\mathcal{A}_6$ & $0.08^{+0.11}_{+0.07}$ & - & $3.47^{+0.99}_{-0.87}$ & $1.92^{+0.38}_{-0.34}$ & $8.45^{+2.19}_{-1.94}$ & $0.64^{+0.13}_{-0.12}$ & $-2.03^{-2.03}_{-1.80}$ & $-0.15^{-0.09}_{+0.07}$ \\
        \hline
        $\mathcal{A}_7$ & $-3.69^{-4.22}_{-3.19}$ & $-144.0^{+27.5}_{-25.1}$ & $8.24^{-2.25}_{+1.98}$ & $-340^{-49.9}_{+46.4}$ & $16.7^{+4.35}_{-3.84}$ & $-79.8^{-11.1}_{-10.3}$ & $-4.02^{-4.02}_{-3.56}$ & $-5.58^{-3.54}_{+3.12}$ \\
        \hline
        $\mathcal{A}_8$ & $8.76^{+10.1}_{+7.47}$ & - & $0.16^{+0.045}_{-0.040}$ & $787^{+125}_{-116}$ & $0.40^{+0.10}_{-0.09}$ & $211^{+33.0}_{-30.6}$ & $-0.99^{-0.99}_{-0.87}$ & $-6.56^{-2.38}_{+2.13}$ \\
        \hline
        $\mathcal{A}_9$ & $-0.22^{+0.32}_{-0.14}$ & $-5.21^{+3.41}_{-3.00}$ & $-8.24^{+2.25}_{-1.98}$ & $-8.62^{+5.50}_{-4.85}$ & $-16.7^{-4.35}_{+3.84}$ & $-2.79^{-1.77}_{-1.56}$ & $4.02^{+4.02}_{-3.56}$ & $-160^{-22.1}_{+20.6}$ \\
        \hline
        $\mathcal{A}_{10}$ & $-0.23^{-0.31}_{-0.15}$ & - & $-0.33^{-0.09}_{+0.08}$ & $-20.5^{-7.51}_{+6.72}$ & $-0.80^{-0.21}_{+0.18}$ & $-6.56^{-2.38}_{-2.13}$ & $1.97^{+1.97}_{-1.74}$ & $846^{+132}_{-122}$ \\
        \hline
        $\mathcal{A}_{11}$ & $8.76^{+10.1}_{+7.47}$ & - & $0.16^{+0.045}_{-0.040}$ & $787^{+125}_{-116}$ & $0.40^{+0.10}_{-0.09}$ & $211^{+33.0}_{-30.6}$ & $-0.99^{+0.99}_{-0.87}$ & $-6.56^{-2.38}_{+2.13}$ \\
        \hline
        $\mathcal{A}_{12}$ & $0.59^{+0.75}_{+0.45}$ & $-5.25^{+1.07}_{+0.97}$ & $38.0^{+10.2}_{-9.00}$ & $-1.52^{+0.21}_{-0.13}$ & $76.8^{+19.7}_{-17.4}$ & $-0.35^{+0.10}_{-0.07}$ & $-18.2^{-18.2}_{-16.1}$  & $-0.51^{-0.33}_{+0.28}$ \\
        \hline
        $\mathcal{A}_{13}$ & $0.59^{+0.69}_{+0.49}$ & - & $1.52^{-0.41}_{-0.36}$ & $51.5^{+8.98}_{-8.27}$ & $3.69^{+0.95}_{-0.84}$ & $16.5^{+2.85}_{-2.63}$ & $-8.95^{+8.95}_{-7.92}$ & $-2.31^{-1.17}_{+1.04}$ \\
        \hline
        $\mathcal{A}_{14}$ & $-0.07^{+0.09}_{-0.04}$ & - & $-1.52^{-0.41}_{+0.36}$ & $-3.57^{-1.82}_{+1.61}$ & $-3.69^{-0.95}_{+0.84}$ & $-1.16^{-0.58}_{-0.52}$ & $8.95^{+8.95}_{-7.92}$ & $32.9^{+5.71}_{-5.26}$ \\
        \hline
        $\mathcal{A}_{15}$ & $0.08^{+0.11}_{+0.07}$ & - & $3.50^{+0.94}_{-0.82}$ & $1.95^{+0.39}_{-0.35}$ & $8.48^{+2.17}_{-1.92}$ & $0.64^{+0.13}_{-0.12}$ & $-2.03^{-2.03}_{-1.80}$ & $-0.15^{-0.09}_{+0.07}$ \\
        \hline
        $\mathcal{A}_{16}$ & $-0.62^{-0.78}_{-0.47}$ & $\approx 0.00$ & $-38.0^{-10.2}_{+9.00}$ & $-0.76^{-0.50}_{+0.41}$ & $-76.8^{-19.7}_{+17.4}$ & $-0.26^{-0.17}_{-0.14}$ & $18.2^{+18.2}_{-16.1}$ & $-0.70^{+0.20}_{-0.14}$ \\
        \hline
        $\mathcal{A}_{17}$ & $-0.07^{+0.09}_{-0.04}$ & - & $-1.52^{+0.41}_{+0.36}$ & $-3.57^{-1.82}_{+1.61}$ & $-3.69^{-0.95}_{+0.84}$ & $-1.16^{-0.58}_{-0.52}$ & $8.95^{-8.95}_{-7.92}$ & $32.9^{+5.71}_{-5.26}$ \\
        \hline
        $\mathcal{A}_{18}$ & $0.59^{+0.69}_{+0.49}$ & - & $1.52^{+0.41}_{-0.36}$ & $51.5^{+8.98}_{-8.27}$ & $3.69^{+0.95}_{-0.84}$ & $16.5^{+2.85}_{-2.63}$ & $-8.95^{-8.95}_{-7.92}$ & $-2.31^{-1.17}_{+1.04}$ \\
        \hline
        $\mathcal{A}_{19}$ & $-0.14^{-0.18}_{-0.11}$ & - & $-7.00^{-1.88}_{+1.65}$ & $-0.46^{-0.26}_{+0.22}$ & $-16.9^{-4.34}_{+3.84}$ & $-0.15^{-0.09}_{-0.07}$ & $4.05^{+4.05}_{-3.59}$ & $2.56^{+0.52}_{-0.47}$ \\
        \hline
        $\mathcal{A}_{20}$ & $0.08^{+0.11}_{+0.07}$ & - & $3.50^{+0.94}_{-0.82}$ & $1.95^{+0.39}_{-0.35}$ & $8.48^{+2.17}_{-1.92}$ & $0.64^{+0.13}_{-0.12}$ & $-2.03^{-2.03}_{-1.80}$ & $-0.15^{-0.09}_{+0.07}$ \\
        \hline
        $\mathcal{A}_{21}$ & - & $34.3^{+6.57}_{-6.00}$ & - & - & - & - & - & -  \\
        \hline
        $\mathcal{A}_{22}$ & - & $-1.24^{-0.81}_{+0.72}$ & - & - & - & - & - & -  \\
        \hline
        $\mathcal{A}_{23}$ & - & $1.24^{+0.81}_{-0.72}$ & - & - & - & - & - & - \\
        \hline
        $\mathcal{A}_{24}$ & - & $-34.3^{-6.57}_{+6.01}$ & - & - & - & - & - & -  \\
        \hline
        $\mathcal{A}_{25}$ & - & $1.25^{+0.25}_{-0.23}$ & - & - & - & - & - & -  \\
        \hline
        $\mathcal{A}_{26}$ & - & $\approx 0.00$ & - & - & - & - & - & -  \\
        \hline
        $\mathcal{A}_{27}$ & - & $\approx 0.00$ & - & - & - & - & - & -  \\
        \hline
        $\mathcal{A}_{28}$ & - & $-1.25^{-0.26}_{-0.23}$ & - & - & - & - & - & - \\
        \hline
    \end{tabular}
    }
    \caption{Values for the constants $\mathcal{A}_j$ for $B_{s}\to K^{\ast}\mu^{+}\mu^{-}$ observables in [$q^2_{\text{min}}$, 6] GeV$^2$ used in the expressions given in Eq. (\ref{Expressions}).}
    \label{tabcof1}
\end{table}

\begin{table}[H]
    \centering
    \renewcommand{\arraystretch}{1.5}
    \scalebox{0.785}{
    \begin{tabular}{||>{\columncolor{green!5}}c|c|c|c||>{\columncolor{green!5}}c|c|c|c|c|}
        \hline
        \textbf{Coeff.}  & $\langle I_{4} \rangle $ & $\langle I_{5} \rangle $ & $\langle I_{6s} \rangle $ & \textbf{Coeff.} & $Br_D$ & $A_{FB-D} \times 10^{-4}$ & $f_{L-D} \times 10^{-4}$ & $I_{i-D}$ \\
        \hline
        $\mathcal{A}_1$  & $23.5^{+8.47}_{-7.25}$ & $36.1^{+6.90}_{-6.38}$    & $7.31^{-0.92}_{-0.68}$    & $\mathcal{B}_1$  & 1 & $102^{+20.9}_{-18.9}$ & $51.2^{+10.5}_{-9.45}$ & $447^{+105}_{-93.7}$ \\
        \hline
        $\mathcal{A}_2$  & $-19.0^{-4.70}_{+4.18}$ & $-8.61^{-16.5}_{+1.52}$   & $-1.74^{+0.22}_{-0.16}$   & $\mathcal{B}_2$  & - & $-6.52^{-1.82}_{+1.59}$ & $-3.26^{-0.91}_{+0.79}$ & $-91.3^{-22.4}_{+19.9}$ \\
        \hline
        $\mathcal{A}_3$  & $2.26^{+0.56}_{-0.50}$ & -                        & -                       & $\mathcal{B}_3$  & - & $0.78^{+0.22}_{-0.19}$ & $0.38^{+0.11}_{-0.09}$ & $10.9^{+2.67}_{-2.37}$ \\
        \hline
        $\mathcal{A}_4$  & $19.0^{+4.70}_{-4.18}$ & $8.61^{+16.5}_{-1.52}$    & $0.93^{+0.58}_{-0.52}$   & $\mathcal{B}_4$  & - & $5.90^{+1.72}_{-1.49}$ & $2.95^{+0.86}_{-0.75}$ & $72.4^{+19.5}_{-17.2}$ \\
        \hline
        $\mathcal{A}_5$  & $-4.53^{-1.12}_{+0.99}$ & -                        & -                        & $\mathcal{B}_5$  & - & $-1.41^{-0.41}_{+0.36}$ & $-0.70^{-0.21}_{+0.18}$ & $-17.3^{+4.65}_{+4.11}$ \\
        \hline
        $\mathcal{A}_6$  & $2.26^{+0.56}_{-0.49}$  & -                        & -                       & $\mathcal{B}_6$  & - & $0.78^{+0.22}_{-0.19}$ & $0.38^{+0.11}_{-0.09}$ & $10.9^{+2.67}_{-2.37}$ \\
        \hline
        $\mathcal{A}_7$  & $168^{+34.6}_{+31.5}$ & $-308^{-61.5}_{+56.3}$& $-280^{+48.3}_{-44.6}$    & $\mathcal{B}_7$  & - & $-386^{-65.1}_{+60.0}$ & $-193^{-32.5}_{+30.0}$ & $-453^{+65.0}_{+60.7}$ \\
        \hline
        $\mathcal{A}_8$  & $7.92^{+1.65}_{-1.50}$ & -                        & -                       & $\mathcal{B}_8$  & - & $573^{+98.2}_{-90.4}$ & $287^{+49.1}_{-45.2}$ & $1076^{+171}_{-158}$ \\
        \hline
        $\mathcal{A}_9$  & $-168^{-34.6}_{-31.5}$ & $-308^{-61.5}_{+56.3}$& $-10.7^{-6.65}_{-5.95}$   & $\mathcal{B}_9$  & - & $-2.31^{-0.98}_{+0.85}$ & $-1.16^{-0.49}_{+0.43}$ & $-27.9^{+11.7}_{10.3}$ \\
        \hline
        $\mathcal{A}_{10}$  & $-15.9^{-3.29}_{+2.99}$ & -                     & -                        & $\mathcal{B}_{10}$  & - & $-2.17^{-0.96}_{+0.85}$ & $-1.09^{-0.48}_{+0.42}$ & $-28.3^{+10.3}_{+9.20}$ \\
        \hline
        $\mathcal{A}_{11}$ & $7.92^{+1.65}_{-1.50}$  & -                     & -                       & $\mathcal{B}_{11}$  & - & $573^{+98.2}_{-90.4}$ & $287^{+49.1}_{-45.2}$ & $1076^{+171}_{-158}$ \\
        \hline
        $\mathcal{A}_{12}$ & $6.79^{+2.26}_{-1.95}$ & $-17.1^{+3.48}_{+3.17}$ & $-21.5^{-4.19}_{-3.83}$   & $\mathcal{B}_{12}$  & - & $6.28^{+1.81}_{-1.58}$ & $3.14^{+0.90}_{-0.79}$ & $73.3^{+19.6}_{-17.3}$ \\
        \hline
        $\mathcal{A}_{13}$  & $36.8^{-7.56}_{+6.89}$ & -                     & -                        & $\mathcal{B}_{13}$  & - & $6.03^{+1.19}_{-1.09}$ & $3.01^{+0.59}_{-0.55}$ & $72.8^{+13.0}_{-11.9}$ \\
        \hline
        $\mathcal{A}_{14}$  & $-36.8^{+7.56}_{-6.89}$ & -                     & -                        & $\mathcal{B}_{14}$   & - & $-0.58^{-0.25}_{+0.22}$ & $-0.29^{-0.12}_{+0.11}$ & $-8.43^{+3.38}_{+2.99}$ \\
        \hline
        $\mathcal{A}_{15}$  & $2.26^{+0.56}_{-0.49}$  & -                     & -                        & $\mathcal{B}_{15}$ & - & $0.79^{+0.21}_{-0.18}$ & $0.39^{+0.10}_{-0.09}$ & $10.9^{+2.67}_{-2.37}$ \\
        \hline
        $\mathcal{A}_{16}$  & $-6.79^{+2.26}_{+1.95}$ & $-17.1^{+3.48}_{+3.17}$ & $\approx 0.00$            & $\mathcal{B}_{16}$  & - & $-7.56^{-2.04}_{+1.79}$ & $-3.78^{-1.02}_{+0.89}$ & $-76.4^{+20.0}_{+17.7}$ \\
        \hline
        $\mathcal{A}_{17}$  & $-36.8^{-7.56}_{+6.89}$  & -                     & -                       & $\mathcal{B}_{17}$  & - & $-0.58^{-0.25}_{+0.22}$ & $-0.29^{-0.12}_{+0.11}$ & $-8.43^{+3.38}_{+2.99}$ \\
        \hline
        $\mathcal{A}_{18}$  & $36.8^{-7.56}_{-6.89}$ & -                     & -                        & $\mathcal{B}_{18}$  & - & $6.03^{+1.19}_{-1.09}$ & $3.01^{+0.59}_{-0.55}$ & $72.8^{+13.0}_{-11.9}$ \\
        \hline
        $\mathcal{A}_{19}$  & $-4.53^{-1.12}_{+0.99}$ & -                     & -                        & $\mathcal{B}_{19}$  & - & $-1.42^{-0.39}_{+0.34}$ & $-0.71^{-0.19}_{+0.17}$ & $-17.3^{-4.62}_{+4.08}$ \\
        \hline
        $\mathcal{A}_{20}$  & $2.26^{+0.56}_{-0.50}$ & -                     & -                        & $\mathcal{B}_{20}$  & - & $0.79^{+0.21}_{-0.18}$ & $0.39^{+0.10}_{-0.09}$ & $10.9^{+2.67}_{-2.37}$  \\
        \hline
        $\mathcal{A}_{21}$ & - & $73.4^{+14.7}_{-13.4}$   & $66.7^{+11.5}_{-10.6}$   & $\mathcal{B}_{21}$ & - & - & - & -\\
        \hline
        $\mathcal{A}_{22}$  & - & $-73.4^{-14.7}_{-13.4}$  & $-2.54^{-1.59}_{-1.42}$ & $\mathcal{B}_{22}$  & - & - & - & -\\
        \hline
        $\mathcal{A}_{23}$ & - & $73.4^{+14.7}_{+13.4}$   & $2.54^{+1.59}_{-1.42}$   & $\mathcal{B}_{23}$  & - & - & - & -\\
        \hline
        $\mathcal{A}_{24}$ & - & $-73.4^{-14.7}_{+13.4}$  & $-66.7^{-11.5}_{-10.6}$ & $\mathcal{B}_{24}$ & - & - & - & -\\
        \hline
        $\mathcal{A}_{25}$ & - & $4.07^{+8.29}_{-0.76}$ & $5.12^{+0.99}_{-0.91}$   & $\mathcal{B}_{25}$  & - & - & - & -\\
        \hline
        $\mathcal{A}_{26}$  & - & $-4.07^{-8.29}_{+0.76}$& $\approx 0.00$            & $\mathcal{B}_{26}$  & - & - & - & -\\
        \hline
        $\mathcal{A}_{27}$  & - & $4.07^{+8.29}_{+0.76}$ & $\approx 0.00$            & $\mathcal{B}_{27}$  & - & - & - & -\\
        \hline
        $\mathcal{A}_{28}$  & - & $-4.07^{-8.29}_{+0.76}$& $-5.12^{-0.99}_{-0.91}$ & $\mathcal{B}_{28}$  & - & - & - & -\\
        \hline
    \end{tabular}
    }
    \caption{Values for the constants $\mathcal{A}_j$ and $\mathcal{B}_j$ for $B_{s}\to K^{\ast}\mu^{+}\mu^{-}$ observables in [$q^2_{\text{min}}$, 6] GeV$^2$ used in the expressions given in Eq. (\ref{Expressions}).}
    \label{tabcof2}
\end{table}


\end{document}